\documentclass[reprint,showpacs,preprintnumbers,amsmath,amssymb,aps,prd,nofootinbib,floatfix,longbibliography]{revtex4-1}

\usepackage{bbold}
\usepackage{hyperref}

\usepackage{graphicx}

\usepackage{color}
\def\comment#1{{#1}}
\def\highlight#1{{#1}}

\newcommand{\appsection}[1]{\section{\uppercase{#1}}}

\begin{document}

\title{Gravitational redshift in quantum-clock interferometry}

\author{Albert Roura}
%\email{albert.roura@uni-ulm.de}
\affiliation{Institut f\"ur Quantenphysik, Universit\"at Ulm, Albert-Einstein-Allee 11,
89081 Ulm, Germany}

\date{\today}
\begin{abstract}
The creation of delocalized coherent superpositions of quantum systems experiencing different relativistic effects is an important milestone in future research at the interface of gravity and quantum mechanics. This could be achieved by generating a superposition of quantum clocks that follow paths with different gravitational time dilation and investigating the consequences on the interference signal when they are eventually recombined.
Light-pulse atom interferometry with elements employed in optical atomic clocks is a promising candidate % / technique
for that purpose, but suffers from major challenges including its insensitivity
%the insensitivity of such interferometers
to the gravitational redshift in a uniform field.
All these difficulties can be overcome with a novel scheme presented here which is based on initializing the clock when the spatially separate superposition has already been generated and performing a doubly differential measurement
%\comment{that compares the outcomes for different initialization times.}
\comment{where the differential phase shift between the two internal states is compared
for different initialization times.}
This can be exploited to test the universality of the gravitational redshift (UGR) with delocalized coherent superpositions of quantum clocks and it is %will be
argued that its experimental implementation should be feasible with a new generation of 10-meter atomic fountains that will soon become available.
Interestingly,
the approach also offers significant advantages for more compact set-ups based on guided interferometry or hybrid configurations.
Furthermore, in order to provide % / lay down
a solid foundation for the analysis of the various interferometry schemes and the effects that can be measured with them, a general formalism for a relativistic description of atom interferometry in curved spacetime is developed. It can deal with freely falling atoms, but also include the effects of external forces and guiding potentials, and can be applied to a very wide range of situations.
% / phenomena)) beyond quantum-clock interferometry
\comment{As an important ingredient for quantum-clock interferometry, suitable diffraction mechanisms for atoms in internal-state superpositions are investigated too.} % / explored.
Finally, the relation of the proposed doubly-differential measurement scheme to other experimental approaches and to tests of the universality of free fall (UFF) is discussed in detail.
\end{abstract}


\maketitle

\section{Introduction}
\label{sec:introduction}

In this article a general formalism
%In this article I will develop a general formalism capable of
describing relativistic effects in atom interferometry for atoms propagating in curved spacetime is developed.
This is then exploited in Sec.~\ref{sec:LPAI_redshift} to present a novel scheme for quantum-clock interferometry which is sensitive to gravitational-redshift effects and whose experimental implementation should be within reach of the 10-meter atomic fountains of 
%feasible to implement with 10-meter atomic fountains
%employing Sr or Yb atoms that will soon become available.
Sr and Yb atoms that will soon become available
in Stanford and HITec (Hanover) respectively.
%\comment{Moreover, besides light-pulse atom interferometers, the scheme can also be used in more compact set-ups based on guided interferometry or hybrid configurations.}

%central result ... Sec.~\ref{sec:LPAI_redshift} ...

Remarkable advances in atom interferometry have enabled the creation of macroscopically delocalized quantum superpositions with atomic wave packets separated up to half a meter \cite{kovachy15b}. Nevertheless, in all cases realized so far the differences in the dynamics of the two wave packets of the superposition can be entirely described in terms of Newtonian mechanics.
%It is true that
And while the impressive precision of atomic clocks based on optical transitions enable the measurement of the gravitational redshift for height differences as little as one centimeter, %but
this is achieved by comparing two independent clocks.
In contrast, creating delocalized coherent superpositions of quantum systems experiencing different relativistic effects remains an important milestone in future research at the interface of gravity and quantum mechanics.

This could be achieved by generating a superposition of quantum clocks that follow paths with different gravitational time dilation and investigating the consequences on the interference signal when they are eventually recombined.
More specifically, the proper-time differences between the two interferometer arms imprint which-way information on the internal state of the clock which reduces the visibility of the observed interference \cite{zych11}.
%causes a decrease of 
Both optical atomic clocks \cite{chou10,matveev13,pruttivarasin15,ludlow15,poli14} and light-pulse atom interferometers \cite{kasevich91,peters99,rosi14,bouchendira13,parker18,fray04,bonnin13,schlippert14,zhou15,rosi17b,asenbaum17} have demonstrated their ability to carry out high-precision measurements in a wide range of applications. Therefore, a natural possibility is to perform light-pulse atom interferometry with the same atomic species employed in optical atomic clocks and prepare them in a superposition of the two internal states involved in the clock transition.
Unfortunately, as explained in Sec.~\ref{sec:light-pulse_interf}, quantum-clock interferometers based on this kind of set-up suffer from major challenges, including their insensitivity to gravitational time dilation in uniform fields and the differential recoil for the two internal states.
Furthermore, even if they were sensitive to the gravitational redshift, the parameter ranges typically attainable would lead to rather small changes of \comment{visibility (also known %referred to
as interferometer contrast)} which would be very difficult to measure, partly because other effects leading to contrast fluctuations and contrast reduction would mask such small changes.
%especially given that there are many other effects that lower the visibility and would mask the signal of interest.

In this article I will present a promising scheme for quantum-clock interferometry which overcomes all these difficulties and is sensitive to the gravitational redshift in a uniform field. The key idea is to consider an adjustable time for clock initialization
%, which is applied when the spatially separate superposition has already been generated,
and perform
%%clock-initialization times %It is based on 
a doubly differential measurement comparing the outcomes for different initialization times (defined with respect to the laboratory frame).  More specifically, one measures the differential phase-shift between the two internal states for a given initialization time and then performs a second differential phase-shift measurement for a different initialization time but leaving everything else unchanged. The difference between the two measurements is directly related to the different gravitational time dilation experienced by wave packets at different heights.

%By considering dilaton models as a consistent framework for parametrizing violations of the equivalence principle, it will be shown that this scheme can be employed to test the universality of the gravitational redshift (UGR) with spatially delocalized quantum superpositions.
This scheme can be employed to test the \comment{UGR}
%\comment{universality of the gravitational redshift (UGR)}
with spatially delocalized quantum superpositions as will be shown with the example of dilaton models, which provide a consistent framework for parametrizing violations of the equivalence principle.
Moreover, it will be argued that this kind of experiments should be feasible to implement with the 10-meter atomic fountains employing Sr or Yb atoms that will soon become available.
%Sr and Yb atoms that will soon become available
%in Stanford and HITec (Hanover) respectively.
%
Interestingly, besides light-pulse atom interferometers, the scheme can also
be used in more compact set-ups based on guided interferometry or hybrid configurations.

In order to lay down %provide
a solid foundation for the analysis of the various quantum-clock interferometry schemes and the relativistic effects that can be measured with them, several sections and appendices will be devoted to the formulation and derivation of a general formalism for a relativistic description of atom interferometry in curved spacetime.
A similar result for the propagation phase in the relativistic case had been previously obtained based on a semicalssical ansatz and restricted to freely falling particles \cite{dimopoulos08a,linet76}. Here we will provide instead a clean derivation of both the propagation phase and the full wave packet evolution which is valid not only for freely falling particles but also in presence of external forces and guiding potentials.
Furthermore, the formalism
will be applied to extensions of general relativity involving dilaton models and
to the discussion of related experimental approaches as well as the effect of gravity gradients on the proper-time difference between the two interferometer arms.

Note that although in the paper we will mainly focus on examples of nearly uniform gravitational fields, this %the above
formalism is applicable to %is valid for %holds for
general spacetimes and can also
be employed, for example, for a detailed investigation of the effects of gravitational waves on matter waves.
Furthermore, as an additional by-product the proposed diffraction mechanisms for atoms in internal-state superpositions can be %additionally
exploited in tests of the \comment{UFF} %\comment{universality of free fall (UFF)}
with superposition states such as those reported in Ref.~\cite{rosi17b} but involving optical rather than hyperfine transitions.

The rest of the paper is organized as follows. After introducing the basic aspects of the quantum-clock model in Sec.~\ref{sec:clock_model}, the general formalism describing the evolution of atomic wave packets in curved space time is presented in Sec.~\ref{sec:propagation}. It can deal with freely falling atoms, but also with external forces and even guided propagation. Moreover, it can be integrated into a relativistic description of full atom-interferometer sequences, which is then employed in Sec.~\ref{sec:clock_interferometry} to discuss important aspects of quantum-clock interferometry. There the main limitations of light-pulse atom interferometers in this context, including their insensitivity to the gravitational redshift in a uniform field, are explained and compared to the case of guided interferometry. %interferometers
Next, the novel scheme based on a doubly differential measurement comparing different initialization times and which overcomes these difficulties is presented in Sec.~\ref{sec:LPAI_redshift}. It is argued that its implementation should be feasible in the 10-meter atomic fountains employing Sr or Yb atoms that will soon become available, and it is shown that it can be additionally used in guided and hybrid interferometry, where it offers significant advantages too.
In Sec.~\ref{sec:UGR_UFF} dilaton models are considered as a consistent framework for investigating violations of the equivalence principle and it is shown that the proposed quantum-clock interferometry scheme can directly test the UGR.
%gravitational redshift (UGR).
Furthermore, the relation to other approaches and to violations of the UFF
%universality of free fall (UFF)
is also discussed in detail. 
%close connection between violations of UGR and the universality of free fall (UFF) and the
Finally, this is followed by the conclusions in Sec.~\ref{sec:conclusion}.

%On the other hand,
The technical details for a number of important issues are addressed in several Appendices.
The Fermi-Walker frame and the associated coordinates are presented in Appendix~\ref{sec:fermi-walker}. They are exploited in Appendix~\ref{sec:wp_propagation} to derive the evolution of atomic wave packets in a general curved spacetime. This is first done for freely falling atoms and then including the effects of external forces and guiding potentials. A relativistic description for the state evolution in an atom interferometer is provided in Appendix~\ref{sec:AI}, where the effect of the laser pulses is analyzed and the connection between the separation phase and the proper-time differences in different frames for open interferometers is elucidated.
The two-photon pulse for the clock initialization and the implications for the proposed quantum-interferometry scheme are investigated in Appendix~\ref{sec:two-photon_pulse}, whereas the possible diffraction mechanisms for atoms in internal-state superpositions are considered in Appendix~\ref{sec:diffraction_superpos}.
The effects of gravity gradients on the proper-time difference for light-pulse atom interferometers are analyzed in Appendix~\ref{sec:gravity_gradient}, where we show that the measurements of tidal-force effects on delocalized quantum superpositions reported in Ref.~\cite{asenbaum17} can be alternatively interpreted in terms of such proper-time differences.
Finally, Appendix~\ref{sec:QFTtoQM} outlines how the formulation based on single-particle relativistic quantum mechanics employed throughout the paper can be derived from quantum field theory (QFT) in curved spacetime and \comment{establishes} under what conditions this is possible.

%... remarks on notation ...

Throughout the paper we use the Einstein summation convention for repeated indices and the (+, +, +) sign conventions of Ref.~\cite{misner73}, which includes a positive signature for the metric.
Greek indices range over space and time while Latin indices denote spatial components only. Moreover, vector and matrix notation with vectors denoted by boldface characters is often employed for the spatial components.

%\pagebreak[4]
%\newpage
\vspace{0.5cm}

\section{Quantum-clock model}
\label{sec:clock_model}

\subsection{Two-level atom}
\label{sec:2-level_atom}

%Relativistic field-theory action suitable for second quantization $\rightarrow$ Appendix

%Relativistic action for two particles with different masses.

%(Spontaneous decay neglected... )

\comment{As a model for the quantum clock} we will consider atoms characterized by their center-of-mass (COM) motion and their internal structure, represented by the two electronic energy levels $| \mathrm{g} \rangle$ and $| \mathrm{e} \rangle$ that will play a relevant role in our analysis. In absence of electromagnetic radiation driving transitions between the two levels, the Hamiltonian operator $\hat{H}$ governing the dynamics of such a two-level atom consists of two contributions, one for each internal state:
%comprises the two Hamiltonians for the evolution of the atom in each internal state:
\begin{equation}
\hat{H} \, = \, \hat{H}_1 \otimes | \mathrm{g} \rangle \langle \mathrm{g} | \,+\, \hat{H}_2 \otimes | \mathrm{e} \rangle \langle \mathrm{e} |
\label{eq:hamiltonian1} ,
\end{equation}
where $\hat{H}_1$ and $\hat{H}_2$ are the Hamiltonian operators for the COM dynamics of an atom in the $| \mathrm{g} \rangle$ and $| \mathrm{e} \rangle$ internal states.
%respectively.
They are associated with the classical actions
\begin{align}
S_n \big[ x^\mu(\lambda) \big] &= -m_n c^2 \int d \tau \nonumber \\
&= -m_n c \int d\lambda \, \sqrt{-g_{\mu\nu} \frac{dx^\mu}{d\lambda}\frac{dx^\nu}{d\lambda}}
\quad \text{with}\ n = 1,2
\label{eq:action1}\, ,
\end{align}
describing the motion of relativistic massive particles in a spacetime with metric $g_{\mu\nu}$.
The parameters $m_1 = m_\text{g}$ and $m_2 = m_\text{g} + \Delta m$ correspond to the rest mass of an atom in the ground and excited states respectively, and $\Delta m = \Delta E / c^2$ is directly related to the energy difference $\Delta E$ between the two internal states.

Several remarks are in order. %concerning Eqs.~\eqref{eq:hamiltonian1}--\eqref{eq:action1}.
Firstly, as explicitly indicated in Eq.~\eqref{eq:action1}, the classical action is proportional to the proper time along the worldline $x^\mu(\lambda)$ corresponding to the classical spacetime trajectory. \comment{Secondly, although the action is reparametrization invariant, i.e.\ invariant under changes of the worldline parameter $\lambda$, throughout the paper we will typically fix this freedom by choosing the parameter to coincide with the time coordinate within the %various %reference frames
coordinate system under consideration in each case,} so that the spacetime trajectory $x^\mu(t) = \big(c\, t, \mathbf{x}(t) \big)$ is entirely determined by its spatial part $\mathbf{x}(t)$.
\comment{Thirdly, we will assume that the lifetime of the excited state $| \mathrm{e} \rangle$ is much longer than the total evolution time, so that spontaneous decay can be neglected throughout out the analysis.}

%\comment{(fixing of the reparametrization invariance)}
%
%\comment{(internal energy included)
%(spontaneous decay neglected ...)}

For non-relativistic COM motion in a weak gravitational field \highlight{generated by Newtonian sources (with non-relativistic motion)}, the action in Eq.~\eqref{eq:action1} reduces to
%\begin{align}
%S_j \big[ \mathbf{x}(t) \big] &= \int dt \, \left(  -m_j c^2 + \frac{1}{2} m_j \mathbf{v}^2
%- m_j \,U(\mathbf{x}) \right) \nonumber \\
%&= \int dt \, \left(  -m_j c^2 + \frac{1}{2} m_j \mathbf{v}^2 \right. \nonumber \\
%&\qquad\qquad\quad \left. - m_j \, \Big( U_0 + \mathbf{g} \cdot \mathbf{x}
%- \frac{1}{2} \mathbf{x}^\mathrm{T} \Gamma \mathbf{x} \Big) \right)
%\label{eq:action2} ,
%\end{align}
\begin{equation}
S_n \big[ \mathbf{x}(t) \big] = \int dt \, \left(  -m_n c^2 + \frac{1}{2} m_n \mathbf{v}^2
- m_n \,U(\mathbf{x}) \right)
\label{eq:action2} ,
\end{equation}
where we will often consider an expansion up to quadratic order of the gravitational potential $U(\mathbf{x})$ around a given point $\mathbf{x}_0$,
\begin{equation}
U(\mathbf{x}) = U_0 - \mathbf{g} \cdot (\mathbf{x} - \mathbf{x}_0)
- \frac{1}{2} (\mathbf{x} - \mathbf{x}_0)^\mathrm{T} \, \Gamma \, (\mathbf{x} - \mathbf{x}_0)
\label{eq:grav_potential1} ,
\end{equation}
in terms of the gravitational acceleration $\mathbf{g}$ at that point and the gravity gradient tensor $\Gamma$.
In deriving Eq.~\eqref{eq:action2} a metric of the form
\begin{equation}
%ds^2 =
g_{\mu\nu}\, dx^\mu dx^\nu = \, - \big( c^2 + 2\hspace{0.2ex}U(\mathbf{x}) \big) \, dt^2
+ \big(1 - 2\hspace{0.2ex}U(\mathbf{x}) / c^2 \big) \, d\mathbf{x}^2
\label{eq:metric1} ,
\end{equation}
has been considered and terms of higher order in $\mathbf{v}^2 / c^2$ and $U(\mathbf{x})$ have been neglected. In fact, the dependence on $U(\mathbf{x})$ of the spatial components of the metric, which gives rise to terms of order $(\mathbf{v}^2 / c^2) \big( U(\mathbf{x})/c^2 \big)$, does not contribute at this order either.
\comment{Eqs.~\eqref{eq:action2}--\eqref{eq:grav_potential1} can be immediately %straightforwardly
generalized to time-dependent gravitational fields such as those sourced %generated
by a time-dependent mass distribution: one simply needs to include a time-dependence for the gravitational potential $U(\mathbf{x},t)$ as well as the expansion coefficients $U_0 (t)$, $\mathbf{g} (t)$ and $\Gamma (t)$.
Moreover, one can also include the effects of non-gravitational external forces by adding an external potential as explained in Sec.~\ref{sec:propagation_forces}.}

\comment{Note that whereas Eq.~\eqref{eq:action2} is restricted to the special case of non-relativistic motion and weak fields, the formalism that will be presented in Sec.~\ref{sec:propagation} and Appendices~\ref{sec:fermi-walker}--\ref{sec:AI} holds for arbitrary relativistic motion in a general curved spacetime.}

%\comment{... time dependence ...} of the gravitational potential $U(\mathbf{x},t)$ as well as the expansion coefficients $U_0 (t)$, $\mathbf{g} (t)$ and $\Gamma (t)$.

%\highlight{Comment on relativistic definition of COM \cite{toros,...} ...}

\subsection{Clock initialization and read-out}
\label{sec:clock_initialization}

Atomic clocks are typically operated in such a way that the phases accumulated by the two internal states differ by a term proportional to the proper time $\tau$ and the energy difference $\Delta E$. This is indeed the case for freely falling atoms launched in an atomic fountain or for atoms trapped in an optical lattice with a magic wavelength, as further discussed in the next section.

The transitions between the two internal states are driven by coherent electromagnetic radiation that resonantly couples both states. In particular, when starting with atoms in the ground state, the clock is \emph{initialized} by applying a pulse with suitably chosen amplitude and duration, a so-called $\pi/2$ pulse, that creates an equal amplitude superposition of the two states:
\begin{equation}
| \mathrm{g} \rangle \to \frac{1}{\sqrt{2}} \Big( | \mathrm{g} \rangle - i\, e^{i \varphi_1} e^{-i \omega t_1} | \mathrm{e} \rangle \Big)
\label{eq:initialization1} ,
\end{equation}
where $t_1$ characterizes the time when the pulse is applied in terms of the time coordinate $t$ for the reference frame naturally associated with the pulse generation, $\varphi_1$ is the pulse phase and $\omega$ is the angular frequency of the pulse in that frame%
\footnote{Note that for two-photon processes like those considered in Sec.~\ref{sec:LPAI_redshift} it corresponds to the sum of the single-photon frequencies: $\omega = \omega_1 + \omega_2$.}.
After evolving for some proper time $\Delta \tau = (\tau - \tau_1)$, the state becomes
\begin{equation}
\big| \Phi (\tau) \big\rangle = \frac{e^{- i m_\text{g} c^2 \Delta\tau / \hbar}}{\sqrt{2}}
\left( | \mathrm{g} \rangle - i\, e^{i \varphi_1} e^{-i \omega t_1}
e^{- i \Delta E \, \Delta\tau / \hbar} | \mathrm{e} \rangle \right)
\label{eq:evolution1} .
\end{equation}

Finally, the clock is typically read out by applying at some time $t_2$ a second $\pi/2$ pulse that recombines the two internal states and leads to a quantum state with the following amplitudes:
\begin{equation}
\begin{aligned}
\left| \langle \mathrm{g} \big| \Phi (\tau_2) \big\rangle \right| &= %&\propto
\frac{1}{\sqrt{2}} \left| 1 - e^{- i \delta \varphi} e^{i \omega \Delta t} e^{- i \Delta E \, \Delta\tau / \hbar} \right| , \\
\left| \langle \mathrm{e} \big| \Phi (\tau_2) \big\rangle \right| &= %&\propto
\frac{1}{\sqrt{2}} \left| 1 + e^{- i \delta \varphi} e^{i \omega \Delta t} e^{- i \Delta E \, \Delta\tau / \hbar} \right| ,
\end{aligned}
\label{eq:read-out1} 
\end{equation}
with $\delta \varphi = \varphi_2 - \varphi_1$, $\Delta t = t_2 - t_1$ and $\Delta \tau = \tau_2 - \tau_1$.
The resulting interference leads to oscillations in the number of atoms in the ground and excited state as a function of the pulse frequency $\omega$, which is the basis of the Ramsey spectroscopy method. It can be used to tie the pulse frequency $\omega$ to the energy difference $\Delta E$ between the two atomic levels and this is, in turn, employed to stabilize the clock's local oscillator to which the pulse frequency is referenced.
Whenever $\Delta \tau = \Delta t$, for example for atoms trapped in an optical lattice and at rest in the reference frame where the interrogating laser pulses are generated, the angular frequency $\omega$ can be directly linked to the energy $\Delta E$ of the atomic transition divided by $\hbar$.
In general, however, the number of atoms in each state after the read-out pulse oscillates as a function of $\big( \omega - (\Delta E / \hbar)(\Delta \tau / \Delta t) \big)$, i.e.\
\begin{equation}
P_{| \mathrm{g} \rangle} = 1 - P_{| \mathrm{e} \rangle}
= P \Big( \omega - (\Delta E / \hbar)(\Delta \tau / \Delta t) \Big)
\label{eq:read-out2} ,
\end{equation}
so that $\omega$ is linked to $\Delta E / \hbar$ times the redshift factor $\Delta \tau / \Delta t$.

Note that in contrast to the usual operation of atomic clocks based on Ramsey spectroscopy,
in Secs.~\ref{sec:clock_interferometry}-\ref{sec:UGR_UFF} we will consider the state evolution as a function of proper time \emph{without} applying the final read-out pulse. Instead, we will be interested in the decrease of the quantum overlap between clock states evolving along two interferometer branches with different proper times.

%\highlight{recoil ...} %star

\section{Free/guided propagation and gravitational redshift}
\label{sec:propagation}

\comment{The evolution of an atomic wave packet can be conveniently described in terms of its \emph{central trajectory}, which satisfies the classical equation of motion, and a \emph{centered wave packet} accounting for its expansion dynamics. This has been previously established for non-relativistic atoms \cite{borde92,antoine03b,hogan08,roura14}. Its generalization to the relativistic case for matter waves propagating in a general curved spacetime is derived in Appendix~\ref{sec:wp_propagation} and the main results are presented in the next subsections.}

\subsection{Free propagation of the quantum clock}
\label{sec:free_propagation}

As explained in Appendix~\ref{sec:free_prop}, the state evolution for an atom \emph{freely falling} in a gravitational field can be naturally described in terms of the \emph{Fermi normal coordinates} associated the spacetime geodesic $X^\mu (\lambda)$ followed by the central position of the atomic wave packet, which is at rest in that frame. The time coordinate $\tau_\text{c}$ in this comoving frame coincides with the proper time along the central trajectory, whose Fermi coordinates are simply $X^\mu(\tau_\text{c}) = \big(c\, \tau_\text{c}\, , \mathbf{0} \big)$. \comment{In turn,} the phase accumulated by the wave packet, which is given by the action in Eq.~\eqref{eq:action1} evaluated along the central trajectory, reduces to
\begin{equation}
S_n \big[ X^\mu(\lambda) \big] = -m_n c^2 \int d \tau_\text{c}
\label{eq:action_free1}\ .
\end{equation}
Moreover, in this frame the evolution of the centered wave packet $\big| \psi_\text{c}^{(n)} (\tau_\text{c}) \big\rangle$ for each internal state is governed by the Schr\"odinger equation with the Hamiltonian operator
\begin{equation}
\hat{H}_\text{c}^{(n)} = \frac{1}{2m_n}\, \hat{\mathbf{p}}^2
- \frac{m_n}{2}\, \hat{\mathbf{x}}^\text{T} \, \Gamma (\tau_\text{c})\, \hat{\mathbf{x}}
\label{eq:hamiltonian_free1}\ ,
\end{equation}
where the gravity gradient tensor $\Gamma (\tau_\text{c})$ is directly related to the certain components of the spacetime curvature in Fermi coordinates evaluated on the central trajectory: \comment{$\Gamma_{ij} (\tau_\text{c}) = - c^2 R_{0i0j} (\tau_\text{c},\mathbf{0})$.}
%((More specifically, one has
%\comment{$\Gamma_{ij} (\tau_\text{c}) = - c^2 R_{0i0j} (\tau_\text{c},\mathbf{0})$.}))
%
These results are obtained after two approximations discussed in \comment{Appendix~\ref{sec:free_prop}} which are valid for a sufficiently small \highlight{wave packet size $\Delta x$} compared to the characteristic \highlight{curvature radius}%
\footnote{\comment{As explained in Appendices~\ref{sec:fermi-walker} and \ref{sec:wp_propagation}, in some cases it may be necessary to include small anharmonicities associated with higher multipoles of the gravitational field, which are dominated by contributions from the local mass distribution.}}
and for centered wave packets with non-relativistic momentum components.

%\comment{approxim ... non-relativistic momentum components (no contribution from $0i$ and $ij$ metric components) ...}

\highlight{While these coordinates are particularly convenient when calculating the evolution of the centered wave packet, other suitable coordinate systems can be employed to find the explicit result for the central trajectory $X^\mu (\lambda)$ as a solution of the classical equation of motion associated with the action in Eq.~\eqref{eq:action1}. One can then calculate the proper time and the propagation phase along the trajectory by evaluating Eq.~\eqref{eq:action1} for that solution.}

%\highlight{((Remarks on the calculation of the central trajectory and the corresponding action in any other (possibly more convenient) coordinate system ...))}

Fermi coordinates have been previously used for studying the expansion of a freely falling BEC in a Schwarzschild spacetime  \cite{gabel18} as well as the effect of gravitational waves in an atom interferometer \cite{roura04} or in the evolution of a trapped BEC \cite{hartley18}.
%\highlight{Somewhere else ??}

\subsection{Propagation in presence of external forces}
\label{sec:propagation_forces}

In a realistic situation, however, there are small but non-vanishing \emph{external forces} acting on a freely falling atom (e.g.\ due to spurious magnetic field gradients) and the central position of the atomic wave packets no longer follows a spacetime geodesic but an accelerated trajectory $X^\mu (\lambda)$. Fortunately, the formalism presented above for the freely falling case can be generalized to this situation, as shown in Appendix~\ref{sec:external_forces}, by considering the \emph{Fermi-Walker coordinates} associated with the accelerated trajectory. In terms of these coordinates, detailed in Appendix~\ref{sec:fermi-walker}, the central trajectory is given by $X^\mu(\tau_\text{c}) = \big(c\, \tau_\text{c}\, , \mathbf{0} \big)$ and the time coordinate $\tau_\text{c}$ coincides with the proper time along the trajectory, with four-velocity $U^\mu = dX^\mu / d\tau = (c, \mathbf{0} )$ and non-vanishing acceleration $U^\nu \, \nabla_\nu U^\mu = \big(0 , \mathbf{a}(\tau_\text{c}) \big)$.
%which exhibits a non-vanishing
%\comment{acceleration $D^2X^\mu / d\tau^2 = \big(0 , \mathbf{a}(\tau_\text{c}) \big)$.}

For simplicity we will model the external forces acting on the atom by adding to the right-hand side of Eq.~\eqref{eq:action1} the proper-time integral, with a minus sign, of a potential $V_n (x^\mu)$. The classical action that corresponds to the phase accumulated by the wave packets becomes then
\begin{equation}
S_n \big[ X^\mu (\lambda) \big] = -m_n c^2 \int d \tau_\text{c}
\,-\, \int d \tau_\text{c}\, V_n ( \tau_\text{c},\mathbf{0} )
\label{eq:action_accel1}\ ,
\end{equation}
and $V_n ( \tau_\text{c},\mathbf{x} )$ is here the potential characterizing the external forces evaluated in the Fermi-Walker frame, where the wave packet is at rest.
The gradient of this potential is directly related to the acceleration of the central trajectory as follows:
\begin{equation}
\mathbf{a} (\tau_\text{c}) = - \frac{(\boldsymbol{\nabla} V_n) ( \tau_\text{c},\mathbf{0} )}
{m_n + V_n ( \tau_\text{c},\mathbf{0} ) /  c^2}
\label{eq:accel1}\ .
\end{equation}
%, which follows from the equation of motion for $x^\mu(\lambda)$ derived from the
%classical action in Eq.~\eqref{eq:action1} plus the potential term.
On the other hand, the second derivatives of the potential contribute to the expansion dynamics of the wave packet.
Indeed, for locally harmonic potentials (which can be well approximated by a quadratic function within the size of the wave packet) the dynamics of the centered wave packets $\big| \psi_\text{c}^{(n)} (\tau_\text{c}) \big\rangle$ in the Fermi-Walker frame is governed by a Schr\"odinger equation with the Hamiltonian operator
\begin{equation}
\hat{H}_\text{c}^{(n)} = \frac{1}{2m_n}\, \hat{\mathbf{p}}^2
+ \frac{1}{2}\, \hat{\mathbf{x}}^\text{T} \Big( \mathcal{V}^{(n)} (\tau_\text{c})
- m_n \Gamma (\tau_\text{c}) \Big)\, \hat{\mathbf{x}}
\label{eq:hamiltonian_accel1}\ ,
\end{equation}
where $\mathcal{V}_{ij}^{(n)} (\tau_\text{c})
= \, \partial_i \partial_j V_n (\tau_\text{c},\mathbf{x}) \, \big|_{\mathbf{x} = \mathbf{0}}$.

\comment{As done in the previous subsection, the two approximations based on the small wave-packet size compared to the curvature radius and the non-relativistic momenta of the centered wave packet have been used when deriving Eq.~\eqref{eq:hamiltonian_accel1}. Moreover, in this case an additional approximation discussed in Appendix~\ref{sec:external_forces} and relying on the condition $|\mathbf{a} \cdot \Delta \mathbf{x}|/c^2 \ll 1$ has also been employed.}

\subsection{Guided propagation}
\label{sec:guided_propagation}

The approach introduced in the previous subsection in order to account for external forces can describe the effect  not only of small spurious forces but also of the much stronger ones employed for guiding the propagation of the atomic wave packets.
As a simple illustration we will analyze the example of a static trapped configuration in an approximately uniform  gravitational field.

Let us consider a harmonic trapping potential 
\begin{align}
V_n (\mathbf{x}) &= V^{(n)}_0
+ \frac{1}{2}\, (\mathbf{x} - \mathbf{x}_0 )^\text{T}\, \mathcal{V}^{(n)} (\mathbf{x} - \mathbf{x}_0 ) \nonumber \\
&=  V^{(n)}_0 + \frac{1}{2}\, m_n\, (\mathbf{x} - \mathbf{x}_0 )^\text{T}\, \Omega_n^2\, (\mathbf{x} - \mathbf{x}_0 )
\label{eq:trapping1} ,
\end{align}
where we have introduced for later convenience the frequency matrix $\Omega_n$ defined by $m_n\, \Omega_n^2 = \mathcal{V}^{(n)}$,
and let us assume that we are in a regime where the classical action is well approximated by the non-relativistic expression in Eq.~\eqref{eq:action2}, from which one can obtain %solve for
the central trajectory and calculate the propagation phase.
For a static central trajectory, the kinetic term does not contribute and the combination of gravitational and external potentials becomes
\begin{align}
m_n\, U (\mathbf{x}) + V_n (\mathbf{x}) &=
m_n \big(U_0 - \mathbf{g} \cdot \Delta\mathbf{x}_n \big) + V^{(n)}_0 + \Delta V^{(n)}_0
%+ \frac{1}{2}\, m_n\, \mathbf{g} \cdot \Delta\mathbf{x}_n
 \nonumber \\
&\quad + \frac{1}{2}\, m_n\, (\mathbf{x} - \bar{\mathbf{x}}_0 )^\text{T}\, \Omega_n^2\, (\mathbf{x} - \bar{\mathbf{x}}_0 )
\label{eq:trapping2} ,
\end{align}
where $\bar{\mathbf{x}}_0 = \mathbf{x}_0 + \Delta\mathbf{x}_n$ with $\Delta\mathbf{x}_n = \Omega_n^{-2} \mathbf{g}$ and $\Delta V^{(n)}_0 = 
(m_n / 2)\, \Delta\mathbf{x}_n^\text{T}\, \Omega_n^2\, \Delta\mathbf{x}_n =
m_n\, \mathbf{g} \cdot \Delta\mathbf{x}_n / 2$.

Thus, we see that the effect of the uniform gravitational field is to shift the equilibrium position by $\Delta\mathbf{x}_n$ from the minimum of the external potential at $\mathbf{x} = \mathbf{x}_0$. The shift will be equal for both internal states (i.e.~$\Delta\mathbf{x}_1 = \Delta\mathbf{x}_2 \equiv \Delta\mathbf{x}$) if one adjusts $\mathcal{V}^{(n)}$ for the two states so that $\Omega_1 = \Omega_2$. Such a choice also guarantees that the COM energy for the two states is the same ($\hbar\, \Omega_1 / 2 = \hbar\, \Omega_2 / 2$) if the trapped wave packets find themselves in the ground state. This implies that the difference between the phases accumulated by the two internal states with centered wave packets in the ground state of the trapping potential is independent of this external potential and given by
\begin{equation}
\big( S_2 - S_1 \big) / \hbar = \Delta m\, c^2 \Delta\tau / \hbar
\label{eq:trapped_phase_diff1}\ ,
\end{equation}
provided that $V^{(1)}_0 = V^{(2)}_0$.
%Eq.~\eqref{eq:trapped_phase_diff1}
The result follows after neglecting the contribution from $\Delta V^{(2)}_0 - \Delta V^{(1)}_0$, which amounts to $\Delta m\, \mathbf{g} \cdot \Delta\mathbf{x}$ and could be resolved by the most precise optical atomic clocks to date only for $\Delta x \gtrsim 1\, \text{cm}$, whereas the spatial shifts typically induced are much smaller than that.

This simple example closely resembles the situation for atomic clocks based on optical transitions of neutral atoms trapped in an optical lattice generated by counter-propagating beams with a ``magic'' wavelength \cite{ludlow15}. For instance, for sufficiently deep blue-detuned 3-D lattices, the periodic optical potential can be approximated near every minimum by Eq.~\eqref{eq:trapping1} with $V^{(n)}_0 = 0$.
Strictly speaking, one would actually need to tune the laser wavelength slightly away from the ``magic'' wavelength, for which $\mathcal{V}^{(1)} = \mathcal{V}^{(2)}$.
%, by a small amount.
Indeed, in order to have $\Omega_1 = \Omega_2$, a relative difference between $\mathcal{V}^{(1)}$ and $\mathcal{V}^{(2)}$ of order $\Delta m / m$ is necessary.
In principle, this would be implicitly taken into account in the experimental implementation when the laser wavelength of the optical lattice is calibrated by requiring that the transition frequency between the two clock states, $| \mathrm{g} \rangle$ and $| \mathrm{e} \rangle$, becomes independent of the intensity of the lattice beams, i.e.\ independent of the amplitude of the optical potential  \cite{ludlow15}.
However, current set-ups, with frequencies $\Omega_1$ and $\Omega_2$ ranging from tens of kHz to a few MHz and $\Delta m / m \sim 10^{-11}$, are not sensitive to this effect yet because relative differences between $\Omega_1$ and $\Omega_2$ of order $\Delta m / m$ imply changes of the order of $10^{-5}\, \text{Hz}$ or less in the frequency
%$\omega_0 = \Delta E / 2$
of the clock transition, which is below the best precisions achieved so far.
Therefore, at this level it is also possible to use red-detuned 1-D lattices at a ``magic'' wavelength, with $\mathcal{V}^{(1)} = \mathcal{V}^{(2)}$ and $V^{(1)}_0 = V^{(2)}_0 < 0$.

For simplicity we have  employed Eq.~\eqref{eq:action2}, valid for weak gravitational fields and non-relativistic COM motion. However, the previous considerations can be straightforwardly generalized to the fully relativistic case by working in the Fermi-Walker frame introduced in the previous subsection and making use of Eqs.~\eqref{eq:action_accel1}--\eqref{eq:hamiltonian_accel1}; see Appendix~\ref{sec:waveguide_minimum} for further details.
In doing so, one takes into account that a static central position in a \comment{time-independent} gravitational field corresponds to an accelerated spacetime trajectory with $\mathbf{a} = -\mathbf{g}$.
Notice also that the gravity gradient contributes to the dynamics of the centered wave packets, governed by the Hamiltonian in Eq.~\eqref{eq:hamiltonian_accel1}. This is usually a rather small effect, but it can be easily included by subtracting the gravity-gradient tensor $ \Gamma$ from the frequency matrices $\Omega_1^2$ and $\Omega_2^2$.

%\subsection{Non-relativistic motion in a uniform gravitational field}

\subsection{Example: gravitational redshift in atomic-fountain clocks}
\label{sec:atomic_fountain}

\begin{figure}[h]
\begin{center}
\includegraphics[width=7.5cm]{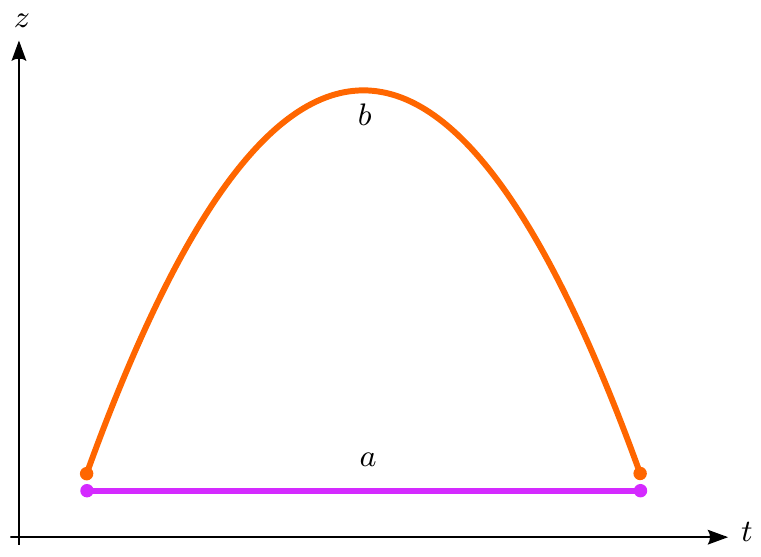}
%%\vspace{-1.0ex}
\end{center}
\caption{Spacetime trajectories in a uniform gravitational field %corresponding, respectively, to
for a two-level atom trapped at constant height ($a$) and for a freely falling one ($b$). They correspond to the central trajectories of the atomic wave packets.}
\label{fig:atomic_fountain}
\end{figure}

As a simple application illustrating a number of aspects introduced in the previous subsections, we will consider a pair of clocks based on atomic wave packets following two different trajectories in a uniform gravitational field, as shown in Fig.~\ref{fig:atomic_fountain}. One corresponds to free fall along the vertical direction and the other to atomic wave packets held at a static position by a trapping potential. Moreover, we will assume that the trapping potential fulfills the conditions discussed in Sec.~\ref{sec:guided_propagation} and Eq.~\eqref{eq:trapped_phase_diff1} holds. Therefore, if the two-level atoms are initialized when the two trajectories first coincide and read out when they coincide again, the difference between the two clocks will correspond to %be given by
the proper-time difference between the two trajectories.
%difference of the proper times along the two trajectories.
\highlight{(Here we assume that the recoil from the initialization and read-out pulses can be neglected, either because it is small or because the atoms are tightly confined.)}

The proper times along the two spacetime trajectories can be calculated by means of the fully relativistic expression in Eq.~\eqref{eq:action1}, but for weak fields and non-relativistic velocities Eq.~\eqref{eq:action2} is a good approximation.
The phase shift between the two clock states for the trapped atoms is then given by
\begin{equation}
\phi_2 - \phi_1 = - \left( \Delta m\, c^2 / \hbar \right) \left(1 + U_0 / c^2 \right) \Delta t
\label{eq:static1} ,
\end{equation}
where $U_0 = U(\bar{\mathbf{x}}_0)$ is the value of the gravitational potential at the central position of the trapped wave packets.
On the other hand, evaluating Eq.~\eqref{eq:action2} for the freely falling trajectory (parallel to $\mathbf{g}$) in a uniform gravitational field yields
\begin{equation}
S_n = - m_n c^2 \left(1 + U_0 / c^2 \right) \Delta t - \comment{\frac{1}{24}} m_n g^2 \Delta t^3
\label{eq:fountain1} ,
\end{equation}
where $g = |\mathbf{g}|$ and $\Delta t$ is the time interval between the first and second intersections of the freely falling spacetime trajectory with the static one at $\mathbf{x} = \bar{\mathbf{x}}_0$. The $\Delta t^3$ phase contribution for uniform force fields as well as possible ways of measuring it with atom interferometry have been investigated in Ref.~\cite{zimmermann17} and its connection with the relativistic time dilation for a freely falling particle has been pointed out in Ref.~\cite{anastopoulos18}.
From Eq.~\eqref{eq:fountain1} one can immediately obtain the following phase difference between the internal states, which determines the outcome of the atomic clock's read-out:
\begin{equation}
\phi_2 - \phi_1 = - \left( \frac{\Delta m\, c^2}{\hbar} \right)
\left[ \left(1 + U_0 / c^2 \right) \Delta t
+ \frac{1}{24} \frac{g^2}{c^2} \Delta t^3 \right]
\label{eq:fountain2} .
\end{equation}

The term proportional to $g^2 \Delta t^3$, which can be interpreted as the proper-time difference between the two trajectories in Fig.~\ref{fig:atomic_fountain}, can be measured by comparing the read-outs of the static and freely falling clocks, determined respectively by Eqs.~\eqref{eq:static1} and \eqref{eq:fountain2}. As explained in Sec.~\ref{sec:clock_initialization}, in practice one actually determines the transition frequency in a \highlight{Ramsey spectroscopy} measurement and the resulting frequencies for the two clocks are proportional to the corresponding redshift factor $\Delta \tau / \Delta t$ in each case, which differ by $(g^2 / c^2)\, \Delta t^2 / 24$.
For $\Delta t = 1\, \text{s}$ this amounts to a relative frequency difference $\Delta \nu / \nu \sim 5 \times 10^{-17}$. While this precision is feasible for static clocks based on optical transitions of cold atoms trapped in magic-wavelength optical lattices, it is about an order of magnitude more demanding than the highest precision achievable to date with atomic clocks based on microwave transitions of cold atoms freely falling in atomic fountains.
Improvements in the latter would therefore be necessary in order to see this effect when comparing the two%
\footnote{Although we have considered the same $\Delta m$ in Eqs.~\eqref{eq:static1} and \eqref{eq:fountain2}, the same conclusions apply when the static clock has a different~$\Delta m'$: one simply needs to take into account the 
%fixed
constant factor $(\Delta m' / \Delta m)$ when comparing the frequencies of the two spectroscopic measurements.}.
Alternatively, in larger atomic fountains such as Stanford's 10-meter tower \cite{dickerson13}, where times in excess of $\Delta t = 2\, \text{s}$ can be reached, the resulting frequency difference would increase by an order of magnitude and become comparable to the current sensitivity of microwave-based clocks.  
%(In fact, there are much larger special and general relativistic effects associated with different rotation velocities for different latitudes and with different laboratory heights to which atomic clocks are sensitive, but they affect in the same way the pair of clocks being compared here.)

As a matter of fact, there are much larger special and general relativistic time-dilation effects to which atomic clocks are sensitive, but they would affect in the same way the two clocks being compared here. They are associated with different Earth rotation velocities for different latitudes (corresponding to differences of the order of $10^2\, \text{m/s}$) and with laboratory height differences of the order of $10^2$ or $10^3\, \text{m}$.

%\highlight{... recoil neglected ...} %star

The example analyzed in this subsection involves independent atoms (in a superposition of internal states) propagating along the two trajectories and is equivalent to comparing classical clocks following those trajectories.
In contrast, we will next consider a quantum superposition for each single atom of %atomic
wave packets following two spatially separated paths.

\section{Quantum-clock interferometry}
\label{sec:clock_interferometry}

\subsection{Proper time and quantum-clock interferometry}
\label{sec:proper_time}

\begin{figure}[h]
\begin{center}
\includegraphics[width=8.0cm]{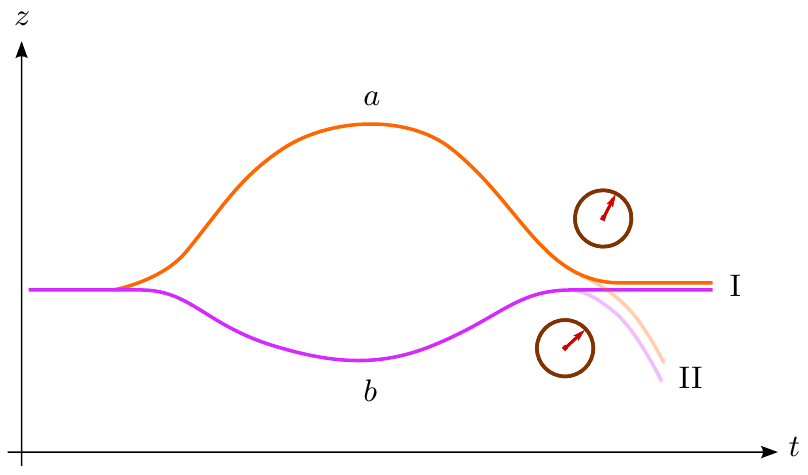}
%\vspace{-1.0ex}
\end{center}
\caption{Central trajectories for the interfering wave packets of a quantum clock at exit port~I. Non-trivial effects arise when the proper time along the two interferometer branches ($a$ and $b$) differ. Analogous results hold for exit port~II.}
\label{fig:clock_interferometry}
\end{figure}

Let us consider an atom interferometer with the central trajectories of the atomic wave packets propagating along the different interferometer branches schematically depicted in Fig.~\ref{fig:clock_interferometry}. If we assume for simplicity that the evolution of the centered wave packets along the two interferometer arms ($a$ and $b$) is approximately the same, the state at the first exit port (I) is given by
\begin{equation}
| \psi_\text{I} \rangle = \frac{1}{2} \Big( e^{i \phi_a} | \psi_a \rangle
+ e^{i \phi_b} | \psi_b \rangle \Big)
\approx \frac{1}{2} \big( 1 + e^{i \, \delta\phi} \big) \, e^{i \phi_a} | \psi_a \rangle
\label{eq:exit_port1} ,
\end{equation}
where the phase shift $\delta \phi = \phi_b - \phi_a$ is the difference between the phases accumulated along the two branches by the interfering wave packets. These phases include the propagation phase described in Sec.~\ref{sec:propagation} for both free and guided propagation, corresponding to Eqs.~\eqref{eq:action_free1} and \eqref{eq:action_accel1},
\highlight{as well as the laser phases associated with any laser pulses employed to diffract the atomic wave packets.}
\comment{Further details can be found in Appendix~\ref{sec:AI}, where the description of a full atom interferometer including relativistic effects is provided.}

%\comment{The latter are essentially given by ...}
%\footnote{The corrections arising from a more realistic treatment of the diffraction process can be included a posteriori and will not be essential for the discussion of the effects that are primarily of interest in this article.}

From Eq.~\eqref{eq:exit_port1} the following probability for each atom to be detected in exit port I is immediately obtained:
\begin{equation}
\langle \psi_\text{I} | \psi_\text{I} \rangle
= \frac{1}{2} \big(1 + \cos \delta\phi \big)
\label{eq:exit_port2} ,
\end{equation}
which exhibits oscillations as a function of $\delta \phi$ due to the interference of the wave packets propagating along the two interferometer arms. \comment{Thus, the phase shift $\delta \phi$ can be experimentally obtained by measuring the oscillations of the fraction of atoms detected in each exit port.}
The expression for the second exit port (II) is completely analogous to Eq.~\eqref{eq:exit_port2} but with a minus sign before the cosine function.

These results also hold for the two-level atom introduced in Sec.~\ref{sec:clock_model} if one starts with an initial state $| \psi_0 \rangle \otimes | \text{g} \rangle$ which is a tensor product of the state $| \psi_0 \rangle$ for the COM and the internal state $| \text{g} \rangle$. However, the situation changes if one initializes the clock as described in Sec.~\ref{sec:clock_initialization},
%and implying
\begin{equation}
| \psi_0 \rangle \otimes | \mathrm{g} \rangle \to | \psi_0 \rangle \otimes
\frac{1}{\sqrt{2}} \Big( | \mathrm{g} \rangle - i\, e^{i \varphi_0} | \mathrm{e} \rangle \Big)
\label{eq:initialization} ,
\end{equation}
before the COM state is split into a coherent superposition of wave packets following different central trajectories that are eventually recombined.
\comment{Provided that the external potentials for the two internal states fulfill the conditions discussed in Sec.~\ref{sec:guided_propagation}, so that their effect on the evolution of the two internal states is the same, the state at the exit port is given by}
%... equal potentials ... evolution of centered wave packets approximately the same for both internal states ...}
\begin{equation}
| \Psi_\text{I} \rangle = \frac{1}{2} \Big( e^{i \phi_a} | \psi_a \rangle \otimes | \Phi_a \rangle
+ e^{i \phi_b} | \psi_b \rangle  \otimes | \Phi_b \rangle \Big)
\label{eq:exit_port3} ,
\end{equation}
with 
\begin{equation}
\begin{aligned}
| \Phi_a \rangle &= \frac{1}{\sqrt{2}} \Big( | \mathrm{g} \rangle - i\, e^{- i \Delta E \, \Delta\tau_a / \hbar}
e^{i \varphi_0}  | \mathrm{e} \rangle \Big) , \\
| \Phi_b \rangle &= \frac{1}{\sqrt{2}} \Big( | \mathrm{g} \rangle - i\, e^{- i \Delta E \, \Delta\tau_b / \hbar}
e^{i \varphi_0}  | \mathrm{e} \rangle \Big) ,
\end{aligned}
\label{eq:clock_states1}
\end{equation}
where $\Delta\tau_a$ and $\Delta\tau_b$ are the proper-time intervals for the central trajectory of each interferometer branch.
\comment{In deriving Eq.~\eqref{eq:exit_port3} it has been implicitly assumed that the central trajectories are the same for the two internal states. This assumption can be relaxed when analyzing separately the evolution of the two internal states as explained in the next subsection. In fact, the coincidence of the central trajectories for the two internal state and the implications otherwise will be discussed in Secs.~\ref{sec:diff_recoil} and \ref{sec:residual_recoil} for light-pulse interferometers as well as Secs.~\ref{sec:guided_interf} and \ref{sec:extensions} for guided interferometry.}

%\comment{... implicitly, same central trajectories ... discussed in subsections below and Sec.~\ref{sec:LPAI_redshift}.}

%\comment{Furthermore,}
From Eqs.~\eqref{eq:exit_port3}--\eqref{eq:clock_states1} and if we assume that $| \psi_a \rangle \approx | \psi_b \rangle$  as done above, the probability for each atom to be detected in exit port~I becomes
\begin{equation}
\langle \Psi_\text{I} | \Psi_\text{I} \rangle
= \frac{1}{2} \big(1 + C \cos \delta\phi' \big)
\label{eq:exit_port4} ,
\end{equation}
with
\begin{equation}
C = \big| \langle \Phi_b | \Phi_a \rangle \big|  = \cos \left( \frac{\Delta E}{2 \hbar} \, (\Delta\tau_b - \Delta\tau_a) \right)
\label{eq:contrast1} .
\end{equation}
\comment{Hence, proper-time differences between the two interferometer branches imply a decrease of the quantum overlap between the clock states in the different branches
and leads to a reduced visibility of the interference signal.
As pointed out in Ref.~\cite{zych11}, this visibility reduction can be understood as a consequence of the entanglement between the quantum state of the atom's COM motion and the clock state, which carries \emph{which-way} information.}

%entanglement ... which-way information ... references

\comment{Note that $\langle \Phi_a | \Phi_b \rangle = C \exp \! \big[\! - i\, \Delta E\, (\Delta\tau_b - \Delta\tau_a) / 2 \hbar \big]$ is a complex number and one needs to take into account that its phase also contributes to the phase shift $\delta\phi'$ which determines the detection probability for port~I %through Eq.~\eqref{eq:exit_port4}
and is given by}
\begin{equation}
\delta\phi' = \delta\phi - \frac{\Delta E}{2 \hbar} \, (\Delta\tau_b - \Delta\tau_a)
%\Delta E \, (\Delta\tau_b - \Delta\tau_a) / 2 \hbar
\label{eq:phase_shift1} .
\end{equation}
\comment{Employing Eqs.~\eqref{eq:action_free1} or \eqref{eq:action_accel1} for the computation of the propagation phases that contribute to $\delta \phi$, one gets}
\begin{equation}
\delta\phi' = - \frac{(m_1 + m_2)\, c^2}{2 \hbar} \, (\Delta\tau_b - \Delta\tau_a)
 \,+\, \delta\phi_\text{pot} \,+\, \delta\phi_\text{laser} \,
%\Delta E \, (\Delta\tau_b - \Delta\tau_a) / 2 \hbar
\label{eq:phase_shift2} ,
\end{equation}
\comment{where $\delta\phi_\text{laser}$ and $\delta\phi_\text{pot}$ contain, respectively, the laser phases and the contributions of the external potential to the propagation phases.
This result is valid for closed atom interferometers, whereas for open ones the extra term $\delta\phi_\text{sep}$ discussed in Appendix~\ref{sec:separation_phase} needs to be included.
Remember also that it has been assumed that $\delta\phi_\text{laser}$ and $\delta\phi_\text{pot}$ are the same for both internal states, an assumption that will be relaxed and critically analyzed in the forthcoming subsections.}

\subsection{Time-dilation effects and differential phase-shift measurements}
\label{sec:time_dilation}

By \comment{separately analyzing} the evolution and interference of the wave packets for each internal state \comment{(and making use of the formalism laid out in Appendix~\ref{sec:AI} for the description of a full atom interferometer),} it is possible to have an exact treatment that can go beyond the assumptions made when deriving Eq.~\eqref{eq:exit_port3}.
Furthermore, this provides an alternative interpretation of the loss of contrast found in the previous subsection that can be exploited to devise schemes capable of measuring this effect with a much higher sensitivity.

We will illustrate this alternative interpretation by re-deriving under the same assumptions the results obtained in the previous subsection. Analogously to Eq.~\eqref{eq:exit_port1}, if one takes $| \psi_0 \rangle \otimes | \text{g} \rangle$ as the initial state, the state in the first exit port is given by
\begin{equation}
\big| \Psi_\text{I}^{(1)} \big\rangle = \frac{1}{2} \left( 1 + e^{i \, \delta\phi^{(1)}} \right) \, e^{i \phi_a^{(1)}}
\big| \psi_a^{(1)} \big\rangle \otimes | \text{g} \rangle
\label{eq:exit_port5} .
\end{equation}
Similarly, for the initial state $| \psi_0 \rangle \otimes | \text{e} \rangle$ one has
\begin{equation}
\big| \Psi_\text{I}^{(2)} \big\rangle = \frac{1}{2} \left( 1 + e^{i \, \delta\phi^{(2)}} \right) \, e^{i \phi_a^{(2)}}
\big| \psi_a^{(2)} \big\rangle \otimes | \text{e} \rangle
\label{eq:exit_port6} .
\end{equation}
%with $\delta\phi^{(2)} = \delta\phi^{(1)} + (\Delta E / 2 \hbar) (\Delta\tau_b - \Delta\tau_a)$.
with
\begin{equation}
\delta\phi^{(2)} = \delta\phi^{(1)} - (\Delta E / \hbar) (\Delta\tau_b - \Delta\tau_a)
\label{eq:phase_shift3} .
\end{equation}
Therefore, if one initializes the clock state according to Eq.~\eqref{eq:initialization}, the state in exit port I becomes, by linearity,
\begin{equation}
| \Psi_\text{I} \rangle =
\frac{1}{\sqrt{2}} \left( \big| \Psi_\text{I}^{(1)} \big\rangle
- i\, e^{i \varphi_0} \big| \Psi_\text{I}^{(2)} \big\rangle \right)
\label{eq:exit_port7} ,
\end{equation}
and the probability for each atom to be detected in this port independently of the internal state is
\begin{align}
\langle \Psi_\text{I} | \Psi_\text{I} \rangle &=
\frac{1}{2} \left( \big\langle \Psi_\text{I}^{(1)} \big| \Psi_\text{I}^{(1)} \big\rangle
+ \big\langle \Psi_\text{I}^{(2)} \big| \Psi_\text{I}^{(2)} \big\rangle \right) \nonumber\\
&= \frac{1}{4} \left( 2 + \cos \delta\phi^{(1)} + \cos \delta\phi^{(2)} \right) \nonumber\\
&= \frac{1}{2} + \frac{1}{2} \cos \! \left( \frac{\delta\phi^{(2)} - \delta\phi^{(1)}}{2} \right)
\cos \! \left( \frac{\delta\phi^{(1)} + \delta\phi^{(2)}}{2} \right)
\label{eq:exit_port8} .
\end{align}
When combined with Eq.~\eqref{eq:phase_shift3}, it is clear that we recover the results of Eqs.~\eqref{eq:exit_port4}--\eqref{eq:phase_shift1} after taking into account that $\delta\phi$ corresponds to $\delta\phi^{(1)}$.

Interestingly, Eq.~\eqref{eq:exit_port8} shows that the loss of contrast in quantum-clock interferometry caused by unequal proper times can be naturally interpreted as the result of a \emph{dephasing} in the interference signal for the two internal states, whose oscillations as a function of the proper-time difference are proportional to the atom's rest mass. The mass difference between the two internal states gives then rise to
\comment{a beating-like behavior as a function of the proper-time difference.}
%the beating-like behavior \highlight{((observed / depicted)) in Fig.~\ref{fig:...}.}

More importantly, this immediately suggests a method for measuring the effect with much higher sensitivity. The key point is to use a state-selective detection that can discriminate betweeen the two internal states and determine the number of atoms in each state (rather than the total atom number) that reach port~I and those that reach port~II. This can then be used to infer both $\delta\phi^{(1)}$ and $\delta\phi^{(2)}$. In principle, the phase-shift difference $\big( \delta\phi^{(2)} - \delta\phi^{(1)} \big)$ contains the same information as the contrast reduction, which is entirely determined by the first cosine factor on the right-hand side of Eq.~\eqref{eq:exit_port8}. In practice, however, \emph{differential phase-shift measurements} of this kind can be performed with much higher precision (\comment{potentially} reaching a few mrad per shot) because a number of systematic effects and the main noise sources affect equally both phase shifts and are highly suppressed in the differential measurement \comment{(including any effects that take place before the initialization pulse).}
This common-mode %noise
rejection is particularly effective when both internal states are simultaneously addressed by the same laser pulses.
%Furthermore, 
\comment{Instead, the corresponding decrease of contrast would be much smaller because it depends quadratically on the phase-shift difference, as follows from perturbatively expanding the cosine for small arguments (e.g.\ a phase-shift difference of $1\, \text{mrad}$ implies a contrast reduction of $10^{-6}$).
%Instead, it would be much harder to measure a decrease of order $10^{-5}$ or $10^{-6}$ of the contrast, %(with a value close to one), %close to one by a small amount,
%partly because other effects leading to contrast fluctuations and contrast reduction would mask such small changes.
Furthermore, it is much harder to measure a small decrease of contrast because other effects leading to contrast fluctuations and contrast reduction would mask such small changes.}
On the other hand, the much higher sensitivity of the differential phase-shift measurement will be exploited in Sec.~\ref{sec:LPAI_redshift} to propose feasible experiments involving parameter ranges that can be achieved in existing facilities or new facilities \comment{that will soon become available.}
%\comment{that are soon to become available.}
%currently under construction.

Incidentally, the same result for the detection probability in port~I, as given by Eq.~\eqref{eq:exit_port8}, holds if one considers the incoherent mixture 
\begin{equation}
\hat{\rho} = | \psi_0 \rangle \langle \psi_0 | \otimes \frac{1}{2}
\Big( | \mathrm{g} \rangle \langle \mathrm{g} | + | \mathrm{e} \rangle \langle \mathrm{e} | \Big)
\label{eq:incoherent_mixture} ,
\end{equation}
rather than the coherent superposition in Eq.~\eqref{eq:initialization} as the initial state.

\subsection{Light-pulse interferometers}
\label{sec:light-pulse_interf}

\subsubsection{Insensitivity to gravitational time dilation}

%... two-photon rather than single-photon ...

Unfortunately, standard light-pulse atom interferometers cannot directly measure the effect of gravitational time dilation \comment{in a uniform gravitational field}. This is because when described in the laboratory frame, the different gravitational time dilation experienced by the atoms along the two interferometer branches due to height differences is exactly compensated by the differences in the special-relativistic time dilation along the two branches due to velocity changes caused by the gravitational field. As a result, the proper-time difference $(\Delta\tau_b - \Delta\tau_a)$ is independent of the gravitational acceleration $\mathbf{g}$.

\begin{figure}[h]
\begin{center}
\includegraphics[width=8.0cm]{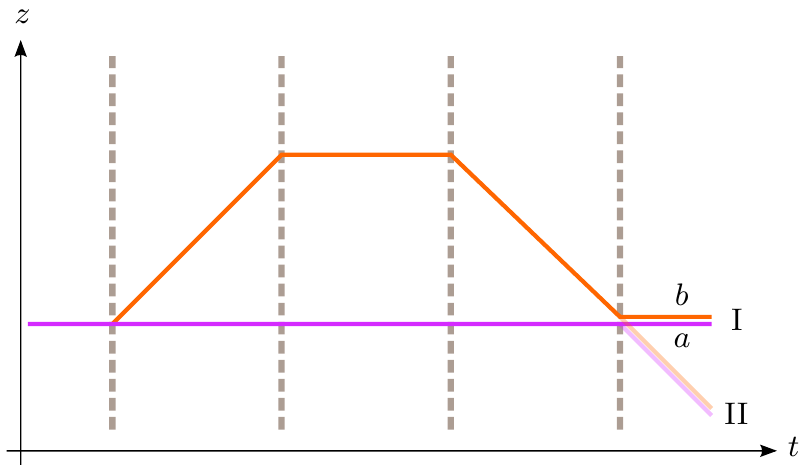}
%\vspace{-1.0ex}
\end{center}
\caption{Central trajectories for a Ramsey-Bord\'e interferometer in a uniform gravitational field as seen in the freely falling frame. In this frame the spacetime trajectories are straight lines and the propagation phases are manifestly independent of the gravitational acceleration $g$.}
\label{fig:ramsey-borde}
\end{figure}

This point can be checked by explicit calculation in the laboratory frame, but can be seen much more easily by considering a freely falling frame and taking into account that the proper times $\Delta\tau_a$ and $\Delta\tau_b$ are invariant geometric quantities independent of the particular coordinate system employed. Indeed, in a freely falling frame the central spacetime trajectories between laser pulses are straight lines, as shown in the \comment{example} depicted in Fig.~\ref{fig:ramsey-borde}. Therefore, the proper-time difference, which is entirely determined in that frame by the momentum transfer $\hbar k_\text{eff}$ from each laser pulse and the time between pulses, is clearly independent of $\mathbf{g}$.
In contrast, the expression for each laser phase involves additional terms that depend explicitly on $\mathbf{g}$ and arise
%when changing from the laboratory frame to the freely falling one.
due to the change of reference frame, namely from the laboratory frame to the freely falling one.
When woking in the freely falling frame, these terms are entirely responsible for the dependence on $\mathbf{g}$ of the total phase-shift $\delta\phi$ for a light-pulse atom interferometer in a uniform gravitational field.

Strictly speaking, one should take into account that the change from the laboratory frame to the freely falling frame will introduce small changes in the pulse timing that depend on $\mathbf{g}$, but these will be suppressed by $1/c$. More specifically, it will typically lead to changes of the pulse timing in the freely falling frame of the order of $\delta T \sim (g T / c)\, T$, which imply extra contributions to the phase shift of the order of
\begin{equation}
\delta\phi_\text{timing} \sim m v_\text{rec}^2 \delta T / \hbar
= (v_\text{rec} / c)\, m g\, \Delta z\, T / \hbar
\label{eq:timing} ,
\end{equation}
where $v_\text{rec} = \hbar k_\text{eff} / m$ is the recoil velocity induced by the momentum transfer from the laser pulse and $\Delta z = v_\text{rec} T$ is the characteristic spatial separation between the interferometer arms.
Hence, such contributions are suppressed by a factor $(v_\text{rec} / c) \sim 10^{-10}$ compared to those that we are interested in.

Finally, it should be noted that the above conclusions concerning the insensitivity of light-pulse atom interferometers to gravitational time dilation apply to \emph{closed} interferometers, i.e.\ those where the central trajectories of the two interfering wave packets at each exit port coincide. The case of %situation for
\emph{open} interferometers
%, whose phase-shift depends on the central position and velocity of the
%initial wave packet ,
will be discussed in Sec.~\ref{sec:ff_frame}.
Nevertheless, it is worth pointing out here that the small changes in the timing of the laser pulses mentioned in the previous paragraph, if not corrected for, will lead to a relative displacement $\delta z \sim v_\text{rec}\, \delta T$ between the interfering wave packets.
This implies a contribution to the interferometer's phase shift that depends on the initial central velocity $v_0$ \cite{roura14,roura17a} and is given by $m v_0\, \delta x / \hbar \sim (v_0 / c)\, m g\, \Delta z\, T / \hbar$, which is again largely suppressed for typical values of $v_0$.

%... gravity gradients ...

\subsubsection{Differential recoil}
\label{sec:diff_recoil}

Attempts to implement quantum-clock interferometry using light-pulse atom interferometers suffer %, moreover,
from a \comment{serious difficulty} due to the different recoil experienced by the internal states. If one starts with atomic wave packets with vanishing mean velocity, a momentum transfer of $\hbar k_\text{eff}$ upon diffraction by a laser pulse, leads to the following recoil velocities, which depend on the internal state:
\begin{equation}
v_\text{rec}^{(1)} = \frac{\hbar k_\text{eff}}{m_1} \equiv v_\text{rec}
\qquad
v_\text{rec}^{(2)} = \frac{\hbar k_\text{eff}}{m_2}
\approx v_\text{rec} \left(1 - \frac{\Delta m}{m_1} \right)
\label{eq:recoil_vel} ,
\end{equation}
where we have taken into account that the recoil velocities are non-relativistic and in the last equality we have neglected terms of higher order in $\Delta m / m_1$.
These recoil velocities gives rise to different paths for atoms in different internal states, so that one can no longer speak of a well defined central trajectory for the quantum clock.

\begin{figure}[h]
\begin{center}
\includegraphics[width=8.0cm]{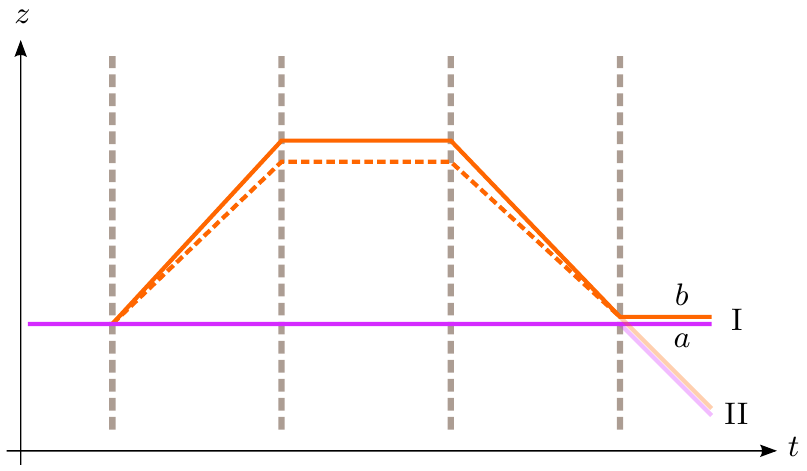}
%\vspace{-1.0ex}
\end{center}
\caption{The mass difference $\Delta m$ leads to a differential recoil for the two internal states. As a consequence, the central trajectories of the diffracted wave packets (orange lines) differ slightly for the ground state (continuous line) and the excited one (dashed line).}
\label{fig:ramsey-borde_diff_recoil}
\end{figure}

Although the differences are rather small, they can be relevant.
Indeed, the small changes of proper time associated with such path differences imply phase-shift changes of the same order as the quantum-clock effects discussed in Sec.~\ref{sec:clock_interferometry} and that we are interested in.
This point can be clearly illustrated with the example of a Ramsey-Bord\'e interferometer in absence of gravity (or in a freely falling frame), whose central trajectories are depicted in Fig.~\ref{fig:ramsey-borde_diff_recoil}.
Such an interferometer is sensitive to special-relativistic time dilation effects: the differences in the central velocities on the two branches give rise to a non-vanishing proper-time difference $(\Delta\tau_b - \Delta\tau_a) = 2T \big( 1/\gamma(v_\text{rec}) -1 \big) \approx - T \, (v_\text{rec}/c)^2$.
Employing Eq.~\eqref{eq:action2}, one gets the following result for the phase shift associated with the internal state~$|g\rangle$:
\begin{equation}
\delta\phi^{(1)} =  (2T) \, m_1 v_\text{rec}^2 / 2 \hbar
\,+\, \delta\phi^{(1)}_\text{laser}
\label{eq:recoil_ps1}\, .
\end{equation}
When calculating the phase shift for the internal state~$|e\rangle$, on the other hand, one needs to evaluate the action along a slightly different trajectory implied by the recoil difference (dashed line in Fig.~\ref{fig:ramsey-borde_diff_recoil}):
\begin{align}
\delta\phi^{(2)} &=  (2T) \, (m_1 + \Delta m) \big( v_\text{rec}^{(2)} \big)^2 / 2 \hbar
\,+\, \delta\phi^{(2)}_\text{laser}  \nonumber \\
&\approx (2T) \, (m_1 + \Delta m)\, v_\text{rec}^2 / 2 \hbar
- (2T) \, \Delta m\, v_\text{rec}^2 / \hbar
\nonumber \\
&\quad\ + \delta\phi^{(2)}_\text{laser}
\label{eq:recoil_ps2}\, .
\end{align}
Because of that, the differential phase shift
\begin{align}
\delta\phi^{(2)} - \delta\phi^{(1)} &\approx (2T / 2 \hbar) \, \Delta m \, v_\text{rec}^2
\Big( 1 - 2 \Big)
+ \delta\phi^{(2)}_\text{laser} - \delta\phi^{(1)}_\text{laser}
\nonumber \\
&=  - (\Delta E / \hbar) (\Delta\tau_b - \Delta\tau_a) \Big( 1 - 2 \Big)
\nonumber \\
&\quad\ + \delta\phi^{(2)}_\text{laser} - \delta\phi^{(1)}_\text{laser}
\label{eq:recoil_diffps}\, .
\end{align}
involves an additional contribution (second term inside the big parentheses) of the same order as the result obtained in Sec.~\ref{sec:clock_interferometry} and given by Eq.~\eqref{eq:phase_shift1}, which corresponds to the first term inside the parentheses.

It should be noted that in the previous example the laser-phase contribution $\delta\phi^{(2)}_\text{laser} - \delta\phi^{(1)}_\text{laser}$ would vanish if the trajectories for the two internal states were identical. However, due to the recoil difference one has instead $\delta\phi^{(2)}_\text{laser} - \delta\phi^{(1)}_\text{laser} = - 2\, k_\text{eff} \big( v_\text{rec}^{(2)} - v_\text{rec}^{(1)} \big) T / \hbar \approx 2\, \Delta m\, v_\text{rec}^2 T / \hbar$.
Interestingly, this contribution exactly cancels the extra contribution found in Eq.~\eqref{eq:recoil_diffps},
\comment{a fact that can be easily understood by considering momentum eigenstates rather than wave packets in position representation.}

%\footnote{Note that $\delta\phi^{(2)}_\text{laser} - \delta\phi^{(1)}_\text{laser}$
%%the total contribution of the laser phases to the differential phase shift
%would vanish if the trajectories for the two internal states were identical.
%However, due to the recoil difference one has
%$\delta\phi^{(2)}_\text{laser} - \delta\phi^{(1)}_\text{laser} =
%- 2\, k_\text{eff} \big( v_\text{rec}^{(2)} - v_\text{rec}^{(1)} \big) T / \hbar
%= 2\, \Delta m\, v_\text{rec}^2 T / \hbar$.}

The shortcomings associated with the differential recoil are circumvented by the scheme for measuring the gravitational redshift that will be presented in Sec.~\ref{sec:LPAI_redshift}.
Other alternatives addressing these difficulties are the use of (partially) reflecting potentials for the beam-splitting and deflection processes \cite{giese18} or the use of guided interferometry. Implementing the former is problematic due to wave-packet distortions as well as the difficulty of achieving sufficiently long interferometer times and will not be considered here. Guided interferometry, on the other hand, will be briefly discussed next.

\subsection{Guided interferometers}
\label{sec:guided_interf}

The main goal of this subsection is to show that guided interferometers can in principle be sensitive to %gravitational time-dilation effects
the gravitational redshift in a uniform gravitational field.
\comment{However, the implementation details will be discussed only briefly and a thorougher investigation is left for future work.}

\begin{figure}[h]
\begin{center}
\includegraphics[width=8.0cm]{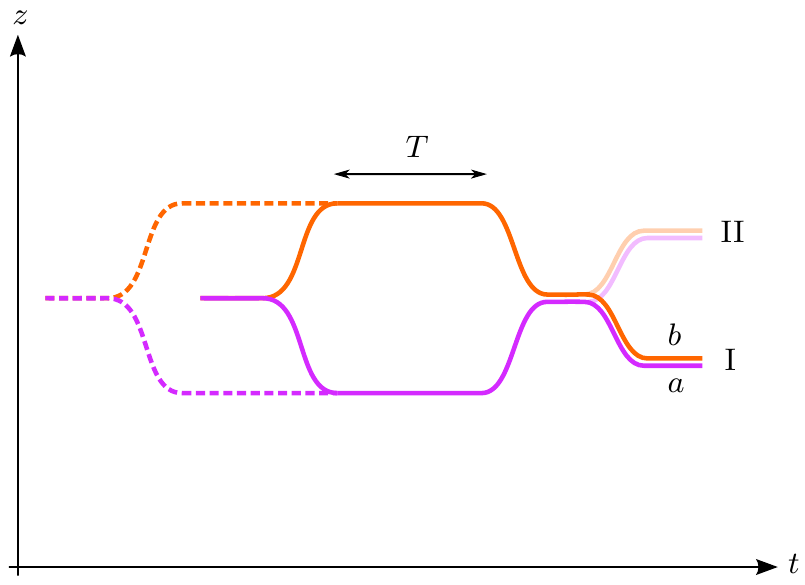}
%\vspace{-1.0ex}
\end{center}
\caption{Central trajectories in the laboratory frame for a guided interferometer where the wave packets are held at different constant heights for some time $T$.
The effect of the gravitational redshift on the differential phase shift can be identified by comparing the outcome with another interferometer where the holding time $T$ is extended while leaving everything else unchanged (dashed lines).}
\label{fig:guided_interferometry}
\end{figure}

%In order to illustrate the key idea, 
As a simple description of the waveguide for the atomic wave packets we will consider the potential analyzed in Sec.~\ref{sec:guided_propagation} and given by Eq.~\eqref{eq:trapping1} but with a time-dependent position $\mathbf{x}_0 (t)$ of the minimum.
If we assume for simplicity that the guiding potential is steep enough (i.e.\ that the relevant eigenvalues of the matrix $\Omega^2$ are large enough), the wave packet's central trajectory can be approximated by $\mathbf{x}_0 (t)$ when evaluating the propagation phase %using
through Eq.~\eqref{eq:action_accel1}. Provided that the conditions discussed in Sec.~\ref{sec:guided_propagation} and relating the potentials for the two internal states are fulfilled,   the difference between the phases accumulated by the internal states is then given by Eq.~\eqref{eq:trapped_phase_diff1} with the proper time interval $\Delta\tau$ evaluated along the spacetime trajectory defined by $\mathbf{x}_0 (t)$.
In particular, given a guided interferometer in a uniform gravitational field with the trajectories $\mathbf{x}_0 (t)$ for the two branches depicted in Fig.~\ref{fig:guided_interferometry}, there will be the following contribution from the static segments to the differential phase shift:
\begin{equation}
\delta\phi^{(2)}_\text{static} - \delta\phi^{(1)}_\text{static} = \Delta m \, g \, \Delta z \, T / \hbar
\label{eq:phase_shift_static} ,
\end{equation}
where $\Delta z$ \comment{is the spatial separation between the two branches
%((is the spatial separation between the two branches // corresponds to the difference
%of the minimum positions $\mathbf{x}_0$ for the static segments of the two branches projected))
along the direction of $\mathbf{g}$.}
This contribution to the differential phase shift can be extracted from the full differential phase-shift measurement by comparing the outcome of experiments with different values of $T$ but leaving everything else unchanged (as long as the contributions from the beam-splitting and recombination parts remain the same despite the changes in $T$). 

The actual central trajectories will differ slightly from $\mathbf{x}_0 (t)$ and when calculating the proper time along them, this will lead to deviations from the result in Eq.~\eqref{eq:phase_shift_static}. Moreover, differences between the central trajectories for the two internal states, even small ones, can be particularly critical because they can completely mask the contribution in Eq.~\eqref{eq:phase_shift_static}.
The amplitude of the oscillations around $\mathbf{x}_0$ in the static segments can be reduced, besides using steep guiding potentials, by employing optimal control techniques \cite{hohenester07,de_chiara08}
%and related techniques \cite{robin}
to select a detailed time-dependence of $\mathbf{x}_0 (t)$ during the beam-splitting part that minimizes the amplitude of those oscillations.
Furthermore, the scheme that will be presented in Sec.~\ref{sec:extensions}  can be very helpful to avoid differences between the central trajectories of the two internal states in the static segments as well as guaranteeing that the contributions from the recombination part are the same when comparing the outcomes for different values of the intermediate time $T$.

In order to investigate the oscillations around $\mathbf{x}_0 (t)$, it is convenient to work in the accelerated frame where the position of the minimum of the potential is at rest at all times, as done in Appendix~\ref{sec:waveguide_minimum}. Within a fully relativistic treatment this corresponds to the Fermi-Walker frame associated with the spacetime trajectory $X_0^\mu (t) = \big( c\, t, \mathbf{x}_0 (t) \big)$, where it becomes $X_0^\mu (\tau_\text{c}) = \big( c\, \tau_\text{c}, \mathbf{0} \big)$. For non-relativistic motion in a uniform gravitational field its acceleration in Fermi-Walker coordinates reduces to $\mathbf{a}(t) = - \mathbf{g} + \ddot{\mathbf{x}}_0 (t)$. \comment{As shown in Appendix~\ref{sec:waveguide_minimum},} this leads to a potential of the \comment{same form as the right-hand side of Eq.~\eqref{eq:trapping2}} but with the replacement $\mathbf{g} \to \mathbf{g} - \ddot{\mathbf{x}}_0 (t)$, which implies a time-dependent shift $\Delta\mathbf{x}_n (t)$ of the equilibrium position in this frame.
In addition to analyzing and minimizing the amplitude of the oscillations around $\mathbf{x}_0$ in the static segments,
%which can be minimized by suitably tailoring the time dependence of $\mathbf{x}_0 (t)$ in the beam-splitting part,
this Fermi-Walker frame is well suited to studying the corrections to the propagation phase that arise from the deviations of the central trajectory away from $\mathbf{x}_0 (t)$.
\comment{Indeed, calculating in this frame the non-relativistic classical action for these deviations directly provides the corrections that would ensue if one were to calculate the %\comment{((exact))}
propagation phase by evaluating Eq.~\eqref{eq:action_ext1} for the actual central trajectory rather than $X_0^\mu (t)$.}
%, which \comment{reduces to $- m_n c^2 \Delta \tau_\text{c} / \hbar$ in Fermi-Walker coordinates.}

Guided atom interferometers have been implemented using waveguides based on magnetic fields \cite{wang05,qi17}, rf-dressed potentials \cite{berrada13,navez16}, optical potentials \cite{mcdonald13a,kueber16,akatsuka17} (including ``painted'' potentials \cite{ryu15}) and accelerated optical lattices \cite{clade09,mueller09,kovachy10,mcdonald13b,hilico15}.
Among these, optical potentials and accelerated optical lattices seem particularly promising for quantum-clock interferometry because one can achieve potentials for both internal states which are identical to a very high degree by employing a ``magic'' wavelength \cite{akatsuka17}.

\highlight{It should be stressed that some of the interferometry schemes referred to in the previous paragraph involve a combination of guiding potentials and laser pulses. The experiments of Refs.~\cite{berrada13,ryu15} are examples of purely guided interferometry, to which the considerations in this subsection would directly apply.
%On the other hand,
In contrast, these will not necessarily hold for hybrid interferometers. For instance, the atom interferometers of Refs.~\cite{charriere12,zhang16}, briefly discussed in Sec.~\ref{sec:extensions} below and based on a combination of several laser pulses and an optical lattice, are actually insensitive to the gravitational redshift.}

%\comment{... entirely guided ... only Refs.~\cite{...}}

%\comment{(... guided interferometers offer the possibility ... artificial acceleration ... stronger signal ...)}

In any case, although guided atom interferometers have great potential as compact sensors with long interrogation times, they are still at an earlier development stage compared to light-pulse atom interferometers, which have already proven their maturity for high-precision experiments.
Motivated by this, in the next section we will introduce a scheme for quantum-clock interferometry that while being based on light-pulse interferometry, is sensitive to the gravitational redshift in a uniform field.

%\section{Gravitational-redshift measurements with quantum clocks
%in light-pulse atom interferometry}
%\section{Gravitational-redshift measurement with light-pulse AI\lowercase{s}}
\section{Gravitational-redshift measurement with light-pulse\\ atom interferometry}
\label{sec:LPAI_redshift}

\subsection{Light-pulse quantum-clock interferometry scheme sensitive to the gravitational redshift}
\label{sec:LPAI}

\begin{figure}[h]
\begin{center}
\includegraphics[width=13.0cm]{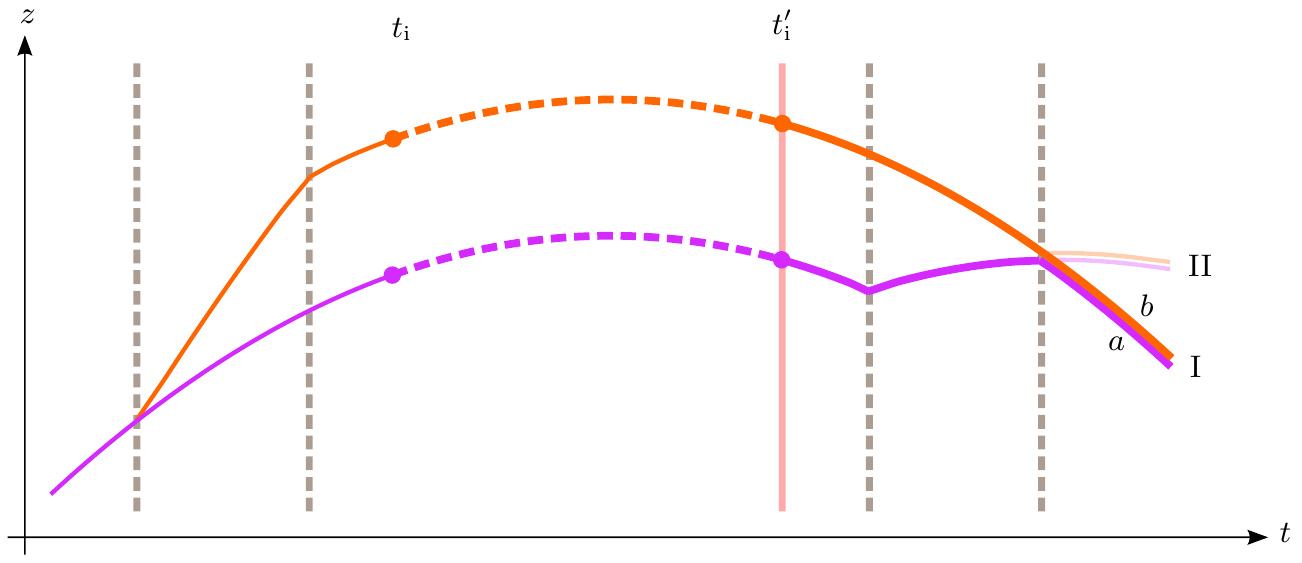}
%\vspace{-1.0ex}
\end{center}
\caption{Central trajectories for a reversed Ramsey-Bord\'e interferometer in a uniform gravitational field. A doubly differential measurement comparing the outcomes for different initialization times ($t_\text{i}$ and $t'_\text{i}$) is directly related to the proper-time difference between the two dashed worldline segments and is sensitive to gravitational time-dilation effects.}
\label{fig:doubly_differential}
\end{figure}

The scheme is based on a \comment{reversed} Ramsey-Bord\'e interferometer, summarized in Fig.~\ref{fig:doubly_differential}, where a pair of $\pi/2$ pulses separated by a time $T'$ are applied to prepare a superposition of two atomic wave packets propagating along the vertical direction with the same velocity but separated by a distance $\Delta z$. After letting the wave packets propagate freely for a longer time $T$, they are finally recombined by applying a second pair of $\pi/2$ pulses separated by a time $T'$.
The key novel idea is to initialize the quantum clock at some \emph{adjustable} time after the first pair of $\pi/2$ pulses by means of a suitable pulse involving a pair of counter-propagating laser beams with angular frequency $\omega_0 = \Delta E / 2\hbar$ which is further described
%((described / discussed))
in Appendix~\ref{sec:two-photon_pulse}. By choosing appropriately the duration and intensity of this pulse, one can create an equal-amplitude superposition of internal states, as given by Eq.~\eqref{eq:initialization}, while leaving the COM motion essentially unchanged thanks to the cancellation of the momentum transfer from both laser beams -- see, however, the discussion in Sec.~\ref{sec:residual_recoil} below for some subtle details.

By using a state-selective detection, one can separately determine the fraction of atoms in each exit port for each internal state and extract the corresponding phase shifts $\delta\phi^{(1)}$ and $\delta\phi^{(2)}$, from which the differential phase-shift $\delta\phi^{(2)} - \delta\phi^{(1)}$ can be obtained. These measurements need to be repeated for different initialization times but leaving everything else unchanged. The difference between the differential phase-shift measurements for different initialization times $t_\text{i}$ and $t'_\text{i}$ contains then very valuable information. Indeed, it is directly related to the proper-time difference between the two interferometer arms for the time interval between the two initialization times:
\begin{align}
&\big( \delta\phi^{(2)}(t'_\text{i}) - \delta\phi^{(1)}(t'_\text{i}) \big)
- \big( \delta\phi^{(2)}(t_\text{i}) - \delta\phi^{(1)}(t_\text{i}) \big)
\nonumber \\
&\quad = \frac{\Delta E}{2 \hbar} \, (\Delta\tau_b - \Delta\tau_a)
= \Delta m \, g \, \Delta z \, (t'_\text{i} - t_\text{i}) / \hbar
\label{eq:doubly_diff_ps}\, ,
\end{align}
%where $\Delta z$ is the separation between the two arms and
where the arguments of the phase shifts correspond here to the initialization times $t_\text{i}$ and $t'_\text{i}$ (as time coordinates in the laboratory reference frame) and where the approximation for non-relativistic velocities and weak gravitational fields leading to Eq.~\eqref{eq:action2} was used in the last equality. \comment{The proper times $\Delta\tau_a$ and $\Delta\tau_b$ correspond to the dashed segments of the central trajectories depicted in Fig.~\ref{fig:doubly_differential}.}

\onecolumngrid
\vspace{1.0ex}

%\pagebreak[4]
\twocolumngrid

In principle one could have tried to compare the interference contrast $C$ for state-independent detection obtained at different times in order to measure the loss of contrast described in Sec.~\ref{sec:proper_time}. (The vibration noise of the retro-reflection mirror naturally provides a uniform random phase-shift distribution for repeated shots, so that the contrast can be determined through a suitable statistical analysis of the distribution of outcomes \cite{geiger11}.)
However, the alternative method based on the \emph{doubly differential measurement} presented above and encoded in Eq.~\eqref{eq:doubly_diff_ps} has %numerous
clearly many advantages.
Firstly, as already pointed out in Sec.~\ref{sec:time_dilation}, important systematic effects and noise sources are highly suppressed in differential phase-shift measurements through common-mode rejection and much higher sensitivities than in a direct contrast measurement \comment{can be achieved.}
Secondly, subtracting the differential phase shifts for different initialization times while leaving everything else unchanged provides further immunity \comment{over the whole duration of the interferometer} to unwanted effects that are independent of the internal state as well as to any unwanted effects (even state-dependent ones) that take place \comment{before the earliest or after the latest of the two initialization times} and are hence common to both differential phase-shift measurements. %
%\footnote{\highlight{... not simultaneous measurements ... stability ... systematics .. drifts... main noise sources already eliminated in differential phase shift ...}}. %star
Finally, as shown by Eq.~\eqref{eq:doubly_diff_ps}, the gravitational time dilation can be directly read out from the measurement. \comment{This can be exploited to test the universality of the gravitational redshift in this context as explained in Sec.~\ref{sec:UGR_UFF}.}

%\highlight{(at which the initialization pulse is applied ... comment on initialization time ... Appendix)} %star

\subsection{Description in the freely falling frame}
\label{sec:ff_frame}

It is instructive %((interesting / instructive / illuminating))
to reanalyze in a freely falling frame the quantum-clock interferometry scheme just proposed, especially given that the insensitivity of standard light-pulse atom interferometers to gravitational time dilation argued in Sec.~\ref{sec:light-pulse_interf} could be most clearly seen in such frames.

\begin{figure}[h]
\begin{center}
\includegraphics[width=12.5cm]{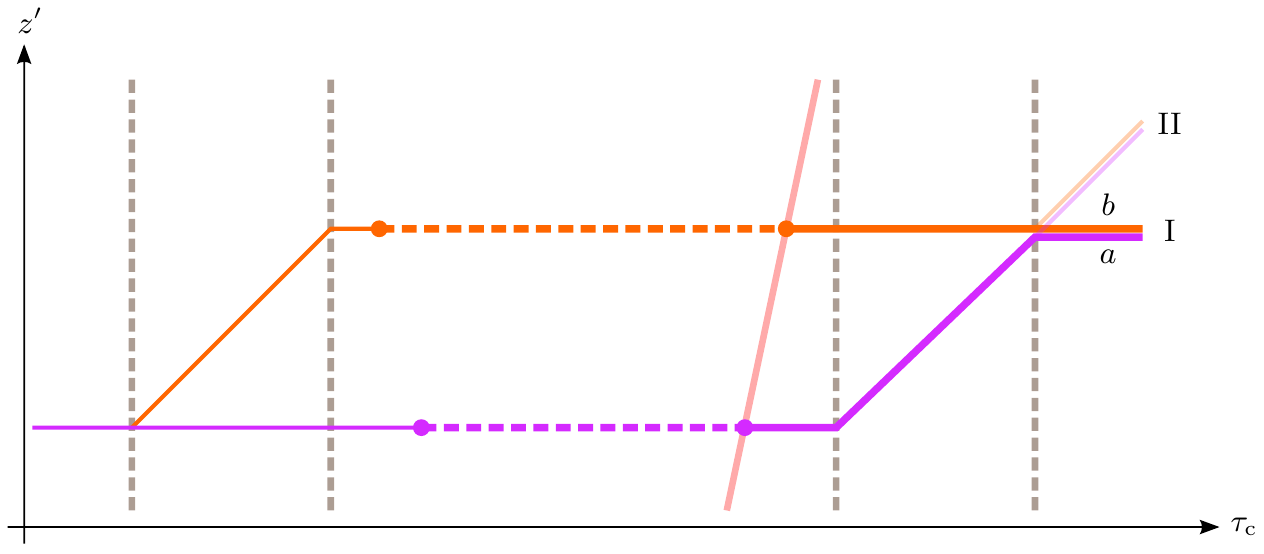}
%\vspace{-1.0ex}
\end{center}
\caption{In the freely falling frame the proper-time difference between the dashed segments in the doubly differential measurement is a consequence of the lack of simultaneity for the spatial hypersurfaces associated with the initialization pulses.}
\label{fig:doubly_differential_ff}
\end{figure}

Fig.~\ref{fig:doubly_differential_ff} displays the central trajectories of the interferometer in a freely falling frame, more specifically in the frame where the trajectories are at rest after the first pair of Bragg pulses. The key point is that while the constant-phase hypersurfaces for the initialization pulse correspond to constant-time hypersurfaces in the laboratory frame, they are no longer hypersurfaces of simultaneity in the freely falling frame: they appear as \comment{tilted straight lines} %(...)
in the 1+1 spacetime diagram of Fig.~\ref{fig:doubly_differential_ff}.
This means that their intersection points with the two central trajectories will exhibit the following time difference in the freely falling frame:
\begin{equation}
\delta \tau_\text{c} = -v(t) \, \Delta z / c^2 = g\, (t - t_\text{ap}) \, \Delta z / c^2
\label{eq:time_shift1} ,
\end{equation}
where \comment{$v(t) = -g\, (t - t_\text{ap})$} is the relative velocity between the freely falling frame and the laboratory frame, and we have again considered for simplicity the regime of weak gravitational fields and non-relativistic velocities, so that terms suppressed by higher powers of $1/c^2$ have been neglected.
The time at which the apex of the central trajectories is reached has been denoted by $t_\text{ap}$.
\comment{Alternatively, one can obtain the time difference in Eq.~\eqref{eq:time_shift1} from the fact that the effective phase factor for the two-photon transition driven by the initialization pulse, which is spatially independent in the laboratory frame, becomes
$\exp \left( -i\, \bar{\omega}_\text{c} (\tau_\text{c} - \tau_\text{c}^\text{(i)})
+ i\, \bar{\mathbf{k}}' \cdot (\mathbf{x}' - \mathbf{x}'_\text{i}) \right)$
in the comoving frame as explained in Appendix~\ref{sec:two-photon_pulse}.}

\onecolumngrid

\twocolumngrid

From Eq.~\eqref{eq:time_shift1} it is clear that the proper time elapsed along the two interferometer arms between initialization pulses at laboratory times $t_\text{i}$ and $t'_\text{i}$ will differ by
\begin{equation}
\delta \tau_\text{c}(t'_\text{i}) - \delta \tau_\text{c}(t_\text{i})
= \big( v(t_\text{i}) - v(t'_\text{i}) \big) \Delta z / c^2 = g\, (t'_\text{i} - t_\text{i}) \Delta z / c^2
\label{eq:time_shift2} ,
\end{equation}
from which the differential-phase-shift difference immediately follows:
\begin{align}
&\big( \delta\phi^{(2)}(t'_\text{i}) - \delta\phi^{(1)}(t'_\text{i}) \big)
- \big( \delta\phi^{(2)}(t_\text{i}) - \delta\phi^{(1)}(t_\text{i}) \big)
\nonumber \\
&\quad = \frac{\Delta E}{2 \hbar} \, (\Delta\tau_b - \Delta\tau_a)
= \Delta m \, g \, \Delta z \, (t'_\text{i} - t_\text{i}) / \hbar
\label{eq:doubly_diff_ps2}\, ,
\end{align}
where we have simply taken into account that $(\Delta\tau_b - \Delta\tau_a) = \delta \tau_\text{c}(t'_\text{i}) - \delta \tau_\text{c}(t_\text{i})$ and made use of Eq.~\eqref{eq:time_shift2}. This \linebreak[4]
result for the differential-phase-shift difference agrees with the result obtained in the laboratory frame, given by Eq.~\eqref{eq:doubly_diff_ps}.
%
%\qquad \comment{... check signs ...}

\subsubsection*{Open interferometers}

\begin{figure}[h]
\begin{center}
\includegraphics[width=8.0cm]{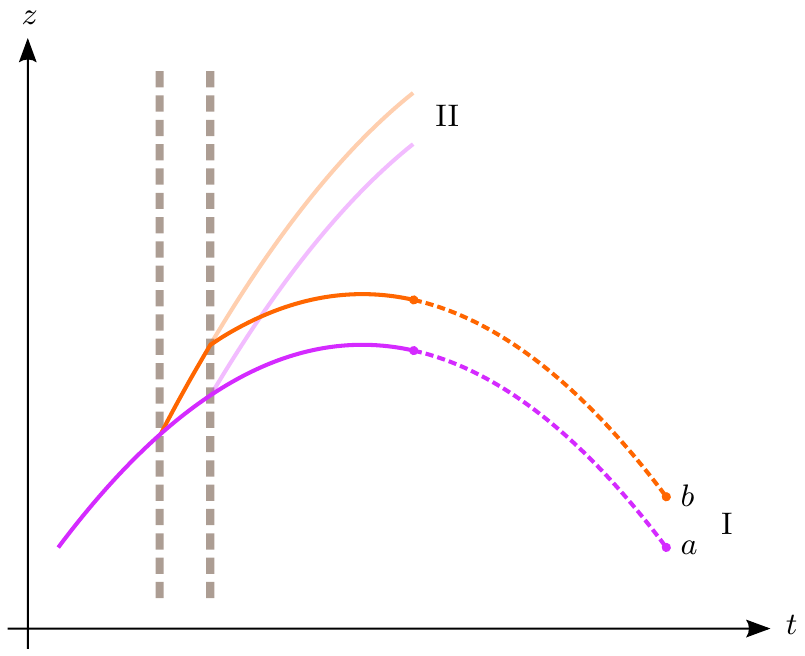}
%\vspace{-1.0ex}
\end{center}
\caption{Central trajectories for an open interferometer in the laboratory frame. The proper-time difference between the extended wordlines as the detection time is delayed (dashed lines) increases due to the gravitational time dilation, but this is exactly compensated by the growth of the separation phase as explained in Appendix~\ref{sec:separation_phase}.}
%\comment{...similarly to the doubly differential scheme above...}}
\label{fig:ramsey_lab}
\end{figure}

After this rederivation in the freely falling frame
we are in a good position to generalize the argument of Sec.~\ref{sec:light-pulse_interf} to \comment{open interferometers.}
In the laboratory frame different detection times at the exit port of an open interferometer, as depicted in Fig.~\ref{fig:ramsey_lab}, lead to changes of the proper-time difference between the interferometer branches analogous to those in Eq.~\eqref{eq:time_shift2}. One could therefore be tempted to conclude that it implies a differential phase shift which depends on $g$ and is sensitive to the gravitational redshift in a uniform gravitational field. However, this is not the case because as explained in Appendix~\ref{sec:separation_phase}, the relative displacement between the interfering wave packets gives rise to an additional phase-shift contribution $\delta \phi_\text{sep}$ that exactly cancels those changes in the proper-time difference. In fact, the total phase shift corresponds to the proper-time difference calculated in the freely falling frame where the central trajectories for the exit port under consideration (or at least the mid-trajectory) are at rest and it is independent of $g$. This generalizes to open interferometers the conclusion of Sec.~\ref{sec:light-pulse_interf} about the insensitivity to the gravitational redshift of light-pulse interferometers in a uniform field.
Moreover, the cancelation of any dependence of the total phase shift $\delta \phi'$ on the detection time after the last beam splitter is important for consistency because for %(( / in))
an interferometer such as that of Fig.~\ref{fig:ramsey_lab} the fraction of atoms detected at each exit port, which is entirely determined by $\delta \phi'$ through Eqs.~\eqref{eq:exit_port2b}--\eqref{eq:exit_port2c}, should be independent of the exact detection time.

In contrast, by applying the initialization pulse at some adjustable time between the two pairs of Bragg pulses in a Ramsey-Bord\'e interferometer, the doubly differential scheme above leads to an effectively open interferometer as far as the phase accumulation of the excited state is concerned while avoiding a relative displacement between the interfering wave packets and the associated separation phase $\delta \phi_\text{sep}$.
Of course, one could in principle perform an analogous doubly differential measurement with the open interferometer of Fig.~\ref{fig:ramsey_lab} by considering different initialization times after the last Bragg pulse, but it is far less convenient. As explained in Appendix~\ref{sec:separation_phase}, one could then read out the phase shift $\delta \phi'$ from the %relative
exact location of the interference fringes in the density profile at the exit port. However, in order to enhance the weak signal in Eqs.~\eqref{eq:doubly_diff_ps} and \eqref{eq:doubly_diff_ps2}, one needs a sufficiently large spatial separation $\Delta z$, but this leads to a very small fringe spacing, which is inversely proportional to $\Delta z$, and is further limited by an eventual lack of overlap \comment{between} the envelopes of the two interfering wave packets.
To a certain extent these difficulties can be alleviated by letting the two wave packets expand for a sufficiently long time, but one is then left with a rather dilute density profile leading to a low signal-to-noise ratio that prevents resolving the fringes %one from ...
%being the fringes from being resolved
with high spatial accuracy. It is therefore much better to employ closed interferometers, such as the Ramsey-Bord\'e geometry, which do not suffer from these problems.

%\highlight{slightly open interferometers due to pulse timing (finite speed of light) ...
%$m c^2$ times $1$ versus $v^2/c^2$ and $g/c^2$ ...}

\subsection{Implications of \comment{the residual} recoil}
\label{sec:residual_recoil}

The initialization pulse based on two counter-propagating laser beams with equal frequencies in the laboratory frame drives the transition between the two internal states with no momentum transfer to the COM motion. However,  because the excitation from ground to excited state increases the total inertial mass of the atom by $\Delta m$, an atom with velocity $v$ along the vertical direction when the pulse is applied will experience a velocity change $\Delta v = - (\Delta m / m)\, v$ due to momentum conservation (terms of higher-order in $\Delta m / m$ and $v/c$ have been neglected since both are very small for typical values of $\Delta m$ and $v$ in this context).
Alternatively, one can easily reach the same conclusion by considering the freely falling frame where the wave packets are at rest rather than the laboratory frame. Indeed, in such a frame the angular frequencies of the two counter-propagating beams differ by $\Delta \omega = -(2\, v/c)\, \omega_0 = - v\, \Delta m\, c / \hbar$ due to the Doppler shift with opposite signs for the two beams (to lowest order in~$v/c$). Therefore, the two-photon transition gives rise to a non-vanishing momentum transfer $\hbar\, \Delta \omega / c = - v\, \Delta m$, as explained in Appendix~\ref{sec:two-photon_pulse}, and the wave packets acquire a non-vanishing central velocity $\Delta v = - (\Delta m / m)\, v$.

The residual recoil discussed in the previous paragraph implies a velocity change $\Delta v = - (\Delta m / m)\, v$ for the central trajectory of the excited state after the initialization pulse. However, since this affects in the same way both interferometer branches, it leaves unchanged the contribution to $ \delta\phi^{(2)}$ from the propagation phases accumulated between the first and second pairs of Bragg pulses. This is because the central velocities continue to be equal on the two branches at any instant of time during that period, so that the contributions to the phase shift from the kinetic term in Eq.~\eqref{eq:action2} still cancel out. Similarly, the separation between the slightly modified central trajectories for the two branches continues to be $\Delta z$ and the net contribution to $\delta\phi^{(2)}$ from the gravitational potential in the laboratory frame remains unchanged.
Equivalent conclusions are reached when analyzing the situation in the freely falling frame.

\onecolumngrid

\begin{figure}[h]
\begin{center}
\includegraphics[width=12.5cm]{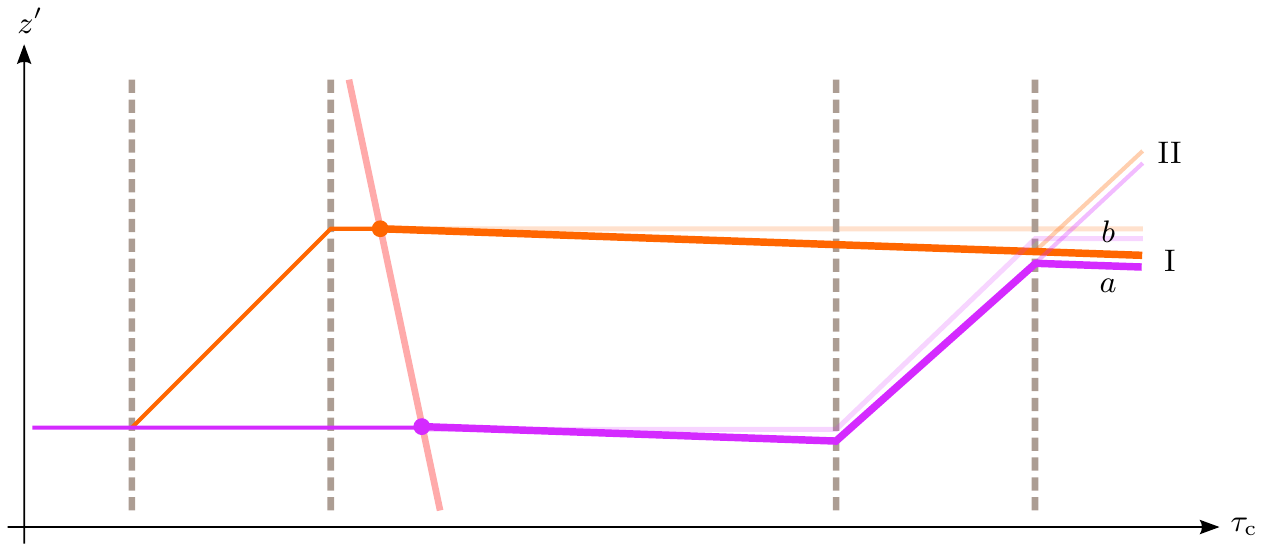}
%\vspace{-1.0ex}
\end{center}
\caption{Central trajectories in the freely falling frame showing how the residual recoil from the initialization pulse leads to slightly modified trajectories for the excited state. Nevertheless, the phase shift as well as the interpretation of the doubly differential measurement in terms of proper-time differences remain unaffected.}
\label{fig:residual_recoil}
\end{figure}

\twocolumngrid

%Finally, we need to determine whether the small change of the central trajectories for the excited state alter the phase-shift contribution from the second pair of Bragg pulses and the free evolution between them.
Furthermore, one can also show that the small change of the central trajectories for the excited state do not alter the net phase-shift contribution from the second pair of Bragg pulses and the free evolution between them.
This point is simpler to analyze in the freely falling frame, where the central trajectories $z_a (t)$ and $z_b (t)$ are modified as follows due to the residual recoil from the initialization pulse:
\begin{equation}
\begin{aligned}
\tilde{z}_a (t) &= z_a (t) + \delta z_3 +\Delta v\, (t-t_3) \, , \\
\tilde{z}_b (t) &= z_b (t) + \delta z_3 +\Delta v\, (t-t_3) \, .
\end{aligned}
\label{eq:pert_traj}
\end{equation}
% *** This signs are for the reversed Ramsey-Bord\'e interferometer ***
In this frame the expression for the phase-shift contribution associated with the second pair of Bragg pulses and the evolution between them, which comprises the laser phases and the kinetic terms, is given by
\comment{
\begin{align}
\delta\phi^{(2)}_\text{2nd\,pair} &= - k_\text{eff}\, \tilde{z}_b (t_3) + k_\text{eff}\, \tilde{z}_a (t_4)
\nonumber \\
&\quad - \frac{m_2}{2 \hbar} \Big( \big( v_\text{rec}^{(2)} + \Delta v \big)^2
- \Delta v^2 \Big) \big( t_4 - t_3 \big)
\label{eq:2_pair_ps} ,
\end{align}
}%
where $t_3$ and $t_4$ are the times of the first and second pulses of this pair (third and fourth Bragg pulses of the full interferometer sequence). 
Substituting the modified trajectories into Eq.~\eqref{eq:2_pair_ps}, we find that any dependence on $\delta z_3$ and $\Delta v$ cancels out.

Therefore, we can altogether conclude that the residual recoil of the initialization pulse has no impact on the total phase shift for the excited state nor the interpretation of the doubly differential measurement as directly reflecting the gravitational redshift between the two interferometer branches.

%\highlight{... remarks on differential recoil from second pair of Bragg pulses ...}

\subsection{Feasibility discussion}
\label{sec:feasibility}

A suitable system for implementing the proposed scheme is the clock transition in neutral atoms typically employed in optical atomic clocks, where the excited state is particularly long lived and $\Delta E$ is of the order of a few eV. As a specific example we will consider $^{87}\text{Sr}$ or $^{88}\text{Sr}$ atoms with a clock transition of wavelength \comment{$\lambda_\text{ph} = 698 \, \text{nm}$} corresponding to \comment{$\Delta m / m = 2 \times 10^{-11}$}. For a branch separation $\Delta z = 1\, \text{cm}$ and initialization times differing by $t'_\text{i} - t_\text{i} = 1\, \text{s}$, the result of the doubly differential measurement amounts to
\begin{align}
&\big( \delta\phi^{(2)}(t'_\text{i}) - \delta\phi^{(1)}(t'_\text{i}) \big)
- \big( \delta\phi^{(2)}(t_\text{i}) - \delta\phi^{(1)}(t_\text{i}) \big)
\nonumber \\
&\quad\quad\quad\quad\quad\quad = \Delta m \, g \, \Delta z \, (t'_\text{i} - t_\text{i}) / \hbar
\,\approx\, 3\, \text{mrad}
\label{eq:doubly_diff_ps3}\, .
\end{align}
With atomic clouds of $N \approx 10^6$ atoms the sensitivity needed for resolving this signal can be achieved in a single shot assuming a phase resolution close to the shot-noise limit $N^{-1/2}$. %$1/\sqrt{N}$.
But even with a much lower phase resolution of $0.1\, \text{rad}$ per shot, the required sensitivity could be reached after averaging $10^3$ measurements.
Measurements of this kind should be possible with a new generation of 10-m atomic fountains capable of performing interferometry with Sr and Yb atoms that will soon become available in Stanford and in Hannover's \comment{HITec} \cite{hitec} respectively. In fact, total interferometer times of up to $2\, \text{s}$ and arm separations of tens of centimeters (up to half a meter \cite{kovachy15b,asenbaum17}) have already been demonstrated in Stanford's first 10-m tower, which operates with Rb atoms \cite{dickerson13}.
On the other hand, alternative configurations with $\Delta z = 1\, \text{mm}$ and $t'_\text{i} - t_\text{i} = 1\, \text{s}$ could be implemented in more compact set-ups with baselines of less than $2\, \text{m}$, where the required sensitivity would be reached for a phase resolution of $10\, \text{mrad}$ per shot after averaging $10^3$ measurements.

Suitable mechanisms for diffraction of atoms in internal-state superpositions, which should act in the same way on both internal states, need to be employed for the second pair of diffraction pulses. Two possibilities are %considered
discussed in some detail in Appendix~\ref{sec:diffraction_superpos}. The first one is Bragg diffraction at a magic wavelength. This guarantees that the Rabi frequency is the same for both internal states, but the required laser power is rather high because these magic wavelengths are far detuned from any transition. The second alternative is based on a sequence of simultaneous pairs of single-photon transitions between the clock states. Interestingly, the lasers required in this case will be readily available in facilities operating with single-photon atom interferometry
%based on single-photon diffraction
such as Stanford's second 10-m tower.
This mechanism is, however, restricted to fermionic isotopes, for which the single-photon transition between the clock states is weakly allowed due to hyperfine mixing \cite{poli14,ludlow15}. %\cite{poli14,ludlow15,??} %star
(The transition also becomes weakly allowed for bosonic isotopes when an external magnetic field is applied \cite{hu17}, but this does not seem a desirable option for precision measurements and long baselines.)

On the other hand, for the first pair of diffraction pulses, which are applied before the initialization pulse, one can make use of efficient diffraction mechanisms acting on the ground state such as Bragg diffraction based on the intercombination transition \cite{del_aguila18}.
Moreover, instead of single pulses it is of course possible to apply a multi-pulse sequence (possibly combined with the use of higher-order Bragg diffraction), which leads to larger momentum transfers so that the targeted arm separation $\Delta z$ can be achieved in shorter times.
\comment{Obviously, when different diffraction processes leading to different effective momentum transfers are employed for the two pairs of diffraction pulses, the time $T'$ between the two pulses in each pair can no longer be the same. Instead, one needs to adjust the timing between first pair of pulses accordingly in order to close the interferometer.}

As usual the frequency difference for the Bragg pulses needs to be chirped linearly in time to keep them on resonance as they fall in Earth's gravitational field \cite{peters01}, and similarly for the individual photon frequencies if single-photon transitions are employed for the second pair of diffraction pulses.
In contrast, for the initialization pulse the frequencies of the two counter-propagating beams should always remain equal (irrespective of the initialization time) so that the constant effective phase corresponds to simultaneity hypersurfaces in the laboratory frame. Moreover, for the two-photon initialization pulse the Doppler effect cancels out at linear order in $v/c$, as explained in Appendix~\ref{sec:two-photon_pulse}, and smaller effects due to the Doppler effect at quadratic order as well as the gravitational redshift of the photons can be compensated with a suitable frequency shift which is identical for both beams
but depends on the initialization time as specified in Appendix~\ref{sec:velocity_redshift}.
%and independent of the initialization time. 

The unwanted effects caused by rotations in atom interferometry, which become particularly relevant for long interferometer times, can be successfully compensated by using a tip-tilt mirror for retro-reflection of the diffraction pulses \cite{hogan08,lan12}. (For the initialization pulse, however, the two counter-propagating beams should be aligned.)
Similarly, the undesirable effects of gravity gradients can be overcome with the method proposed in Ref.~\cite{roura17a} and experimentally demonstrated in Refs.~\cite{d_amico17,overstreet18}, for example through a suitable frequency change for the second pulse of the first pair of Bragg pulses.
\comment{(It is worth pointing out that the phase-shift sensitivity to the initial position and velocity of the atomic wave packet caused by rotations and gravity gradients cancels out to a large degree in the differential phase-shift measurement for the two internal states.)}

%Finally, %Furthermore,
The scheme proposed in Sec.~\ref{sec:LPAI} offers, in addition, the possibility of performing a number of non-trivial checks that can help to identify and calibrate \comment{spurious systematic effects.} For example, changing the initialization time $t_\text{i}$ while keeping the difference $t'_\text{i} - t_\text{i}$ fixed should leave the outcome of the doubly differential measurement unaffected.
Similarly, the outcome should also remain unaltered if the effective momentum transfer of the four diffraction pulses is reversed, or even if it is only reversed for one of the two pairs. (In the latter case it becomes a standard Ramsey-Bord\'e interferometer rather than the reversed configuration, but despite leading to a change of the proper-time difference between the two interferometer arms, the doubly differential measurement still remains unaffected.)
Finally, one can alternatively focus on the conjugate Ramsey-Bord\'e interferometer%
\footnote{The conjugate interferometer arises from the alternative pair of central trajectories after the second beam-splitter pulse.},
which should give equivalent results, by adjusting accordingly the frequencies of the second pair of diffraction pulses \cite{chiow09} and reading out instead its two exit ports, which are spatially well separated from those of the other interferometer.

\subsection{Extension to guided interferometry}
\label{sec:extensions}
%Extensions

\begin{figure}[h]
\begin{center}
\includegraphics[width=8.0cm]{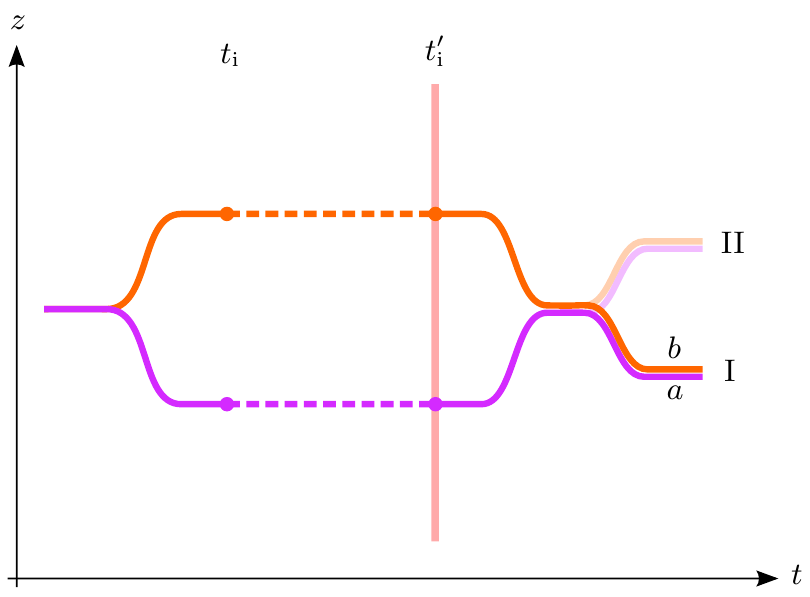}
%\vspace{-1.0ex}
\end{center}
\caption{Central trajectories for a guided interferometer in the laboratory frame. Performing a doubly differential measurements for different initialization times has a number of practical advantages and directly reveals the differences of gravitational time dilation between the two branches.}
\label{fig:doubly_differential_guided}
\end{figure}

The doubly differential measurement technique presented in Sec.~\ref{sec:LPAI} can also be applied to other schemes in quantum-clock interferometry. 
For example, it can be used in a guided interferometer sensitive to the gravitational redshift such as that described %outlined
in Sec.~\ref{sec:guided_interf}.
The essential aspects are sketched in Fig.~\ref{fig:doubly_differential_guided} and are analogous to those of Sec.~\ref{sec:LPAI}, but with the atomic wave packets held at constant height rather than freely falling. The outcome of the doubly differential measurement is again given by Eq.~\eqref{eq:doubly_diff_ps} provided that
%\comment{the guiding potential is the same for both internal states.}
\comment{the guiding potentials for the two internal states fulfill the conditions discussed
in Sec.~\ref{sec:guided_propagation}.}

Remarkably, by employing this technique, one circumvents the difficulty of implementing a beam-splitting process that leads to identical trajectories for the two internal states. Moreover, it is possible to consider interferometers with fixed total time between beam-splitting and recombination, hence avoiding any differences in the phase-shift contribution from the recombination process when applied after different evolution times.
Furthermore, the technique provides immunity to many unwanted noise sources and systematic effects, remaining susceptible only to those acting differently on the two internal states between $t_\text{i}$ and $t'_\text{i}$.
\comment{Incidentally, for a sufficiently steep guiding potential one could even contemplate the possibility of using single-photon initialization pulses.}

In addition, the method can be particularly useful for
%The method can also be particularly useful in
\emph{hybrid} atom-interferometry schemes combining light-pulses and guiding potentials such as those employed in Refs.~\cite{charriere12,zhang16} for gravimetry measurements.
%This corresponds to a modification of the scheme considered in Sec.~\ref{sec:LPAI}
%where the atomic wave packets are held at constant height by means of an optical lattice
%(rather than freely falling) between the two pairs of Bragg pulses.
These correspond to a modification of the \comment{reversed Ramsey-Brod\'e} interferometer
%\footnote{\comment{More specifically, we refer here to the variant of the Ramsey-Brod\'e interferometer where the wave vectors of the second pair of Bragg pulses are the same as %in
%the first pair rather than reversed.}}
in which the atomic wave packets are held at constant height %(rather than freely falling)
for times between the two pairs of Bragg pulses
by means of an optical lattice where they undergo Bloch oscillations.
%\comment{However,
\comment{It should be emphasized that in these hybrid interferometers only the laser phases from the Bragg pulses
%%
%\footnote{(( In light-pulse interferometers with freely falling atoms the frequency differences between the counter-propagating laser beams for the different Bragg pulses are chirped linearly in time in order to compensate their changing velocity %in the laboratory frame
%and maintain the resonance condition. ... scanning ... determine gravitational acceleration $g$ ... jump of the frequency chirping to compensate for integer number of Bloch oscillations during ``holding time'' ... ))}
%%
give rise to a phase-shift contribution that depends on the value of the gravitational acceleration $g$.
In contrast, the phase-shift contribution from the propagation phases, including the Bloch oscillations, is independent of $g$.
This kind of atom interferometers are therefore not sensitive to the gravitational time dilation in a uniform field.}
Nevertheless, the situation is different when they are employed for quantum-clock interferometry and doubly differential measurements comparing the outcomes for different initialization times are performed. The result is then given by Eq.~\eqref{eq:doubly_diff_ps} and reflects the different gravitational redshift experienced by the quantum clocks in the two interferometer branches.

\onecolumngrid
\vspace{1.0ex}

\begin{figure}[h]
\begin{center}
\includegraphics[width=12.5cm]{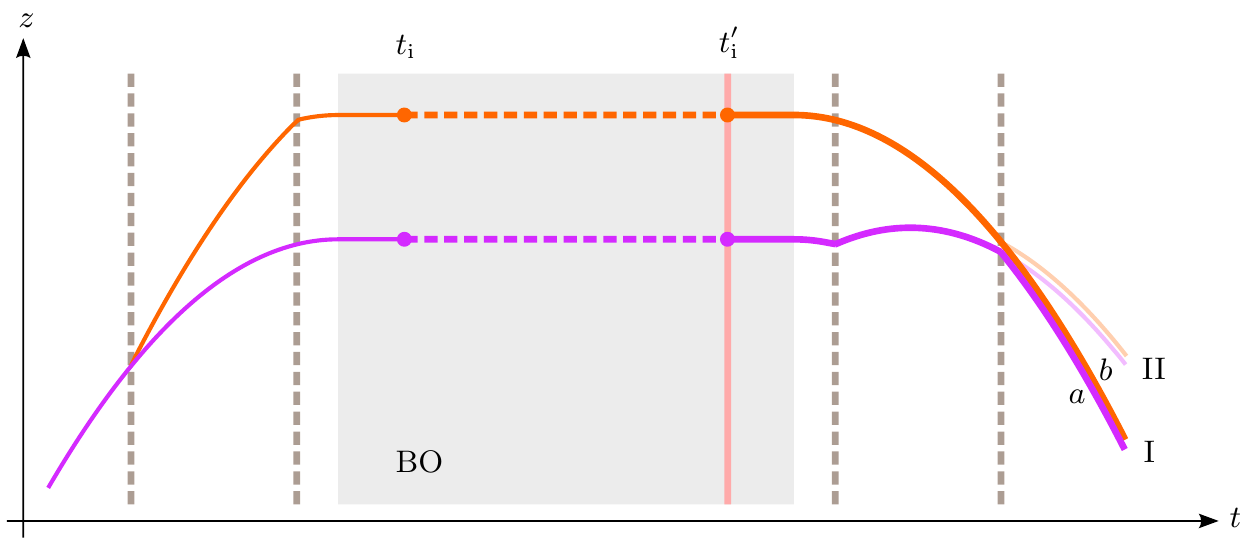}
%\vspace{-1.0ex}
\end{center}
\caption{Central trajectories in the laboratory frame for a hybrid interferometer involving a reversed Ramsey-Bord\'e sequence combined with an optical lattice applied between the two pairs of pulses which \comment{holds the atomic wave packets at constant height} and where they undergo Bloch oscillations (BO). Doubly differential measurements are sensitive to the gravitational redshift in this case too.}
\label{fig:doubly_differential_hybrid}
\end{figure}

\pagebreak[4]

\twocolumngrid

Guided interferometers %(including the hybrid scheme above)
offer an alternative to large atomic fountains and can also reach high sensitivities provided that sufficiently long interferometer times can be achieved. Holding times of 1\,s have already been demonstrated with hybrid schemes employing optical lattices \cite{zhang16} and it is expected that these can be eventually extended to tens of seconds. Since wave-front distortions of the laser beams are one of the major limitations, %in that respect \cite{hasslinger}, ((the use of))
performing atom interferometry inside an optical cavity \cite{hamilton15a} would be advantageous. %\cite{hasslinger}. %star
In this kind of interferometers it is also crucial that the intensity of the optical lattice is the same for both branches, but this requirement can be relaxed in quantum-clock interferometry if a magic-wavelength lattice is employed.

\vspace{-0.2ex} %star

%\section{Testing the equivalence principle: UGR \lowercase{vs.}\ UFF}
\section{Testing the UFF and UGR}
%universality of free fall and gravitational redshift}
\label{sec:UGR_UFF}

Einstein's equivalence principle, which is a cornerstone of general relativity (and metric theories of gravity in general), can be regarded as the combination of three different aspects \cite{will14}: (i)~local Lorentz invariance (LLI), (ii)~UFF, %universality of free fall (UFF),
%also known as weak equivalence principle (WEP),
and (iii)~local position invariance (LPI), also referred to as UGR.
%universality of gravitational redshift (UGR).
In order to illustrate how UGR can be tested with the quantum-clock interferometry scheme presented in Sec.~\ref{sec:LPAI_redshift} and its relation to tests of UFF, we will consider the example of \emph{dilaton models} as a particular framework where violations of the equivalence principle can be consistently parametrized \cite{damour12,damour10a}.

\pagebreak[4]

\subsection{Dilaton models}
\label{sec:dilaton}

%In order to illustrate how UGR can be tested with the quantum-clock interferometry scheme presented in Sec.\ref{sec:LPAI_redshift} and its relation to tests of UFF, we will consider the example of \emph{dilaton models} as a particular framework where violations of the equivalence principle can be consistently parametrized \cite{damour12,damour10a}.
In addition to the spacetime metric the key ingredient of these models is a massless scalar field, the \emph{dilaton} field, that couples non-universally\footnote{Non-universal coupling means here that the dilaton coupling to the fields of the Standard Model cannot be accounted for by considering a redefined spacetime metric.} to the fields of the Standard Model. This massless field mediates a long-range interaction (sometimes referred to as ``fifth force'') that adds to the gravitational interaction and leads to violations of the equivalence principle.

At low energies the coupling of the Standard Model fields to the dilaton implies that the mass of composite particles such as an atom depend on the dilaton field $\varphi (x)$, so that the action governing its COM dynamics needs to be modified from Eq.~\eqref{eq:action1} to
\begin{equation}
S_n \big[ x^\mu(\lambda) \big] = - \int d \tau \, c^2 \, m_n \big( \varphi (x^\mu) \big)
%S_n \big[ x^\mu(\lambda) \big] = - \int d \tau \, c^2 \, m_n \Big( \varphi \big (x^\mu(\lambda) \big) \Big)
%\quad \text{with}\ n = 1,2
%%\nonumber \\
%%&= -m_n c \int m_n \big( \varphi (x^\mu) \big) \, c^2 \,
%%\sqrt{g_{\mu\nu} \frac{dx^\mu}{d\lambda}\frac{dx^\nu}{d\lambda}} \, d\lambda
\label{eq:action_dilaton1} .
\end{equation}
\highlight{Since the value of a scalar field at a spacetime point does not define any preferred direction or rest frame, dilaton models do not give rise to violations of LLI.}
However, they do lead to violations of UFF and UGR because through the dilaton
the mass becomes a function of spacetime and its detailed dependence
on $\varphi (x)$ is species-dependent \cite{damour12,damour10a}.
To see this more explicitly, let us consider the regime of non-relativistic velocities and weak gravitational fields that led to Eq.~\eqref{eq:action2}. Including the corrections that arise from a weak coupling to the dilaton field, it becomes
\begin{widetext}
\begin{equation}
S_n \big[ \mathbf{x}(t) \big] =
\int dt \, \left(  -m_n c^2 \Big( 1 + \bar{\beta}_n\, \varphi(\mathbf{x}) \Big)
+ \frac{1}{2} m_n \mathbf{v}^2 - m_n \,U(\mathbf{x}) \right)
= \int dt \, \left(  -m_n c^2 + \frac{1}{2} m_n \mathbf{v}^2
- m_n \Big( 1 + \bar{\beta}_n\, \bar{\beta}_\text{S} \Big) \,U(\mathbf{x}) \right)
\label{eq:action_dilaton2} ,
\end{equation}
\end{widetext}
%\begin{align}
%S_n \big[ \mathbf{x}(t) \big] &=
%\int dt \, \left(  -m_n c^2 \Big( 1 + \bar{\beta}_n\, \varphi(\mathbf{x}) \Big)
%+ \frac{1}{2} m_n \mathbf{v}^2 - m_n \,U(\mathbf{x}) \right)
%\nonumber \\
%&= \int dt \, \left(  -m_n c^2 + \frac{1}{2} m_n \mathbf{v}^2
%- m_n \Big( 1 + \bar{\beta}_n\, \bar{\beta}_\text{S} \Big) \,U(\mathbf{x}) \right)
%\label{eq:action_dilaton2} ,
%\end{align}
\comment{where $\bar{\beta}_n = (1/m_n)\, (\partial m_n / \partial \varphi)\, \big|_{\varphi = 0}$ and $\bar{\beta}_\text{S}$} is defined analogously for the mass distribution acting as the source of the gravitational field. Moreover, in the second equality we have taken into account that the dilaton field sourced by this mass distribution is given by \comment{$\varphi(\mathbf{x}) = \bar{\beta}_\text{S} \,U(\mathbf{x})$ \cite{damour12}.}
When considering different test masses in the gravitational field of a given source, the dependence on $\bar{\beta}_\text{S}$ is common to all of them and can be absorbed in the definition of a species-dependent parameter $\beta_n \equiv \bar{\beta}_n\, \bar{\beta}_\text{S}$ which directly characterizes the violation of UFF.
Indeed, the E\"otv\"os parameter quantifying the differences in the gravitational acceleration experienced by two different bodies $A$ and $B$ is then given %given in this case
by \comment{$\eta_{A B} \equiv 2 (a_A - a_B)/(a_A + a_B) \approx (\beta_A - \beta_B)$.}

Similarly, the implications on the gravitational redshift of an atomic clock can also be inferred from Eq.~\eqref{eq:action_dilaton2}. If we consider an atom trapped in an optical lattice
\comment{fulfilling the conditions discussed in Sec.~\ref{sec:guided_propagation},}
%\highlight{with a magic wavelength, so that the trapping potential $V(\mathbf{x})$ is the same for both internal states,}
the difference between the phases accumulated by the states $| \mathrm{g} \rangle$ and $| \mathrm{e} \rangle$ is modified, due to the dilaton coupling, from $\Delta E \,\Delta\tau / \hbar$ to
\begin{align}
&  \left[ \Delta m\hspace{0.2ex} c^2 + \Big( m_2 c^2 \big( 1 + \beta_2 \big) - m_1 c^2 \big(1 + \beta_1 \big) \Big) \hspace{0.3ex}
U(\mathbf{x}) \right] \Delta t / \hbar
\nonumber \\
&\quad\quad =  \left[ \Delta m\hspace{0.2ex} c^2 + \Big( \Delta m \, c^2 \,+\,
m_1 c^2 \big( \beta_2 - \beta_1 \big) \Big) \hspace{0.3ex} U(\mathbf{x}) \right] \Delta t / \hbar
\nonumber \\
&\quad\quad = \big( \Delta E \, \Delta t / \hbar \big) \, 
\Big( 1 + \big(1 + \alpha_\text{e-g} \big)\, U(\mathbf{x}) / c^2 \Big)
\label{eq:clock_GR1} ,
\end{align}
%\begin{align}
%&m_2 c^2 (1 + \beta_2)\, U(\mathbf{x})\, \Delta t / \hbar
%- m_1 c^2 (1 + \beta_1)\, U(\mathbf{x})\, \Delta t / \hbar
%\nonumber \\
%&\ = \Delta m \, c^2 \, U(\mathbf{x})\, \Delta t / \hbar
%+ m_1 c^2 (\beta_2 - \beta_1)\, U(\mathbf{x})\, \Delta t / \hbar
%\nonumber \\
%&\ = ( \Delta E \, \Delta t / \hbar ) \, \big(1 + \alpha_\text{e-g} \big) \, U(\mathbf{x}) / c^2
%\label{eq:clock_GR} ,
%\end{align}
where $\mathbf{x}$ is the central position of the atomic wave packet \comment{(which coincides for both internal states)}, the higher-order term proportional to $\Delta m\, \beta_2$ has been neglected in the first equality and we have introduced the parameter $\alpha_\text{e-g}$ specified %defined
below and characterizing the deviation from UGR for an atomic clock based on the transition between the states $| \mathrm{g} \rangle$ and $| \mathrm{e} \rangle$.
%((In particular,))
This means that the times $\Delta \bar{\tau}_a$ and $\Delta \bar{\tau}_b$ measured by two such \comment{static} clocks located at different positions (note that they no longer correspond to the general-relativistic proper time directly calculated from the spacetime metric) are related as follows:
\begin{align}
\frac{\Delta \bar{\tau}_b}{\Delta \bar{\tau}_a} &= 
\frac{1 + \big(1 + \alpha_\text{e-g} \big)\, U(\mathbf{x}_b) / c^2}
{1 + \big(1 + \alpha_\text{e-g} \big)\, U(\mathbf{x}_a) / c^2}
\nonumber \\
&\approx 1 + \big(1 + \alpha_\text{e-g} \big) \Big( U(\mathbf{x}_b) - U(\mathbf{x}_a) \Big) / c^2 
\label{eq:clock_GR2} .
\end{align}
From Eq.~\eqref{eq:clock_GR1} it is clear that the parameter $\alpha_\text{e-g}$ introduced there is given by
\begin{equation}
\alpha_\text{e-g} = \frac{m_1}{\Delta m} \big( \beta_2 - \beta_1 \big)
\label{eq:alpha} ,
\end{equation}
which reveals a close connection between violations of UGR and UFF that will be further discussed in the next two subsections.

%\comment{((... remark on notation ...))}

\comment{Although we have focused for simplicity on the regime of non-relativistic velocities and weak fields, a fully relativistic treatment that goes beyond the weak-field approximation is also possible.
First, one needs to solve the Einstein equations together with the equation of motion for the dilaton field (which constitute in general a coupled system of non-linear partial differential equations) to find the dilaton configuration and spacetime metric generated by the matter sources \cite{damour92,damour93a}.
One can then proceed as done in Sec.~\ref{sec:propagation_forces} by treating the coupling of the test particle (the atoms in our case) to the given dilaton field as an %(additional)
external potential. This will lead to a modification of the central trajectories as well as a slight change, typically rather small, in the evolution of the centered wave packets obtained in the Fermi-Walker frame of the %modified
central trajectories.}

%\comment{remark on fully relativistic treatment: F-W coordinates, interaction with dilaton treated as contribution to external potential $V$ ...}

\subsection{Testing UGR with quantum-clock interferometry}
\label{sec:UGR_test}

Within the framework of the dilaton models considered in the previous subsection
the outcome of the quantum-clock interferometry scheme of Sec.~\ref{sec:LPAI_redshift}  can be easily derived by employing Eq.~\eqref{eq:action_dilaton2} instead of Eq.~\eqref{eq:action2} when computing the propagation phases. In particular, if we focus on uniform gravitational fields, as done in Sec.~\ref{sec:LPAI_redshift}, one simply needs to repeat the analysis with the following state-dependent replacement of the gravitational acceleration: $\mathbf{g} \to (1 + \beta_n)\, \mathbf{g}$. This leads then to the following result for the doubly differential measurement:
%proposed in Sec.~\ref{sec:LPAI}:
\begin{align}
&\big( \delta\phi^{(2)}(t'_\text{i}) - \delta\phi^{(1)}(t'_\text{i}) \big)
- \big( \delta\phi^{(2)}(t_\text{i}) - \delta\phi^{(1)}(t_\text{i}) \big)
\nonumber \\
&\quad =  \Big( m_2 (1 + \beta_2) -  m_1 (1 + \beta_1) \Big) \, g \, \Delta z \,
(t'_\text{i} - t_\text{i}) / \hbar
\nonumber \\
&\quad =  \Big( \Delta m + m_1 (\beta_2 - \beta_1) \Big) \, g \, \Delta z \,
(t'_\text{i} - t_\text{i}) / \hbar
\nonumber \\
&\quad =  \Delta m \, \big(1 + \alpha_\text{e-g} \big) \, g \, \Delta z \,
(t'_\text{i} - t_\text{i}) / \hbar
\label{eq:doubly_diff_ps_dilaton}\, ,
\end{align}
where the higher-order term proportional to $\Delta m\, \beta_2$ has been neglected in the second equality and the parameter $\alpha_\text{e-g}$ specified by Eq.~\eqref{eq:alpha} has been introduced in the last equaltiy.

It should be noted that whenever $\beta_2 \neq \beta_1$, the central trajectories for the wave packets of the two internal states will be slightly different: they will fall with slightly different accelerations. Nevertheless, this does not affect the result because for each internal state the upper and lower trajectory between the two pairs of Bragg pulses have the same velocity at each instant of time (in the laboratory frame) and have constant spatial separation $\Delta z$. Furthermore, the fact that the velocities are slightly different for the two internal states when the first Bragg pulse of the second pair is applied does not lead to any change either. This is because, as already shown in Sec.~\ref{sec:residual_recoil}, the phase-shift contribution from the second pair of pulses plus the free evolution between them is insensitive to \comment{a small change of the incoming} velocity as long as it is the same for the upper and lower trajectory. 

The result of Eq.~\eqref{eq:doubly_diff_ps_dilaton} can still be interpreted in terms of the difference of times $\Delta \bar{\tau}_b$ and $\Delta \bar{\tau}_a$ measured by the clock in the upper and lower branch between the laboratory times $t_\text{i}$ and $t'_\text{i}$, in terms of which the right-hand side of Eq.~\eqref{eq:doubly_diff_ps_dilaton} can be simply written as $\Delta E\, (\Delta \bar{\tau}_b - \Delta \bar{\tau}_a) / \hbar$.
The deviation from UGR in the relation between $\Delta \bar{\tau}_b$ and $\Delta \bar{\tau}_a$ agrees with that found for independent static clocks in Eq.~\eqref{eq:clock_GR2} once we take into account that for a uniform gravitational field $U(\mathbf{x}_b) - U(\mathbf{x}_a) = -\mathbf{g} \cdot (\mathbf{x}_b - \mathbf{x}_a) = g\, \Delta z$.

\comment{Eqs.~\eqref{eq:clock_GR2} and \eqref{eq:alpha} establish a clear connection between UGR tests sensitive to $\alpha_\text{e-g}$ and UFF tests measuring $(\beta_2 - \beta_1)$. In this context the latter can be performed through a differential measurement of Mach-Zehnder interferometers for the two internal states, whereas the UGR tests can be based on either quantum-clock interferometry or the comparison of independent clocks (allowing much higher precision thanks to the far larger height differences possible in this case).
The connection between both kinds of tests illustrates Schiff's conjecture that violations of Einstein's equivalence principle necessarily imply violations of UFF \cite{schiff60,dicke60}, i.e.\ that violations of LLI or UGR can only take place if UFF is also violated. In fact, the relation in Eq.~\eqref{eq:alpha} has been previously derived  on general grounds using an energy-conservation argument~\cite{nordtvedt75}; see also Ref.~\cite{wolf16} for a related derivation.}

\subsection{Relation to other approaches}
\label{sec:other_approaches}

It has been argued in Sec.~\ref{sec:light-pulse_interf} that light-pulse atom interferometers are insensitive to gravitational time dilation in a uniform gravitational field. This is not the case \comment{any more} when inhomogeneities of the gravitational field play a significant role. Indeed, there are a couple of interferometry experiments, either proposed or recently realized, where the proper-time difference between the two branches involves time-dilation effects due to inhomogeneous gravitational fields.

In the first one, proposed in Ref.~\cite{hohensee12}, the central position of the atomic wave packets in the two branches remains each in a different extremum of the gravitational potential for a sufficiently long time $T$ before they are eventually recombined. This gives rise to a proper-time difference $\Delta\tau \approx T\, (U_2 - U_1) / c^2$, where $U_1$ and $U_2$ are the values of the potential at these two extrema (in fact, two saddle points). Because the central position of the wave packets at the extrema experiences no gravitational acceleration, such an experiment has been regarded as a \emph{gravitational} analog of the \emph{scalar Aharonov-Bohm} effect%
\footnote{\comment{Strictly speaking, a gravitational analog of the scalar Aharonov-Bohm effect would require a vanishing spacetime curvature (or a constant gravitational potential in the Newtonian case) in a finite spatial region rather than a point.
However, this is not possible (only approximately) because contrary to electric fields, the gravitational field cannot be %(perfectly) 
screened.
Moreover, one should also be able to change over time the constant value of the potential while the atomic wave packet is within that region.}}.

The second setting corresponds to the experiments reported in Ref.~\cite{asenbaum17}, where  the effect of \emph{tidal forces} on a quantum superposition of spatially separated wave packets was measured for the first time. As explained in Appendix~\ref{sec:gravity_gradient}, by considering the freely falling frame where the initial wave packet is at rest, one can show fairly straightforwardly that the phase-shift contribution $(\hbar\, k^2 / 2 m) \, \Gamma \, T^3$ measured in Ref.~\cite{asenbaum17} is directly related to the proper-time difference between the two interferometer arms.

It is, however, important to keep in mind that the potential differences for the local gravitational field created by the source masses considered in the first proposal or those associated with Earth's gravity gradient are much smaller \comment{(by a factor of $10^{-7}$ or less)} than the potential differences implied by the approximately uniform gravitational field on Earth's surface. The effect is therefore already rather weak for regular atom interferometry and not suitable for a quantum-clock interferometry experiment, which would be further suppressed by the tiny factor $\Delta m / m$. Nevertheless, it is still interesting to discuss how such atom interferometers are related to tests of UFF and UGR, which we do next.

A few years ago it was claimed \cite{mueller10} that gravimetry measurements with light-pulse atom interferometers in a Mach-Zehnder configuration provided, through a suitable reinterpretation, the most precise UGR tests to date. However, it was soon pointed out \cite{wolf11a,schleich13a} that the proper time along the two interferometer arms is the same irrespective of the existence of a uniform gravitational field \comment{(this can again be easily seen in a freely falling frame)}. In fact, the total phase shift is entirely given by the contribution from the laser phases and is sensitive to the acceleration experienced by the central trajectories of the atomic wave packets with respect to the wave fronts of the laser pulses.
When considering deviations from general relativity, for example within the framework of the dilaton models briefly reviewed in Sec.~\ref{sec:dilaton}, these kind of interferometers are directly sensitive to the $\beta_n$ parameter introduced in the paragraph right after Eq.~\eqref{eq:action_dilaton2} and characterizing the deviations from UFF. More specifically, performing a simultaneous measurement for two different species $A$ and $B$ corresponds to a test of UFF where the difference $(\beta_A - \beta_B)$ can be determined, the same combination that can be determined from tests of UFF with macroscopic masses such as comparison of freely falling test masses \cite{niebauer87,touboul17} or torsion-balance experiments \cite{schlamminger08}.

%On the other hand,
In contrast to a Mach-Zehnder interferometer in a uniform gravitational field, the two examples discussed in the first three paragraphs of this subsection do exhibit a non-vanishing proper-time difference due to gravitational time-dilation effects. Nevertheless, they are sensitive to the same parameter $\beta_n$ as UFF tests rather than to a parameter characterizing UGR tests with clocks, such as the parameter $\alpha_\text{e-g}$ in Eqs.~\eqref{eq:clock_GR2}--\eqref{eq:alpha}.

\comment{Finally, it should be stressed %pointed out
that the dilaton models considered in this section do not lead to ``purely quantum'' violations of the equivalence principle in the sense of Ref.~\cite{zych18}.
\highlight{Analyzing the implications of this kind of violations requires instead a separate treatment that will not be pursued here. It is, furthermore, not clear that the phenomenological treatment of Ref.~\cite{zych18} can be naturally generalized beyond the regime of non-relativistic velocities and weak gravitational fields.}
In fact, one expects that such violations will be \highlight{highly} suppressed in low-energy effective field theories accounting for %describing
possible extensions beyond the Standard Model of particle physics plus general relativity, a point to which we plan to return in future work.}

\section{Conclusions}
\label{sec:conclusion}

In this article a general formalism for a relativistic description of atom interferometers in curved spacetime has been %developed %obtained
derived and it has been applied to a detailed investigation of a novel scheme for quantum-clock interferometry %which is
sensitive to gravitational time-dilation effects in uniform fields. It has been shown that this can be exploited to test the UGR with delocalized coherent superpositions of quantum clocks and %it has been
argued that its experimental implementation should be feasible with a new generation of 10-meter atomic fountains that will soon become available in Stanford and HITec (Hanover) \cite{hitec}.

Interestingly, the results obtained here also provide a suitable framework for discussing the interpretation of the experiments reported in Ref.~\cite{margalit15} and %, which were
presented as an analog of the contrast reduction caused by gravitational time dilation in atom interferometry. There the uniform gravitational field was mimicked by a magnetic field gradient, two different spin states corresponded to the two internal states and their different couplings to the magnetic field were the analog of different gravitational masses. A decrease (and eventual revival) of the visibility was indeed observed as one considered increasingly later detection times. %later and later %as a function of the detection time.
However, these experiments are not a good analogy for gravitational effects because the inertial mass is essentially the same for both spin states and the different couplings lead to different accelerations, which would correspond to drastic violations of UFF. Rather than being a minor imperfection of the analogy, this point is actually decisive, as can be clearly seen from the results of Sec.~\ref{sec:ff_frame} and Appendix~\ref{sec:separation_phase}. Indeed, the Ramsey interferometer employed in Ref.~\cite{margalit15} is equivalent to the interferometer depicted in Fig.~\ref{fig:ramsey_lab} and for the gravitational case the separation phase would exactly cancel the difference of propagation phases for the two dashed segments, so that the phase shift remains independent of the detection time and no visibility decrease due to gravitational time-dilation effects in a uniform field would be observable with this set-up%
\footnote{\comment{This discussion applies to atom interferometers in the time domain. For interferometers in the spatial domain, on the other hand, there can be additional effects due to different velocities transverse to the screen for the two internal states \cite{pang16}.}}.
%\footnote{\comment{This discussion applies to atom interferometers in the time domain. For interferometers in the spatial domain, on the other hand, there can be additional effects due to differences between the two internal states in the velocities transverse to the screen \cite{pang16}.}}.
Therefore, although they are an example of which-way information stored in the internal state and leading to loss of visibility, the experiments of Ref.~\cite{margalit15} are not a valid analogy for gravitational time-dilation effects.

The absence, to the best of the author's knowledge, of any other proposals so far for a viable experimental realization (at least in the near future) sensitive to gravitational time-dilation effects in quantum-clock interferometry or even of appropriate analogous experiments, as pointed out in the previous paragraph, makes the scheme presented here particularly valuable. %all the more valuable.

It is worth pointing out that besides being crucial for the proposed scheme, another interesting aspect of the clock-initialization pulses based on two-photon transitions \cite{alden14} discussed in Appendix~\ref{sec:two-photon_pulse} is that they can be used with bosonic isotopes too. This is in contrast with the single-photon transitions between the two clock states, which are weakly allowed for fermionic isotopes but entirely forbidden in the bosonic case unless a strong external magnetic field is applied.
An advantage of being able to work with the bosonic isotopes is that they can be cooled down more easily (even reaching Bose-Einstein condensation) and sufficiently narrow momentum distributions can be reached, which are essential for long interferometer times and high diffraction efficiencies.
On the other hand, of the two diffraction mechanisms for internal-state superpositions analyzed in Appendix~\ref{sec:diffraction_superpos}, Bragg diffraction at a magic wavelength can be employed with both kinds of isotopes but requires a rather large amount of laser power. In that respect, the second method, based on simultaneous pairs of single-photon transitions, constitutes an interesting alternative with appealing properties which can be additionally exploited to perform tests of UFF with atoms in internal-state superpositions. 
However, this requires employing %working with
fermionic isotopes, which should be cooled down through sympathetic cooling, or using bosonic atoms and applying strong magnetic fields during the diffraction pulses, which seems a less viable option for this kind of experiments.
In any case, since many future proposals for atom interferometry \cite{graham13,hogan16,magis}, including a second 10-m atomic fountain in Stanford that will operate with Sr atoms, are based on single-photon transitions between clock states, substantial progress
%on this matter
is expected on this matter in the coming years.

Finally, it should be emphasized that although we have mainly focused on its application to quantum-clock interferometry experiments in nearly uniform gravitational fields, the relativistic description of atom interferometry in curved spacetime developed in the paper can be employed in a very wide range of situations for general spacetimes.
(This may require taking into account the effects on the propagation of the laser pulses due to the spacetime curvature.)
Furthermore, it is not restricted to freely falling particles and can naturally take into account the effect of external forces or even guiding potentials.
In particular, it can be very useful when studying the effects of gravitational waves on matter-wave propagation and matter-wave interferometry, not only for atoms but also for high-mass particles \cite{ kaltenbaek16}.

%\highlight{... UFF test with internal-state superpositions \cite{rosi17b} or entangled particles \cite{geiger18} ...}

%... particles with spin ...
%
%... operator ordering ...
%
%... proper-time difference in Stanford experiments, but can also be explained entirely in terms of Newtonian mechanics (tidal forces) ...

\begin{acknowledgments}
This work has been supported by the German Space Agency (DLR) with funds provided by the Federal Ministry of Economics and Energy (BMWi) under Grant No.~50WM1556 (QUANTUS IV) and by the European Union's Horizon 2020 RISE program under Grant No.~691156 (Q-SENSE).
The author thanks Charis Anastopoulos, Ron Folman, Bei-Lok Hu, Sina Loriani, Yair Margalit, Ernst Rasel, Gabriele Rosi, Dennis Schlippert, Christian Schubert, Wolfgang Schleich and Peter Wolf for interesting discussions.
Conversations with Enno Giese at early stages of this work, where the author shared with him a few results presented here, are also acknowledged.
\end{acknowledgments}

%\vspace{2.0cm}

\newpage
%{.}
%\newpage

\appendix

\appsection{Fermi-Walker coordinates}
\label{sec:fermi-walker}

Given a time-like curve in spacetime, often referred to as a \emph{worldline}, one can always construct a Fermi-Walker frame associated with it. The essential ingredient is a tetrad for each point on the worldline consisting of the normalized tangent vector and a set of three orthonormal spatial vectors orthogonal to the tangent vector. Specifying these three vectors at one point immediately determines their counterparts on the whole curve through the so-called Fermi-Walker transport, which requires that the covariant derivative along the worldline of each one of them is parallel to the tangent vector \cite{hawking73,synge60}.

The Fermi-Walker coordinates associated with this frame comprise the proper time $\tau_\text{c}$ along the worldline, which specifies a point %can be used for labeling the points
on the curve, and three spatial coordinates $\{ x^i \}$ which can be regarded as coefficients of the spatial basis vectors of the tetrad and define an element of the vector space orthogonal to the curve at that point. The direction and modulus of this vector determine a geodesic leaving the worldline point in that direction and the proper distance along the geodesic. In this way the coordinates \comment{$\big( \tau_\text{c}\, , \mathbf{x} \big)$}
%$\big( \tau_\text{c}\, , x^i \big)$
uniquely specify a spacetime point, at least in a finite neighborhood of the worldline. In some cases this can be extended to the whole spacetime and they constitute well-defined global coordinates, but it is not always possible.
Further details about the Fermi-Walker coordinates can be found in Ref.~\cite{synge60}.
%and references therein.

In terms of these coordinates the worldline becomes $X^\mu(\tau_\text{c}) = \big(c\, \tau_\text{c}\, , \mathbf{0} \big)$ with four-velocity $U^\mu = dX^\mu / d\tau = (c, \mathbf{0} )$ and non-vanishing acceleration $U^\nu \, \nabla_\nu U^\mu = \big(0 , \mathbf{a}(\tau_\text{c}) \big)$, where $\nabla_\nu$ denotes the %usual
covariant derivative associated with the metric connection.
The metric, in turn, is given by the following line element:
\begin{equation*}
ds^2 = g_{\mu\nu} dx^\mu dx^\nu = g_{00}\, c^2 d\tau_\text{c}^2
\,+\, 2\, g_{0i}\, c\, d\tau_\text{c}\, dx^i
\,+\, g_{ij}\, dx^i dx^j ,
%\label{eq:FW_metric1}
\end{equation*}
with
\begin{align}
g_{00} &= - \big( 1 + \delta_{ij}\, a^i(\tau_\text{c})\, x^j  / c^2 \big)^2
- R_{0i0j} (\tau_\text{c},\mathbf{0})\, x^i x^j \nonumber \\
& \quad\; + O\big( |\mathbf{x}|^3 \big) , \label{eq:FW_metric2a} \\
g_{0i} &= - \frac{2}{3} R_{0jik} (\tau_\text{c},\mathbf{0})\, x^j x^k
+\, O\big( |\mathbf{x}|^3 \big) ,  \label{eq:FW_metric2b} \\
g_{ij} & = \delta_{ij} - \frac{1}{3} R_{ikjl} (\tau_\text{c},\mathbf{0})\, x^k x^l
+\, O\big( |\mathbf{x}|^3 \big) . \label{eq:FW_metric2c}
\end{align}
Thus, at any point of the worldline the metric coincides with the Minkowski metric, and all the Christoffel symbols vanish except for $\Gamma^i_{00} = a^i / c^2 = \Gamma^0_{0i} = \Gamma^0_{i0}$\,.
%and related permutations of the indices.
This means, in particular, that close to the worldline the equation of motion for  a freely falling particle with a non-relativistic velocity in this frame (i.e.\ $| \dot{\mathbf{x}} | \ll c$ with $\dot{ } \equiv d/d \tau_\text{c}$), which follows from the geodesic equation in this limit, takes the simple form $\ddot{\mathbf{x}} = - \mathbf{a}$ \comment{plus tidal forces that grow linearly with $\mathbf{x}$.}

The Fermi-Walker coordinates are especially well suited when considering a spacetime region close to the worldline. In order to quantify this, it is convenient to introduce a \emph{curvature radius} scale $\ell$ that can be \comment{roughly} defined as $1 / \ell^2 \sim \max | R_{\mu\nu\rho\sigma} |$ in terms of the Riemann-tensor components in an orthonormal basis, and characterizes how strong the spacetime curvature is.
For a weak gravitational field corresponding to Eq.~\eqref{eq:metric1} and slowly varying in time (or comepletely time independent), one has \comment{$1 / \ell^2 \sim | R_{0i0j} | \approx | \partial_i \partial_j U / c^2 |$}. By considering a spherically symmetric %gravitational
source mass, this can be estimated to be $1 / \ell^2 \sim (2\, G M / c^2\, r)\, (1 / r^2) = (r_\text{S} / r)\, (1 / r^2)$, where $r_\text{S}$ is the Schwarzschild radius for that mass. Considering a point near Earth's surface, i.e.\ $r \approx R_\oplus$, and taking into account that $r_\text{S} \approx 9 \, \text{mm}$ for the Earth, we obtain $1 / \ell^2 \sim 10^{-9} \times  (1 / R_\oplus)^2$. Hence, for a spatial size $\Delta x = 1\, \text{mm}$ the relevant ratio becomes $(\Delta x / \ell) \sim 10^{-14}$. The contributions of the curvature terms in Eqs.~\eqref{eq:FW_metric2a}--\eqref{eq:FW_metric2c} are of order $(\Delta x / \ell)^2$ whereas the higher-oder corrections neglected there are suppressed by even higher powers.

More specifically, the higher-order corrections to the expressions of the metric components in Eqs.~\eqref{eq:FW_metric2a}--\eqref{eq:FW_metric2c} involve positive powers of the Riemann tensor, of its covariant derivatives or both. While the size of the Riemann tensor components is characterized by $1/\ell^2$, every derivative can be regarded to contribute with an additional factor $1/\ell'$ characterizing the spacetime variations of the Riemann tensor. %, where the length scale $\ell'$ 
In the above estimate for weak gravitational fields generated by spherically symmetric sources, this corresponds to $\ell' \sim r$. For Earth's gravitational field $\ell' \sim R_\oplus$ is still a rather large length scale, but for nearby masses it can be much smaller and the ratio $(\Delta x / \ell')$ is much less suppressed.
In fact, the derivatives of the gravitational potential $U$ are closely related to the multipole expansion of the gravitational field at any given point, with higher multipoles dominated by the the local mass distribution.
Indeed, for objects with mass densities similar to Earth's density the value of $1/\ell^2$ close enough to the object
%(so that $r$ is ((similar / comparable)) to its own size)
can be comparable to that from Earth's gravitational field, whereas higher derivatives are further suppressed by powers of $1/\ell'$ and their main contributions come from nearby masses.

Next, we briefly discuss two particular cases, corresponding to vanishing acceleration or vanishing curvature, which are especially relevant.

\subsection{Fermi normal coordinates (free fall)}
\label{sec:fermi_coords}

The particular case of vanishing acceleration corresponds to the trajectory of a freely falling particle. The worldline is then a geodesic and the Fermi-Walker transport reduces to the usual parallel transport associated with the metric connection.
Moreover, in this case the Fermi-Walker coordinates coincide with the so-called Fermi normal coordinates and the metric components, which can be obtained by taking $a^i = 0$ in Eqs.~\eqref{eq:FW_metric2a}--\eqref{eq:FW_metric2c}, agree with the well-known result for Fermi coordinates \cite{misner73}.

\subsection{Rindler spacetime (uniform gravitational field)}
\label{sec:rindler}

The metric of a uniform gravitational field in general relativity has the following line element:
\begin{equation}
ds^2 = - \big( 1 - \delta_{ij}\, g^i x^j / c^2 \big)^2\, c^2 d\tau_\text{c}^2
+ \delta_{ij}\, dx^i dx^j ,
\label{eq:rindler}
\end{equation}
and it can be interpreted as the outer gravitational field generated by a homogenous mass distribution on an infinite plane. The spacetime outside this mass distribution is on both sides a vacuum solution of Einstein's equations with planar symmetry, i.e.\ invariant under the \textsf{E(2)} Euclidean group of isometries, which involves two translations and one rotation.

In fact, it has the same form as Rindler spacetime, a region of Minkowski spacetime associated with a rigid congruence of uniformly accelerated observers, and it can be obtained from the Fermi-Walker metric by considering %taking
a vanishing Riemann tensor in Eqs.~\eqref{eq:FW_metric2a}--\eqref{eq:FW_metric2c}. Moreover, by comparison with Eq.~\eqref{eq:FW_metric2a} one can conclude that the acceleration of the static worldline corresponding to $\mathbf{x} = \mathbf{0}$ is $\mathbf{a} (\tau_\text{c}) = - \mathbf{g}$\,, which is time independent.

\appsection{Wave-packet propagation}
\label{sec:wp_propagation}

In order to study the evolution of an atomic wave packet in curved spacetime with possibly relativistic motion, it is convenient to work in a comoving frame where the central position of the wave packet is at rest and its evolution reduces to a non-relativistic problem as long as the velocity spread of the wave packet is much smaller than the speed of light. This is naturally implemented by considering the Fermi-Walker frame associated with the central spacetime trajectory of the wave packet, where its dynamics is particularly simple provided that its spatial width $\Delta x$ is much smaller than the characteristic curvature length scale $\ell$.

\subsection{Free propagation}
\label{sec:free_prop}

Many key aspects of the general case are already present in the case of a freely falling atom, which we will consider first. The central position of the wave packet follows a spacetime geodesic and the corresponding Fermi-Walker coordinates reduce to Fermi normal coordinates.
Substituting Eqs.~\eqref{eq:FW_metric2a}--\eqref{eq:FW_metric2c} with $a^i = 0$ into Eq.~\eqref{eq:action1} and taking into account that $v^i = d x^i / d\tau_\text{c} \ll c$,
the classical action in the Fermi-Walker frame becomes
\begin{equation}
S_n \big[ \mathbf{x}(\tau_\text{c}) \big] \approx \int d\tau_\text{c} \,
\left(  -m_n c^2 + \frac{m_n}{2}\, \mathbf{v}^2
+ \frac{m_n}{2}\, \mathbf{x}^\text{T} \, \Gamma (\tau_\text{c})\, \mathbf{x} \right)
%- \frac{1}{2} m_n\, \Gamma_{ij} (\tau_\text{c}) \, x^i x^j \right)
\label{eq:FW_action1} ,
\end{equation}
where we have introduced the gravity gradient tensor $\Gamma_{ij} (\tau_\text{c}) = - c^2 R_{0i0j} (\tau_\text{c},\mathbf{0})$ and employed matrix notation.
When deriving Eq.~\eqref{eq:FW_action1}, we have taken into account that $d x^0 / d\tau_\text{c} = c$, factored $c^2$ out of the radicand and expanded the resulting square root in powers of $(v^i / c)$ and of the Riemann tensor components times $x^i x^j$.
The terms neglected in Eq.~\eqref{eq:FW_action1} are suppressed by further powers of $(v/c)$, $(\Delta x / \ell)$ and $(\Delta x / \ell')$.
%where $\Delta x$ is the size of the wave packet and $\ell$ is the characteristic curvature length scale.
In particular, the lowest-order corrections are of the order of the kinetic term times $(v/c)^2$ or $(\Delta x / \ell)^2$ and of the order of the gravity-gradient term times $(v/c)$
\comment{or $(\Delta x / \ell')$.}

\comment{One can easily get a quantitative estimate of how small these suppression factors are for typical parameters in atom interferometry. For a velocity spread $\Delta v = 3\, \text{mm/s}$ we have $(\Delta v / c) \sim 10^{-11}$. Similarly, for a wave-packet size $\Delta x = 1\, \text{mm}$, and taking into account that for Earth's gravitational field $\ell \sim 10^{11}\, \text{m}$, one gets $(\Delta x / \ell) \sim 10^{-14}$.
Note, however, that as pointed out in Appendix~\ref{sec:fermi-walker}, the contributions of the local mass distribution can lead to much smaller values of the length scale $\ell'$ characterizing the derivatives of the Riemann tensor, so that the factor $(\Delta x / \ell')$ is much less suppressed. In order to include the corresponding contributions, which are rather small but may eventually become non-negligible, one needs to consider the terms proportional to higher derivatives of the Riemann tensor in Eq.~\eqref{eq:FW_metric2a}. This essentially amounts to considering higher multipoles of the gravitational field in the Fermi-Walker frame and would gives rise to additional terms involving cubic powers of $\mathbf{x}$ and higher in the integrand of Eq.~\eqref{eq:FW_action1}.} \highlight{The effects of such  small anharmonicities due to the gravitational field will be reported elsewhere.}

The Hamiltonian operator associated with the action in Eq.~\eqref{eq:FW_action1} is given by
\begin{equation}
\hat{H}_n \,=\, m_n c^2 + \hat{H}_\text{c}^{(n)}
\label{eq:FW_hamiltonian1} ,
\end{equation}
with
\begin{equation}
\hat{H}_\text{c}^{(n)} = \frac{1}{2m_n}\, \hat{\mathbf{p}}^2
- \frac{m_n}{2}\, \hat{\mathbf{x}}^\text{T} \, \Gamma (\tau_\text{c})\, \hat{\mathbf{x}}
\label{eq:FW_hamiltonian2}\, .
\end{equation}
When computing the corresponding unitary time-evolution operator between comoving times $\tau_1$ and $\tau_2$, the first term on the right-hand said of Eq.~\eqref{eq:FW_hamiltonian1} gives rise to a pure $c$\,--\hspace{0.2ex}number phase factor that can be written as $e^{i\hspace{0.2ex} \mathcal{S}_n / \hbar}$ with %$\exp (i\, S_n / \hbar )$ with
\begin{equation}
\mathcal{S}_n \,= - m_n c^2\, ( \tau_2 - \tau_1) =  - m_n c^2 \int^{\tau_2}_{\tau_1} d\tau_\text{c}
\label{eq:FW_phase1}\, .
\end{equation}
%which can be interpreted as the wave packet's \emph{propagation phase}. Moreover,
Although derived in the Fermi-Walker frame where the wave packet is at rest, the proper time between two spacetime points calculated along the central trajectory $X^\mu(\lambda)$ is an invariant quantity and $\mathcal{S}_n$ can be obtained for an arbitrary coordinate system through Eq.~\eqref{eq:action1}. This phase, which is entirely determined by the central trajectory, can be naturally interpreted as the wave packet's \emph{propagation phase}, whereas a Schr\"odinger equation with the Hamiltonian $\hat{H}_\text{c}^{(n)}$ governs the dynamics of the \emph{centered wave packet} $\big| \psi_\text{c}^{(n)} (\tau_\text{c}) \big\rangle$ in the Fermi-Walker frame.
This useful decomposition of the wave packet evolution in terms of its central trajectory and a centered wave packet constitutes a relativistic generalization of analogous existing results for the non-relativistic case (see for example Refs.~\cite{borde92,antoine03b,hogan08,roura14,zeller16a}).

As argued above, the corrections to the time evolution of the centered wave packet due to the terms neglected in Eq.~\eqref{eq:FW_action1} will typically be very small, but if necessary, they can be computed perturbatively.
More specifically, this can be done by working in the interaction picture and expanding perturbatively the time-ordered exponential of the time integral of the higher-order terms neglected in the Hamiltonian operator of Eq.~\eqref{eq:FW_hamiltonian2}.
\comment{(In doing so, the appropriate operator ordering should be used for the the higher-order corrections to the Hamiltonian involving powers of the $\hat{\mathbf{x}}$ and $\hat{\mathbf{p}}$ operators.)}
These contributions can be organized systematically in terms of powers of $(v/c)$ and $(x^i / \ell)$, but it is usually sufficient to keep the lowest order terms in order to show explicitly (and quantitatively) how small the corrections are.

\subsection{External forces}
\label{sec:external_forces}

The results of the previous subsection can be generalized to the case where the atoms experience external forces and the central trajectory of the wave packet no longer corresponds to a spacetime geodesic. In that case one needs to consider the Fermi-Walker frame for this accelerated trajectory as well as the associated Fermi-Walker coordinates, and the metric components are then given by Eqs.~\eqref{eq:FW_metric2a}--\eqref{eq:FW_metric2c} with $a^i \neq 0$.
Furthermore, the external forces need to be explicitly taken into account. This will be done by adding the contribution of an external potential to the classical action, so that Eq.~\eqref{eq:action1} becomes
\begin{equation}
S_n \big[ x^\mu(\lambda) \big] = -m_n c^2 \int d \tau - \int d \tau\, V_n (x^\mu)
\label{eq:action_ext1}\, .
\end{equation}
\comment{It assumes that the external forces in the Fermi-Walker frame can be satisfactorily characterized through a potential. For most relevant situations involving neutral atoms this is typically the case and one routinely employs magnetic and optical potentials to describe their interactions with magnetic fields or light fields. Otherwise one would need to replace the potential term on the right-hand side of Eq.~\eqref{eq:action_ext1} by a suitable alternative describing the interaction of the atoms with the external forces.}

A useful expression for the classical action in the Fermi-Walker frame analogous to Eq.~\eqref{eq:FW_action1} can also be obtained in this case by proceeding similarly to the derivation in the previous subsection.
First, one groups the two integrals in Eq.~\eqref{eq:action_ext1} into a single one with integrand $- \big(\, m_n c^2 + V_n(x^\mu) \hspace{0.1ex} \big)$. Next, one uses the general expression for the proper time in Eq.~\eqref{eq:action1} specialized to the Fermi-Walker coordinates and with the metric components given by Eqs.~\eqref{eq:FW_metric2a}--\eqref{eq:FW_metric2c}. One can then factor \comment{the curvature-independent term of the metric component $- g_{00}$} times $c^2$ out of the radicand and expand the remaining square root perturbatively in powers of $(v/c)$ and $(x^i / \ell)$, as done in the previous subsection, to obtain
\begin{widetext}
\begin{align}
S_n \big[ \mathbf{x}(t) \big] &\,\approx\, - \int d\tau_\text{c} \,\,
 \Big( m_n c^2 + V_n (\tau_\text{c}, \mathbf{x}) \Big)
\left[  \left( 1 + \mathbf{a}(\tau_\text{c}) \cdot \mathbf{x} / c^2 \right)
- \left( 1 + \mathbf{a}(\tau_\text{c}) \cdot \mathbf{x} / c^2 \right)^{-2}
\left( \frac{1}{2}\, \frac{\mathbf{v}^2}{c^2}
+ \frac{1}{\,2\, c^2}\, \mathbf{x}^\text{T} \, \Gamma (\tau_\text{c})\, \mathbf{x} \right)
\right]  \nonumber \\
&\,\approx\, - \int d\tau_\text{c} \,\,
 \Big( m_n c^2 + V_n (\tau_\text{c}, \mathbf{x}) \Big)
\left[  \left( 1 + \mathbf{a}(\tau_\text{c}) \cdot \mathbf{x} / c^2 \right)
- \left( \frac{1}{2}\, \frac{\mathbf{v}^2}{c^2}
+ \frac{1}{\,2\, c^2}\, \mathbf{x}^\text{T} \, \Gamma (\tau_\text{c})\, \mathbf{x} \right)
\right]
\label{eq:action_ext2} ,
\end{align}
%\begin{equation}
%S_n \big[ \mathbf{x}(t) \big] \approx \int d\tau_\text{c}
%\left[  - \left( m_n c^2 + V_n (\tau_\text{c}, \mathbf{x}) \right)
%\left( 1 + \mathbf{a} \cdot \mathbf{x} / c^2 \right)
%+ \left( 1 + \mathbf{a} \cdot \mathbf{x} / c^2 \right)^{-2}
%\left( \frac{m_n}{2}\, \mathbf{v}^2
%+ \frac{m_n}{2}\, \hat{\mathbf{x}}^\text{T} \, \Gamma (\tau_\text{c})\, \hat{\mathbf{x}} \right)
%\right]
%\label{eq:action_ext2} ,
%\end{equation}
\end{widetext}
where the expansion has been truncated at the same order, $(v/c)^2$ and $(x^i / \ell)^2$, as Eq.~\eqref{eq:FW_action1}. Furthermore, in the second equality we have assumed that $|\mathbf{a} \cdot \mathbf{x}| / c^2 \ll 1$ and neglected terms involving powers of $(\mathbf{a} \cdot \mathbf{x}) / c^2$ times $(v/c)^2$ or $(x^i/ \ell)^2$.
\comment{One can easily check that for typical parameters in atom interferometry this new factor is also very small. Indeed, for an acceleration $a = 10\, \text{m/s}^2$ and a wave-packet size $\Delta x = 1\, \text{mm}$ one has $(a\, \Delta x / c^2) \sim 10^{-19}$.}

Even for the steepest guiding potentials employed one typically has $m_n c^2 \gg V_n \sim m_n \mathbf{v}^2$. Neglecting terms involving the potential times powers of order $(\mathbf{a} \cdot \mathbf{x}) / c^2$, $(v/c)^2$, $(x^i / \ell)^2$ or higher, the action in Eq.~\eqref{eq:action_ext2} becomes
%%
%\begin{align}
%S_n \big[ \mathbf{x}(t) \big] \,\approx\, & \int d\tau_\text{c}
%\, \Bigg[ - m_n c^2 - V_n (\tau_\text{c}, \mathbf{0})
%+ \frac{m_n}{2}\, \mathbf{v}^2
%\nonumber \\
%& \qquad\quad\ - \frac{1}{2}\, \mathbf{x}^\text{T} \Big( \mathcal{V}^{(n)} (\tau_\text{c})
%- m_n \Gamma (\tau_\text{c}) \Big)\, \mathbf{x}
%\nonumber \\
%& \qquad\quad\ - V_\text{anh.}^{(n)} (\tau_\text{c}, \mathbf{x}) \,
%\Bigg]
%\label{eq:action_ext3} ,
%\end{align}
%%
%
\begin{widetext}
\begin{equation}
S_n \big[ \mathbf{x}(t) \big] \,\approx\, \int d\tau_\text{c}
\left[\, - m_n c^2 - V_n (\tau_\text{c}, \mathbf{0})
+ \frac{m_n}{2}\, \mathbf{v}^2
- \frac{1}{2}\, \mathbf{x}^\text{T} \Big( \mathcal{V}^{(n)} (\tau_\text{c})
- m_n \Gamma (\tau_\text{c}) \Big)\, \mathbf{x}
\,-\, V_\text{anh.}^{(n)} (\tau_\text{c}, \mathbf{x}) \,
\right]
\label{eq:action_ext3} ,
\end{equation}
\end{widetext}
where $\mathcal{V}_{ij}^{(n)} (\tau_\text{c})
= \, \partial_i \partial_j V_n (\tau_\text{c},\mathbf{x}) \, \big|_{\mathbf{x} = \mathbf{0}}$ and $V_\text{anh.}^{(n)} (\tau_\text{c}, \mathbf{x})$ corresponds to any anharmonic contributions to the external potential that remain after subtracting the harmonic part (i.e.\ all terms up to quadratic order in $\mathbf{x}$).
Note that the terms linear in $\mathbf{x}$ on the right-hand side of Eq.~\eqref{eq:action_ext3} have cancelled out. This would be true even without the approximations that have been made when deriving Eqs.~\eqref{eq:action_ext2}--\eqref{eq:action_ext3} and it is a consequence of the central trajectory fulfilling the classical equation of motion, which in the Fermi-Walker frame amounts to Eq.~\eqref{eq:accel1}.

The Hamiltonian operator associated with the classical action in Eq.~\eqref{eq:action_ext3} is given by
\begin{equation}
\hat{H}_n \,=\, m_n c^2 + V_n (\tau_\text{c}, \mathbf{0}) + \hat{H}_\text{c}^{(n)}
\label{eq:hamiltonian_ext1} ,
\end{equation}
with
\begin{equation}
\hat{H}_\text{c}^{(n)} = \frac{1}{2m_n}\, \hat{\mathbf{p}}^2
+ \frac{1}{2}\, \hat{\mathbf{x}}^\text{T} \Big( \mathcal{V}^{(n)} (\tau_\text{c})
- m_n \Gamma (\tau_\text{c}) \Big)\, \hat{\mathbf{x}}
%%\hat{H}_\text{c}^{(n)} = \frac{1}{2m_n}\, \hat{\mathbf{p}}^2
%%- \frac{m_n}{2}\, \hat{\mathbf{x}}^\text{T} \, \Gamma (\tau_\text{c})\, \hat{\mathbf{x}}
%\,+\, V_\text{anh.}^{(n)} (\tau_\text{c}, \mathbf{x})
\label{eq:hamiltonian_ext2}\, .
\end{equation}
which is valid for a locally harmonic potential (i.e.\ well approximated by a quadratic function within a region of the size of the wave packet). Otherwise one needs to add the anharmonic contribution $V_\text{anh.}^{(n)} (\tau_\text{c}, \mathbf{x})$
to the right-hand side of Eq.~\eqref{eq:hamiltonian_ext2}.
When computing the unitary time-evolution operator between comoving times $\tau_1$ and $\tau_2$ associated with the Hamiltonian $\hat{H}_n$, the first two terms on the right-hand said of Eq.~\eqref{eq:hamiltonian_ext1} give rise to a pure $c$\,--\hspace{0.2ex}number phase factor that can be written as $e^{i\hspace{0.2ex} \mathcal{S}_n / \hbar}$ with %$\exp (i\, S_n / \hbar )$ with
\begin{equation}
\mathcal{S}_n \,=\, - \int^{\tau_2}_{\tau_1} d\tau_\text{c} \,
\left( m_n c^2 + V_n (\tau_\text{c}, \mathbf{0}) \right)
\label{eq:phase_ext1}\, ,
\end{equation}
and can be interpreted as the wave packet's \emph{propagation phase}.
The Hamiltonian $\hat{H}_\text{c}^{(n)}$, on the other hand, governs the dynamics of the \emph{centered wave packet} $\big| \psi_\text{c}^{(n)} (\tau_\text{c}) \big\rangle$ in the Fermi-Walker frame.

For a locally harmonic potential if one choses a centered wave packet with $\left\langle \hat{\mathbf{x}} \right\rangle = \left\langle \hat{\mathbf{p}} \right\rangle = 0$ at some initial time, the vanishing expectation values will be preserved by the propagation dynamics. This will no longer be the case if one needs to include anharmonic contributions from $V_\text{anh.}^{(n)} (\tau_\text{c}, \mathbf{x})$ which are odd under $x^i \to - x^i$ transformations. However, even in that case one can still use the classical central trajectory to define the Fermi-Walker frame as long as the non-trivial evolution of the expectation values generated by the anharmonic contributions to $\hat{H}_\text{c}^{(n)}$ fulfills the conditions $\left\langle \hat{\mathbf{x}} \right\rangle \ll \ell, \ell'$ and $\left\langle \hat{\mathbf{p}} \right\rangle / m_n \ll c$.
\comment{Exactly the same conclusions would apply to the anharmonicities associated with higher multipoles of the gravitational field discussed in the previous subsection.}

\highlight{Finally, by following the same procedure described at the end of the previous subsection, one can systematically compute the corrections associated with the terms involving higher powers of $(v/c)$, $(x^i / \ell)$ and $(\mathbf{a} \cdot \mathbf{x}) / c^2$ that have not been included in the Hamiltonian $\hat{H}_\text{c}^{(n)}$ given by Eq.~\eqref{eq:hamiltonian_ext2} and governing the dynamics of the centered wave packets.}
%
%\highlight{Moreover, the contributions from $V_n (\tau_\text{c}, \mathbf{0})$ neglected in Eq.~\eqref{eq:action_ext3} can be taken into account by replacing $m_n$ with $m_n + V_n (\tau_\text{c}, \mathbf{0}) / c^2$ in Eq.~\eqref{eq:hamiltonian_ext2}.}

%These contributions can be organized systematically in terms of powers of $(v/c)$ and $(x^i / \ell)$, but it is usually sufficient to keep the lowest order terms in order to show explicitly (and quantitatively) how small the corrections are.

\subsection{Guided propagation around the waveguide minimum}
\label{sec:waveguide_minimum}

As pointed out in Sec.~\ref{sec:guided_interf}, when studying guided interferometry, it is convenient to consider the Fermi-Walker frame associated with the spacetime trajectory $X_0^\mu (t) = \big( c\, t, \mathbf{x}_0 (t) \big)$
of the potential minimum, where it becomes $X_0^\mu (\tau_\text{c}) = \big( c\, \tau_\text{c}, \mathbf{0} \big)$.
The derivation of the action in this frame is very similar to the derivation in the previous subsection, except that the terms linear in $\mathbf{x}$ do not cancel out. This cancellation was a consequence of the central trajectory satisfying the classical equation of motion, which is no longer the case for $X^\mu_0 (t)$. In fact, the linear contribution of the potential vanishes at the minimum, where $\partial V_n / \partial x^i = 0$, and the linear term is entirely given by the acceleration dependence of the $g_{00}$ metric component in Eq.~\eqref{eq:FW_metric2a}.
Therefore, instead of Eq.~\eqref{eq:action_ext3} one has
\begin{widetext}
\begin{equation}
S_n \big[ \mathbf{x}(t) \big] \,\approx\, \int d\tau_\text{c}
\left[\, - m_n c^2 - V_n (\tau_\text{c}, \mathbf{0})
- m_n\, \mathbf{a}(\tau_\text{c}) \cdot \mathbf{x}
+ \frac{m_n}{2}\, \mathbf{v}^2
- \frac{1}{2}\, \mathbf{x}^\text{T} \Big( \mathcal{V}^{(n)} (\tau_\text{c})
- m_n \Gamma (\tau_\text{c}) \Big)\, \mathbf{x}
\,-\, V_\text{anh.}^{(n)} (\tau_\text{c}, \mathbf{x}) \,
\right]
\label{eq:action_ext4} .
\end{equation}
\end{widetext}

By solving the equation of motion that follows from this action, one can calculate the deviations of the actual central trajectory with respect to $X_0^\mu (\tau_\text{c})$. Furthermore, evaluating the action along this solution provides the corrections to the propagation phase due to those deviations. Indeed, whereas the first two terms in the integrand correspond to computing the action along $X_0^\mu (\tau_\text{c})$, the remaining terms account for the extra contributions that arise when evaluating it along the actual central trajectory.
The interest of woking in this frame is that for sufficiently steep guiding potentials the deviations of the central trajectory and the associated corrections to the propagation phase are both small.

Note that for the particular example in which $X_0^\mu (\tau_\text{c})$ is a static trajectory in a time-independent gravitational field one has $\mathbf{a}(\tau_\text{c}) = - \mathbf{g}$ and as long as $V_\text{anh.}^{(n)} (\tau_\text{c}, \mathbf{x})$ can be neglected, it coincides with the case considered in Sec.~\ref{sec:guided_propagation}, where the shift $\Delta \mathbf{x}_n$ of the equilibrium position was determined.
On the other hand, if $X_0^\mu (\tau_\text{c})$ corresponds to some non-relativistic motion around the static trajectory, the acceleration is given instead by $\mathbf{a}(\tau_\text{c}) = - \mathbf{g} + \ddot{\mathbf{x}}_0 (\tau_\text{c})$ with $\ddot{\mathbf{x}}_0 (\tau_\text{c})$ calculated in the Fermi-Walker frame of the static trajectory.
Such a time-dependent acceleration leads to a situation analogous to that in Sec.~\ref{sec:guided_propagation} but with a time-dependent shift $\Delta \mathbf{x}_n (\tau_\text{c})$ of the equilibrium position.

\comment{Besides the trap potentials considered here the approach of this subsection can also be applied to the periodic potentials generated by optical lattices. The Fermi-Walker frame associated with the worldline of one of the potential minima corresponds to working in the comoving frame were the optical lattice is at rest. A non-vanishing acceleration of the worldline gives then rise to Bloch oscillations \cite{peik97,kovachy10}, which can be analytically described in fairly simple terms for the two opposite regimes of shallow and deep lattices.
Further details on this will be provided elsewhere.}

\appsection{Full atom interferometer}
\label{sec:AI}

\begin{figure}[h]
\begin{center}
\includegraphics[width=8.0cm]{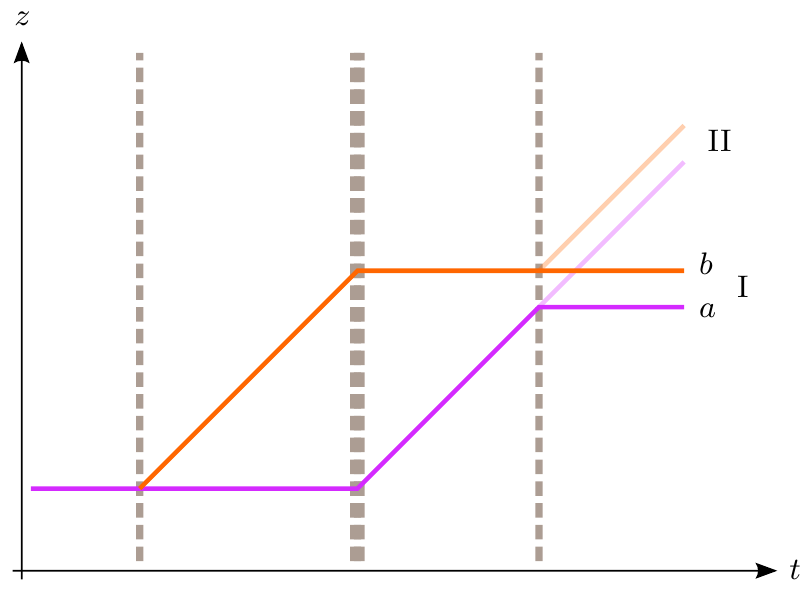}
%\vspace{-1.0ex}
\end{center}
\caption{Schematic representation of the central trajectories of the atomic wave packets in a light-pulse atom interferometer, where branches $a$ and $b$ correspond to the two interfering wave packets in exit port~I. The example displayed here is a Mach-Zehnder interferometer with different times %separations
between the mirror pulse and the initial and final beam-splitter pulses.}
\label{fig:mach-zehnder}
\end{figure}

In order to determine the outcome of an atom interferometer in curved spacetime and including relativistic effects, one can proceed as follows. First, one computes the evolution of the atomic wave packets along each interferometer branch (the different branches for a Mach-Zehnder interferometer are shown in Fig.~\ref{fig:mach-zehnder} as an example).
%; as an example, the different branches for a Mach-Zehnder interferometer are shown in Fig.~\ref{fig:}.
The propagation between laser pulses can be obtained by means of the general formalism introduced in Sec.~\ref{sec:propagation} and derived %/ detailed / developed
in Appendix~\ref{sec:wp_propagation}, whereas the effect of the laser pulses is discussed below. In this way, the resulting sate %at the exit port
of the wave packet evolving along branch $a$ is % therefore ((given by))
$e^{i \phi_a} \big| \psi_\text{c} (\tau_\text{c}) \big\rangle$,
where the centered wave packet $\big| \psi_\text{c} (\tau_\text{c}) \big\rangle$ is a solution of the Schr\"odinger equation with the Hamiltonian operator of Eq.~\eqref{eq:hamiltonian_free1}, or \eqref{eq:hamiltonian_accel1} in presence of external forces, and the phase $\phi_a$ consists of the \emph{propagation phases} for the various segments of the central trajectory associated with that branch as well as the \emph{laser phases} stemming from the different pulses.
Completely analogous expressions hold for the wave packets evolving along the other branches.
%branch $b$.

The state after the last beam splitter can be written as a superposition $| \psi \rangle = | \psi_\text{I} \rangle + | \psi_\text{II} \rangle$ of the states for the two exit ports.
Moreover, if we assume that the central trajectories for the two branches $a$ and $b$ contributing to port~I coincide after the last beam-splitter and that the centered wave packets experience the same evolution along the two branches, the state at this exit port is given by %corresponds to
the following coherent superposition:
\begin{equation}
| \psi_\text{I} \rangle = \frac{1}{2} \big( e^{i \phi_a} + e^{i \phi_b} \big)\, | \psi_\text{c} \rangle
= \frac{1}{2} \big( 1 + e^{i \, \delta\phi} \big) \, e^{i \phi_a} | \psi_\text{c} \rangle
\label{eq:exit_port1b} .
\end{equation}
The interference signal is thus encoded in the phase shift $\delta \phi = \phi_b - \phi_a$, which determines the probability that an atom is detected in port~I:
\begin{equation}
\langle \psi_\text{I} | \psi_\text{I} \rangle
= \frac{1}{2} \big(1 + \cos \delta\phi \big)
\label{eq:exit_port2b} .
\end{equation}
Analogous results hold for the second exit port (II), and the probability that an atom is detected instead in that port is
\begin{equation}
\langle \psi_\text{II} | \psi_\text{II} \rangle
= \frac{1}{2} \big(1 - \cos \delta\phi \big)
\label{eq:exit_port2c} .
\end{equation}
The factor $1/2$ on the right-hand side of Eq.~\eqref{eq:exit_port1b} is the product of two factors $1/\sqrt{2}$ associated respectively with the initial and final beam-splitter pulses. This factor would differ for an interferometer with additional intermediate beam splitters and more than two exit ports, or in the case of unbalanced beam splitters (not leading to equal-amplitude superpositions).

As already pointed out, %mentioned,
for simplicity we have considered above a \emph{closed} interferometer, where the central trajectories of the two interfering wave packets coincide. The case of \emph{open} interferometers, where they no longer coincide, is analyzed below.
On the other hand, the additional assumption that the evolution of the centered wave packets is the same along different branches is a good approximation in many applications, including those explicitly considered here (because the gravity gradients are nearly the same on both branches), but a detailed investigation of the implications %otherwise
when this is not the case will be reported elsewhere.

The results presented in this appendix can be extended to quantum clock-interferometry by including the superscript $(n)$ labeling the internal state and taking it into account the evolution of the different internal states along the interferometer branches, as done in Sec.~\ref{sec:propagation} and Appendix~\ref{sec:wp_propagation}, as well as the effects of the initialization pulse, briefly introduced in Sec.~\ref{sec:clock_initialization} and discussed in detail in Appendix~\ref{sec:two-photon_pulse}.

\subsection{Transformation between different frames}
\label{sec:frame_transf}

Before analyzing the effect of the laser pulses and the case of open interferometers, it is necessary to understand how wave packets transform under frame changes. Let us consider the Fermi-Walker coordinates $\big\{ c\, \tau_\text{c}, \mathbf{x} \big\}$ associated with a worldline $X^\mu_1 (\lambda)$, and a second worldline $X^\mu_2 (\lambda)$ with its corresponding Fermi-Walker coordinates $\big\{ c\, \tau'_\text{c}, \mathbf{x}' \big\}$ that intersects the first one at some point in spacetime. If we choose the origin for both comoving times ($\tau_\text{c}$ and $\tau'_\text{c}$) at the intersection point, the two sets of Fermi-Walker coordinates are related in the neighborhood of that point by the following Lorentz transformation:
\comment{
\begin{align}
\tau'_\text{c} \,&=\, \gamma_v\, \big(\, \tau_\text{c} - \mathbf{v} \cdot \mathbf{x} / c^2 \big)\,
\nonumber \\
& \approx\, \tau_\text{c} + \tau_\text{c} \, (v^2 / 2 c^2) - \mathbf{v} \cdot \mathbf{x} / c^2
\label{eq:lorentz_transf1a} , \\
\mathbf{x}' \,&=\, \mathbf{x}_\perp + \gamma_v\, \big(\, \mathbf{x}_{||} - \mathbf{v} \, \tau_\text{c}\, \big)
\,\approx\, \mathbf{x} - \mathbf{v} \, \tau_\text{c}
\label{eq:lorentz_transf1b} \, ,
\end{align}
}%
%\begin{align}
%\tau'_\text{c} \,&=\, \gamma_v\, \big(\, \tau_\text{c} - \mathbf{v} \cdot \mathbf{x} / c^2 \big)
%\,\approx\, \tau_\text{c} \,+\, \tau_\text{c} \, (v^2 / 2\, c^2) \nonumber \\
%& \qquad \qquad \qquad \qquad \qquad \ - \mathbf{v} \cdot \mathbf{x} / c^2
%\label{eq:lorentz_transf1a} , \\
%\mathbf{x}' \,&=\, \mathbf{x}_\perp + \gamma_v\, \big(\, \mathbf{x}_{||} - \mathbf{v} \, \tau_\text{c}\, \big)
%\,\approx\, \mathbf{x} - \mathbf{v} \, \tau_\text{c}
%\label{eq:lorentz_transf1b} \, ,
%\end{align}
where $U^\mu_2 = \big(c\, \gamma_v, \mathbf{v}\, \gamma_v \big)$ is the four-velocity of %the
worldline $X^\mu_2 (\lambda)$ at the intersection point expressed in terms of the Fermi-Walker frame associated with $X^\mu_1 (\lambda)$,
$\mathbf{x}_{||}$ and $\mathbf{x}_\perp$ denote respectively the parallel and perpendicular projections to $\mathbf{v}$,
and the last equality on the right-hand side of both equations is a good approximation for non-relativistic relative velocities, i.e.\ $v \ll c$. The derivation of Eqs.~\eqref{eq:lorentz_transf1a}--\eqref{eq:lorentz_transf1b} relies on the fact that the metric at the intersection point is the Minkowski metric in both reference frames. Corrections to the Minkowski metric in the neighborhood of the intersection point are small provided that $| \mathbf{a} \cdot \Delta\mathbf{x}| / c^2 \ll 1$ and $(\Delta x)^2 \ll \ell^2$ for both frames and have been neglected, but they can be included if necessary.

In this context, the transformation of wave packets under frame changes takes a particularly simple form in position representation. Indeed, given a wave packet with central trajectory $X^\mu_2 (\lambda)$, one can immediately find how the expressions in the two frames are related near the intersection point by making use of Eqs.~\eqref{eq:lorentz_transf1a}--\eqref{eq:lorentz_transf1b}:
%\begin{align}
%e^{- i \, m c^2 \tau'_\text{c} / \hbar}\,\, \psi_\text{c} \big( \mathbf{x}', \tau'_\text{c} \big)
%\,&\approx\, e^{- i \, m c^2 \tau_\text{c} / \hbar}\,\, e^{i \, m \mathbf{v} \cdot \mathbf{x} / \hbar}\, \nonumber \\
%& \qquad \qquad \, \times\, \psi_\text{c} \big( \mathbf{x} - \mathbf{v} \, \tau_\text{c}, \tau_\text{c} \big)\label{eq:lorentz_transf2} .
%\end{align}
\comment{
\begin{align}
e^{- i \, m c^2 \tau'_\text{c} / \hbar}\,\, \psi_\text{c} \big( \mathbf{x}', \tau'_\text{c} \big)
\,&\approx\, e^{- i \, m c^2 \tau_\text{c} / \hbar}\,\, e^{- i \, m v^2 \tau_\text{c} / 2 \hbar}\, \nonumber \\
& \quad \ \times\, e^{i \, m \mathbf{v} \cdot \mathbf{x} / \hbar}\,
\psi_\text{c} \big( \mathbf{x} - \mathbf{v} \, \tau_\text{c}, \tau_\text{c} \big)\label{eq:lorentz_transf2} .
\end{align}
}%
This result is %restricted to
only valid  for non-relativistic relative velocities because the non-relativistic version of Eqs.~\eqref{eq:lorentz_transf1a}--\eqref{eq:lorentz_transf1b} has been employed in its derivation. Note, however, that although suppressed by $1/c^2$,  \comment{the last two terms on the right-hand side of Eq.~\eqref{eq:lorentz_transf1a} do} give rise to a non-negligible contribution in the non-relativistic limit when multiplied by $m c^2$.

\subsection{Laser kicks and laser phases}
\label{sec:laser_phases}

Here we focus on idealized laser pulses whose finite duration as well as any dispersive effects are neglected and which amount to multiplying %, at a given instant of time,
the wave function for the atom's COM (in position representation) by a factor 
%$e^{i\, \mathbf{k}_\text{eff} \cdot \mathbf{y}}$
$i \exp \hspace{-0.2ex} \big( i\, \mathbf{k}_\text{eff} \cdot \mathbf{y} + i\, \varphi \big)$
expressed in the coordinate system $\{ t, \mathbf{y} \}$ where the laser modes are calculated.
%\highlight{(...)}
Since the wave-packet evolution is best described in a comoving frame defined by its central trajectory, it is convenient to rewrite this factor in the following equivalent form: $\exp \hspace{-0.2ex} \big[ \, i\, \mathbf{k}_\text{eff} \cdot \big( \mathbf{y} - \mathbf{X} (t_j) \big) \big] \comment{\times\,} i \exp \hspace{-0.2ex} \big[ \, i\, \varphi + i\, \mathbf{k}_\text{eff} \cdot \mathbf{X} (t_j) \big]$, where $\mathbf{X} (t_j)$ is the central position of the wave packet when the laser pulse is applied.
These quantities are expressed in the coordinate system considered above for the laser modes. On the other hand, in the Fermi-Walker frame associated with the wave packet's central trajectory the first factor becomes $\exp \hspace{-0.2ex} \big( i\, \tilde{\mathbf{k}}_\text{eff} \cdot \mathbf{x} \big)$, where $\tilde{\mathbf{k}}_\text{eff}$ takes into account the Doppler effect due to the velocity of the central trajectory with respect to the frame of the laser modes. Since in most applications this velocity is small and the laser pulses are based on two-photon processes such as Bragg diffraction where the first-order Doppler effect cancels out, we will consider $\tilde{\mathbf{k}}_\text{eff} \approx \mathbf{k}_\text{eff}$, %below,
but very similar conclusions are reached without this approximation (see below).
\comment{Furthermore, any corrections non-linear in $\mathbf{x}$ to the exponent that may arise from the change of coordinates have been neglected but will be briefly discussed at the end of this subsection.}

The effect of the laser pulse on an atomic wave packet with central trajectory $X^\mu_1 (\lambda)$ and centered wave packet $\psi_\text{c} \big( \mathbf{x}, \tau_\text{c} \big)$ can \comment{then} be easily understood thanks to the results of the previous subsection. If we choose the origin of the comoving time ($\tau_\text{c} = 0$) at the time when the laser pulse is applied, the product $e^{i\, \mathbf{k}_\text{eff} \cdot \mathbf{x}}\, \psi_\text{c} \big( \mathbf{x}, \tau_\text{c} \big)$ coincides with the right-hand side of Eq.~\eqref{eq:lorentz_transf2} for a velocity $\mathbf{v} = \mathbf{k}_\text{eff} / m \equiv \mathbf{v}_\text{rec}$. Therefore, we can conclude that the effect of the pulse was to change the central trajectory of the wave packet to a new trajectory $X^\mu_2 (\lambda)$ with a relative velocity $\mathbf{v}_\text{rec}$ with respect to the first one. And from that point on one can consider the Fermi-Walker frame associated with $X^\mu_2 (\lambda)$, and corresponding to the left-hand side of Eq.~\eqref{eq:lorentz_transf2}, in order to study the propagation of the wave packet.
Note that in the derivation we have assumed that the recoil velocity $\mathbf{v}_\text{rec}$ is non-relativistic, which is always the case in this context.
%, so that Eq.~\eqref{eq:lorentz_transf2} can be applied.
On the other hand, even if the Doppler effect for the pulse's wave vector were not neglected as done above, very similar conclusions would still be obtained. In that case one would need to consider $\mathbf{k}'_\text{eff}$ and the corresponding recoil velocity in the Fermi-Walker frame, but when transforming back to the frame of the laser modes, the change of the central trajectory would correspond to a momentum transfer of $\hbar\hspace{0.2ex} \mathbf{k}_\text{eff}$, in agreement with the expectations from momentum conservation.

Consequently, when calculating the wave-packet propagation along a given branch, the action of each laser pulse can be summarized as follows. First, the central trajectory experiences a momentum kick, as explained above, leading to a velocity change $\varepsilon_j \mathbf{v}_\text{rec}$ with $\varepsilon_j = \pm 1$ depending on whether the direct or inverse transition takes place, and where the index $j$ labels the pulse number.
In addition, the wave packet gets multiplied by a phase factor
%$i \exp \hspace{-0.2ex} \big( i\, \varepsilon_j\, \varphi_j + i\, \varepsilon_j\, \mathbf{k}_\text{eff} \cdot \mathbf{X}(t_j) \big)$,
\begin{equation}
i\, \exp \hspace{-0.2ex} \big[\, i\, \varepsilon_j\, \varphi_j + i\, \varepsilon_j\, \mathbf{k}_\text{eff} \cdot \mathbf{X}(t_j) \big]
\label{eq:laser_phase} ,
\end{equation}
where $\mathbf{X}(t_j)$ is the central position of the wave packet in the coordinate system used for the laser modes. The wave vector $\mathbf{k}_\text{eff}$ is often the same for all pulses, otherwise it needs to be explicitly labeled with the corresponding pulse number.
%It should be noted that
One also needs to take into account that
beam-splitter pulses generate an equal-amplitude superposition of two different wave packets: one following undeflected the original trajectory with no additional phase factor, and the other following the deflected central trajectory and multiplied by the phase factor \eqref{eq:laser_phase}. 

Idealized laser pulses have been considered here in order to concentrate on the key aspects of the associated matter-wave diffraction. Nevertheless, building up on the framework introduced above, a detailed study of the dynamics of the atomic wave packets during the diffraction process is possible. Indeed, one can take into account the finite pulse duration and off-resonant transitions as well as velocity selectivity and dispersion effects (i.e.\ momentum dependence of the diffraction amplitudes) by means of semi-analytical treatments such as those of Refs.~\cite{antoine06b,mueller08b,giese13}, or even numerical simulations, adapted to the Fermi-Walker frame.
\comment{In this respect, it can be useful to consider the Fermi-Walker frame for a trajectory interpolating between the initial and final central trajectories, particularly for large-momentum-transfer (LMT) beam-splitters and mirrors consisting of multiple pulses.}
%In this respect, it can be convenient to consider the Fermi-Walker frame for a ((smooth)) trajectory interpolating between the initial and final central trajectories, which can be particularly useful for large-momentum-transfer (LMT) beam-splitters and mirrors consisting of multiple pulses.
%%involving sequences with multiple pulses.

Furthermore, a more realistic treatment of the laser modes would require considering Gaussian beams rather than plane waves and, more generally, the propagation of electromagnetic waves in curved spacetime as well as the change of coordinates to the relevant Fermi-Walker frame.
The effects of curved spacetime on the laser modes are typically negligible in most applications, but the \comment{influence} of gravitational waves on the propagation of laser beams over long baselines does play a crucial role in proposed gravitational antennas involving a pair of atom interferometers a long distance apart %separated by a large distance
and interrogated by common laser beams.

In any case, all these effects associated with realistic laser pulses have rather limited impact on the doubly-differential measurement scheme presented in Sec.~\ref{sec:LPAI_redshift}. That is because they affect in the same way repeated realizations where only the initialization time is changed and cancel out when taking the difference.

\subsection{Open interferometers and separation phase}
\label{sec:separation_phase}

In \emph{open} interferometers the central trajectories of the two interfering wave packets at the exit port, $X^\mu_a (\lambda)$ and $X^\mu_b (\lambda)$, do not coincide. The separation is typically comparable to or smaller than the wave-packet width $\Delta x$ and much smaller the characteristic curvature radius $\ell$. Therefore, one can naturally consider the Fermi-Walker frame associated with a mid-trajectory where the two central trajectories become $- \delta \mathbf{X} (\tau_\text{c}) / 2$ and $\delta \mathbf{X} (\tau_\text{c}) / 2$ respectively.
It is, nevertheless, instructive to consider also more general frames where the spatial coordinates of the mid-term trajectory $\bar{\mathbf{X}} (t)$ and the associated momentum $\bar{\mathbf{P}} (t)$ do not vanish.
As long as the relative velocity between the different frames is non-relativistic
\highlight{[and approximating the Fermi-Walker metric at the exit port by the Minkowski metric, i.e.\  neglecting the curvature terms in Eqs.~\eqref{eq:FW_metric2a}--\eqref{eq:FW_metric2c}],}
one can make use of the results derived in Sec.~\ref{sec:frame_transf} and in particular of Eq.~\eqref{eq:lorentz_transf2}.
The modulus of the wave-packet superposition at the exit port can then be written in position representation as follows:
\begin{widetext}
\begin{align}
\big| \psi_\text{I} (\mathbf{x},t) \big|
&= \frac{1}{2}\, \Big| \, e^{i \phi_a} \psi_a (\mathbf{x},t) + e^{i \phi_b} \psi_b (\mathbf{x},t) \, \Big|
= \frac{1}{2}\, \Big| \, e^{i \phi_a} e^{i\, \mathbf{P}_a \cdot ( \mathbf{x} - \mathbf{X}_a ) / \hbar}
\psi_\text{c} \big( \mathbf{x} - \mathbf{X}_a, t \big)
+ e^{i \phi_b} e^{i\, \mathbf{P}_b \cdot ( \mathbf{x} - \mathbf{X}_b ) / \hbar}
\psi_\text{c} \big( \mathbf{x} - \mathbf{X}_b, t \big) \, \Big|
\nonumber \\
&= \frac{1}{2}\, \Big| \, e^{i \phi_a} e^{i\, \bar{\mathbf{P}} \cdot \delta \mathbf{X} / 2 \hbar} \,
e^{-i\, \delta \mathbf{P} \cdot ( \mathbf{x} - \bar{\mathbf{X}} ) / 2 \hbar}
\psi_\text{c} \big(\mathbf{x} - \bar{\mathbf{X}} + \delta \mathbf{X} / 2, t \big)
+ e^{i \phi_b} e^{- i\, \bar{\mathbf{P}} \cdot \delta \mathbf{X} / 2 \hbar} \,
e^{i\, \delta \mathbf{P} \cdot ( \mathbf{x} - \bar{\mathbf{X}} ) / 2 \hbar}
\psi_\text{c} \big(\mathbf{x} - \bar{\mathbf{X}} - \delta \mathbf{X} / 2, t \big) \, \Big|
%\nonumber \\
%&= \Big| \, e^{i \phi_a} \psi_\text{c} (\mathbf{x} - \mathbf{X},t) + e^{i \phi_b} \psi_\text{c} (\mathbf{x} - \mathbf{X},t) \, \Big|
\label{eq:exit_port4a} .
\end{align}
\end{widetext}
where we have taken into account that $\bar{\mathbf{X}} = (\mathbf{X}_a + \mathbf{X}_b) / 2$, $\bar{\mathbf{P}} = (\mathbf{P}_a + \mathbf{P}_b) / 2$, $\delta \mathbf{X} = \mathbf{X}_b - \mathbf{X}_a$ and  $\delta \mathbf{P} = \mathbf{P}_b - \mathbf{P}_a$,
and have factored out some common phase factors that do not contribute to the modulus.
%(indeed, ...)
Squaring the right-hand side of Eq.~\eqref{eq:exit_port4a} and integrating over space, one obtains the following representation-free expression for the detection probability in port I:
\begin{equation}
\langle \psi_\text{I} | \psi_\text{I} \rangle
= \frac{1}{2} \big(1 + C \cos \delta\phi' \big)
\label{eq:exit_port4b} .
\end{equation}
where $\delta\phi' = \phi_b - \phi_a + \delta\phi_\text{sep}$ with the separation phase $\delta\phi_\text{sep}$ given by
\begin{equation}
\delta\phi_\text{sep} = - \bar{\mathbf{P}} \cdot \delta \mathbf{X} / \hbar \,
\label{eq:sep_phase1} .
\end{equation}
In turn, the contrast $C$, which characterizes the amplitude of the oscillations as a function of the phase shift, corresponds to
\begin{equation}
C = \Big|\, \big\langle
\psi_\text{c} (t) \big| \hat{\mathcal{D}} (\delta \mathbf{X},\delta \mathbf{P})
\big| \psi_\text{c} (t) \big\rangle \, \Big| \leq 1 \,
\label{eq:contrast} ,
\end{equation}
where $\hat{\mathcal{D}} (\delta \mathbf{X},\delta \mathbf{P})$ denotes the displacement operator
\begin{equation}
\hat{\mathcal{D}} (\delta \mathbf{X},\delta \mathbf{P}) =
\exp \left( i\, \delta \mathbf{P} \cdot \hat{\mathbf{X}} / \hbar - i\, \delta \mathbf{X} \cdot \hat{\mathbf{P}} / \hbar \right)
\label{eq:displacement_op} ,
\end{equation}
and the inequality in Eq.~\eqref{eq:contrast} is saturated only
%and the second equality in Eq.~\eqref{eq:contrast} holds only
when $\delta \mathbf{X} =\delta \mathbf{P} = \mathbf{0}$. Hence, open interferometers lead to a loss of contrast for the oscillations of the integrated atom number in each exit port.
Note also that the expectation value in Eq.~\eqref{eq:contrast} is in general a complex quantity and one needs to add its argument to the phase shift $\delta\phi'$ in Eq.~\eqref{eq:exit_port4b}. However, for symmetric or antisymmetric centered wave packets, i.e.\ those %for which
with $\psi_\text{c} (- \mathbf{x}) = \pm \psi_\text{c} (\mathbf{x})$, the expectation value is \comment{real and there is no additional contribution to $\delta\phi'$ \cite{roura14}.}
%... one needs to replace $\delta\phi$ with $\delta\phi' = \delta\phi + ...$ in Eq.~\eqref{eq:exit_port4b}

%... In general, $\big\langle \psi_\text{c} (t) \big| \hat{\mathcal{D}}(\delta\boldsymbol{\chi}) \big| \psi_\text{c} (t) \big\rangle$ langle  ... rangle = C e i phi

When considering the Fermi-Walker frame associated with the mid-trajectory, we have $\bar{\mathbf{P}} = \mathbf{0}$ and the separation phase vanishes. On the other hand, the non-vanishing separation phase for $\bar{\mathbf{P}} \neq \mathbf{0}$ can be interpreted in terms of the proper-time difference between the two branches together with the relativity of simultaneity in different frames.
Indeed, let us consider simultaneous end points for the central trajectories of the two interfering wave packets in a reference frame where these have a non-vanishing velocity $\bar{\mathbf{v}}$. The proper-time difference between the two branches obtained in this way will differ from the analogous calculation in the Fermi-Walker frame of the mid-trajectory because the two end points considered above will no longer correspond to simultaneous events in this frame.
Making use of Eq.~\eqref{eq:lorentz_transf1a}, one finds that this time difference between the two end points is $\delta \tau_\text{c} \approx - \bar{\mathbf{v}} \cdot \delta \mathbf{X} / c^2$, which corresponds to a phase difference $- m c^2 \delta \tau_\text{c} /\hbar \approx m \mathbf{v} \cdot \delta \mathbf{X} / \hbar = \bar{\mathbf{P}} \cdot \delta \mathbf{X} / \hbar$. But this phase difference exactly cancels out the separation phase in Eq.~\eqref{eq:sep_phase1}, so that the total phase shift $\delta \phi'$ calculated in the frame with $\bar{\mathbf{v}} \neq \mathbf{0}$ coincides with the result obtained in the frame where the mid-trajectory is at rest.
This cancellation is important because the fraction of atoms detected in each port, which is determined by $\delta \phi'$ through Eq.~\eqref{eq:exit_port4b}, must be independent of the reference frame.
\comment{Notice also that although we have implicitly assumed a vanishing $\delta  \mathbf{v}$ in the previous argument, one can straightforwardly show that it also holds for $\delta  \mathbf{v} \neq \mathbf{0}$.}

\begin{figure}[h]
\begin{center}
\includegraphics[width=8.0cm]{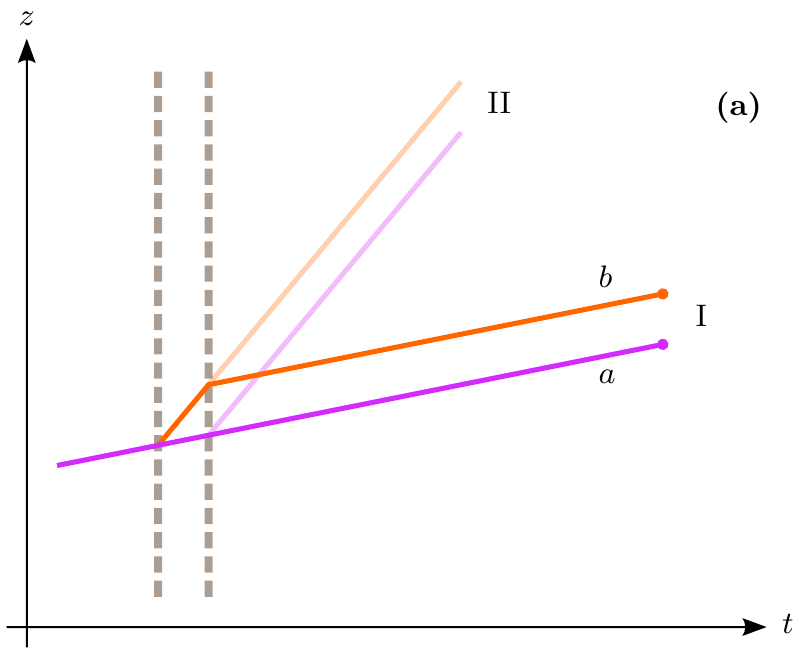}
%%\vspace{-1.0ex}
\end{center}
%\caption{}
%\label{fig:ramsey}
%\end{figure}
\vspace{1.0ex}
%\begin{figure}[h]
\begin{center}
\includegraphics[width=8.0cm]{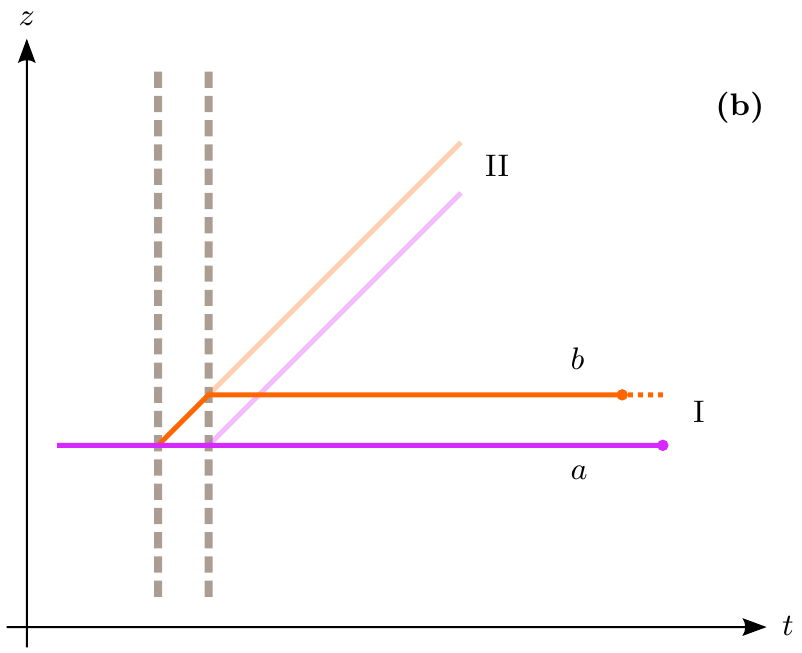}
%%\vspace{-1.0ex}
\end{center}
\caption{Central trajectories at the exit port of an open interferometer.
%Due to the relativity of simultaneity,
Simultaneous detection for non-vanishing central velocities (a) corresponds to %implies
non-simultaneous spacetime events in the comoving frame where the interfering wave packets are at rest (b). The resulting proper-time difference, corresponding to the dashed segment in (b), is exactly compensated by the non-vanishing separation phase $\delta\phi_\text{sep}$ in (a).}
\label{fig:ramsey}
\end{figure}

For simplicity the discussion of the previous paragraph is illustrated in Fig.~\ref{fig:ramsey} with the example of a Ramsey interferometer consisting of two $\pi /2$ pulses, but the argument only depends on the central trajectories at the exit port and holds for any interferometer configuration. 
In general, open interferometers arise from otherwise closed interferometers due to gravity gradients \cite{roura14}, a case which will be considered in Appendix~\ref{sec:gravity_gradient}, or due to changes of the pulse timing such as changing the time between the second and third pulses in a Mach-Zehnder interferometer by $\delta T$ \cite{muentinga13,sugarbaker13,roura14}.
(Rotations also give rise to open interferometers mainly along the transverse direction, i.e.\ along the direction orthogonal to $\mathbf{k}_\text{eff}$ \cite{hogan08,lan12,roura14}.)

We close this subsection with some brief remarks on the fringe pattern of the density profile at the exit ports of an open interferometer. For Gaussian wave packets, expanding BECs within the time-dependent Thomas-Fermi approximation or  \comment{wave packets in general at sufficiently late times} \cite{roura14} the spatial probability density at the exit port is given by
\begin{equation}
\big|\psi_\text{I} (\mathbf{x},t) \big|^2 \approx (1/2) \big[ 1 + C_\text{fr}\, \cos\, (\, \mathbf{k}_\text{fr} \cdot \mathbf{x} + \comment{\delta \phi'}) \big] \, \big|\psi_\text{c} (\mathbf{x},t) \big|^2
\label{eq:fringe_profile1} ,
\end{equation}
where $0 \leq C_\text{fr} \leq 1$, the vector $\mathbf{k}_\text{fr}$ is determined by a linear combination of $\delta \mathbf{X}$ and $\delta \mathbf{P}$ \cite{roura14}, and its modulus $| \mathbf{k}_\text{fr} | = 2 \pi / \lambda_\text{fr}$ is inversely proportional to the fringe spacing $\lambda_\text{fr}$.
In the above expression for the density profile we have assumed that the relative displacement $\delta X$ is small compared to the size of envelope and one can make the approximation $\big|\psi_\text{c} \big(\mathbf{x} - \delta \mathbf{X}/2 , t \big) \big| \, \big|\psi_\text{c} \big(\mathbf{x} + \delta \mathbf{X}/2 , t \big) \big| \approx \big|\psi_\text{c} (\mathbf{x},t) \big|^2$.
It is clear from Eq.~\eqref{eq:fringe_profile1} that  the phase shift \comment{$\delta \phi'$} can be extracted from the relative location of the fringes with respect to the envelope \cite{muentinga13,sugarbaker13,roura14}, or to the fringes of the other atomic cloud in a differential measurement \cite{asenbaum17,overstreet18,williams16}.

%\newpage

\appsection{Two-photon transition for the initialization pulse}
\label{sec:two-photon_pulse}

As seen in the laboratory frame, the initialization pulse employed in the interferometry scheme presented in Sec.~\ref{sec:LPAI_redshift} consists of a pair of counter-propagating equal-frequency laser beams with suitable polarizations and four-dimensional wave vectors $k^\mu_\pm = \big( |\mathbf{k}|, \pm  \mathbf{k} \big)$, where $|\mathbf{k}| = \omega_0 / 2\hspace{0.1ex} c$ corresponds to half the transition frequency between the two clock states \cite{alden14}. This can be naturally implemented with a beam injected along the vertical direction which is retro-reflected by a vibrationally isolated mirror.

\subsection{Atomic wave packet at rest}

For an atom at rest in the laboratory frame the counter-propagating beams resonantly drive the transition between the clock states through a two-photon process with vanishing momentum transfer to the atom's COM motion.
%More
Specifically, each laser beam drives allowed E1 and M1 dipole transitions to a third state which are both far off resonance \cite{alden14}. After adiabatically eliminating the third state, one is effectively left with a two-level system experiencing Rabi oscillations between the two clock states driven by two-photon processes.
%Indeed,
More precisely, the dynamics of the internal state $| \Phi (t) \rangle = g(t)\, | \mathrm{g} \rangle + e(t)\, | \mathrm{e} \rangle$ is governed by the following equation:
\begin{equation}
i
\left( \begin{array}{c}
\dot{e}(t) \\
\dot{g}(t)
\end{array} \right) =
\left( \begin{array}{cc}
\omega_0 & \frac{\Omega}{2}\, e^{-i \omega_0 t}\, e^{i \varphi}   \\
\frac{\Omega}{2}\, e^{i \omega_0 t} e^{-i \varphi} & 0
\end{array} \right)
\left( \begin{array}{c}
e (t) \\
g (t)
\end{array} \right)
\label{eq:two_photon1} .
\end{equation}
The angular Rabi frequency associated with the two-photon process is given by $\Omega = \Omega_1 \Omega_2 / 2 \Delta$, where $\Omega_1$ and $\Omega_2$ are the Rabi frequencies of the allowed off-resonant transitions and $\Delta$ is the detuning of these single-photon processes.
In turn, $\Omega_1$ is proportional to the transition matrix element of the electric-dipole operator and to the amplitude of the electric field, whereas $\Omega_2$ is proportional to the transition matrix element of the magnetic-dipole operator and to the amplitude of the magnetic field.
Therefore, $\Omega$ is overall proportional to the laser intensity and inversely proportional to the detuning $\Delta$.

On the other hand, the time-dependent phase factors in the off-diagonal matrix elements of Eq.~\eqref{eq:two_photon1} arise from contributions to the product of the electric and magnetic fields of the counter-propagating beams which are proportional to
$e^{-i \omega_0 t / 2}\, e^{i \varphi_1}\, e^{i \mathbf{k} \cdot \mathbf{x}}$ and
$e^{-i \omega_0 t / 2}\, e^{i \varphi_2}\, e^{-i \mathbf{k} \cdot \mathbf{x}}$ respectively.
The spatially dependent factors cancel out and the phase oscillations %of the product
are thus independent of the position along the beam, so that the spacetime hypersurfaces of constant phase coincide with the hypersurfaces of simultaneity in the laboratory frame.
%, as shown in Fig.~\ref{fig:}.
In this context the laser phase $\varphi = \varphi_1 + \varphi_2$ has a simple interpretation: $\varphi = \omega_0\, t_\text{i}$ corresponds to a time $t_\text{i}$ when the oscillating phase factor equals one.

In order to obtain the time evolution generated by Eq.~\eqref{eq:two_photon1}, it is convenient to work in a \emph{corotating} internal frame where the off-diagonal matrix elements become time independent. The corresponding change of basis is equivalent in this case to working in the interaction picture, where the evolution associated with the free Hamiltonian and corresponding to the diagonal matrix elements is \comment{absorbed}
%/ included
in the redefinition of the state vectors.
%In order to obtain the time evolution generated by Eq.~\eqref{eq:two_photon1}, it is convenient to work in the \comment{interaction picture}, where the evolution associated with the free Hamiltonian and corresponding to the diagonal matrix elements is ((absorbed / included)) in the redefinition of the state vectors.
By changing from the Schr\"odinger to the interaction picture at time $t_0$, which implies
%$g(t) \to g_\text{I} (t) = g(t)$ and $e(t) \to e_\text{I} (t) = e^{i \omega_0 (t - t_0)} e(t)$,
%\begin{subequations}
%\begin{align}
%g(t) &\,\to\, g_\text{I} (t) = g(t) , \label{eq:two_photon2a} \\
%e(t) &\,\to\, e_\text{I} (t) = e^{i \omega_0 (t - t_0)}\, e(t)
%\label{eq:two_photon2b} ,
%\end{align}
%\end{subequations}
\begin{equation}
\begin{aligned}
g(t) &\,\to\, g_\text{I} (t) = g(t) \, , \\
e(t) &\,\to\, e_\text{I} (t) = e^{i \omega_0 (t - t_0)}\, e(t) \, ,
\end{aligned}
\label{eq:two_photon2}
\end{equation}
Eq.~\eqref{eq:two_photon1} becomes
\begin{equation}
i
\left( \begin{array}{c}
\dot{e}_\text{I} (t) \\
\dot{g}_\text{I} (t)
\end{array} \right) =
\left( \begin{array}{cc}
0 & \frac{\Omega}{2}\, e^{-i \omega_0 (t_0 - t_\text{i})}   \\
\frac{\Omega}{2}\, e^{i \omega_0 (t_0 - t_\text{i})} & 0
\end{array} \right)
\left( \begin{array}{c}
e_\text{I} (t) \\
g_\text{I} (t)
\end{array} \right)
\label{eq:two_photon3} ,
\end{equation}
and can be easily solved.
Let us consider first a square pulse with constant $\Omega$ and pulse duration $\Delta t$.
If we assume that the atoms are initially in the ground state, i.e.\ that $| \Phi (t') \rangle_\text{I} = | \mathrm{g} \rangle$ for any time $t'$ before the pulse, the state at any time $t$ after the pulse is given by
%
%$| \Phi (t) \rangle_\text{I} = \cos \left( \Omega_\text{2ph} t / 2 \right) | \mathrm{g} \rangle
%- i\, e^{-i \omega_0 (t_0 - t_\text{i})} \sin \left( \Omega_\text{2ph} t / 2 \right) | \mathrm{e} \rangle$
%
%\begin{equation}
%| \Phi (t) \rangle_\text{I} = \cos \left( \frac{\Omega_\text{2ph} t}{2} \right) | \mathrm{g} \rangle
%- i\, e^{-i \omega_0 (t_0 - t_\text{i})} \sin \left( \frac{\Omega_\text{2ph} t}{2} \right) | \mathrm{e} \rangle
%\label{eq:two_photon3} .
%\end{equation}
%
\begin{equation}
| \Phi (t) \rangle_\text{I} = \cos \left( \frac{\Omega\, \Delta t}{2} \right) | \mathrm{g} \rangle
- i\, e^{-i \omega_0 (t_0 - t_\text{i})} \sin \left( \frac{\Omega\, \Delta t}{2} \right) | \mathrm{e} \rangle
\label{eq:two_photon4} .
\end{equation}
In particular, for $\Omega\, \Delta t = \pi / 2$ the pulse generates an equal-amplitude superposition of the same form as the initialized state in Eq.~\eqref{eq:initialization1}.
These results will also hold for smooth pulses %where the laser intensity changes continuously in time,
(and for any time-dependent intensity in general)
with the replacement $\Omega\, \Delta t \to \int_{t_1}^{t_2}\! dt' \,\Omega(t')$, valid for pulses with %compact
support within the interval $t_1 < t' < t_2$.
%following replacement in Eq.~\eqref{eq:two_photon4}:

One can then change back to the Schr\"odinger picture by inverting the transformations in Eqs.~\eqref{eq:two_photon2}. The result for a $\pi / 2$ pulse is given by
\begin{equation}
| \Phi (t) \rangle = \frac{1}{\sqrt{2}} \left( | \mathrm{g} \rangle
- i\, e^{-i \omega_0 (t - t_\text{i})} | \mathrm{e} \rangle \right)
\label{eq:two_photon5} ,
\end{equation}
for any time $t$ after the pulse. Eq.~\eqref{eq:two_photon5} shows that
what really matters is the ``laser phase'', encoded in the time $t_\text{i}$, rather than the exact initial time for a square pulse or the exact timing of the envelope $\Omega (t)$ for a smooth one.
\comment{Mirror vibrations} and laser phase noise will lead to fluctuations of $t_\text{i}$ from shot to shot, but this is not a problem for the scheme proposed in Sec.~\ref{sec:LPAI_redshift} because they affect in essentially the same way both interferometer branches.

Note that in order to avoid cumbersome expressions, we have focused on the internal states, but analogous results are obtained by considering the full wave function in the \comment{Fermi-Walker frame} and \comment{substituting} $\tau_\text{c}$ for $t$.
%For static central trajectories in a gravitational field one needs, in addition,
In addition, one needs
to take into account the gravitational redshift of the laser frequency $\omega_0$. The consequences of that will be discussed in the next subsection, where the case of a non-vanishing central velocity in the laboratory frame (e.g.\ for freely falling atoms) is also considered.

\subsection{Atomic wave packet with non-vanishing velocity %central velocity}
and gravitational redshift}
\label{sec:velocity_redshift}

For an atom with a non-vanishing velocity $\mathbf{v}$ parallel to $\mathbf{k}$  in the laboratory frame it is %instead
convenient to work in the comoving frame with coordinates $\big\{ \tau_\text{c}\, , \mathbf{x}' \big\}$, where the atom is at rest and the four-dimensional wave vectors for the two counter-propagating beams become $k^{\mu'}_\pm = \left( |\mathbf{k}'_\pm|, \mathbf{k}'_\pm \right)$ with
%$c\, |\mathbf{k}'_\pm| = \omega_\pm = (\omega_0 / 2) (1 \mp v/c)^{1/2} (1 \pm v/c)^{-1/2}$
%$\mathbf{k}'_\pm = \mathbf{k}\, (1 \mp v/c)^{1/2} (1 \pm v/c)^{-1/2} $
\begin{align}
c\, |\mathbf{k}'_\pm| = \omega_\pm = \frac{\omega_0}{2} \left( \frac{1 \mp v/c}{1 \pm v/c} \right)^{1/2}
\label{eq:two_photon11} .
\end{align}
When deriving the analog of Eq.~\eqref{eq:two_photon1} in this case, the time-dependent phase factor in the off-diagonal matrix elements is given by $e^{-i \bar{\omega}_\text{c} \tau_\text{c}}$ (and its complex conjugate) with
$\bar{\omega}_\text{c} = \omega_+ + \omega_-$ and
%\begin{subequations}
%\begin{align}
%\omega_+' + \omega_-' &= \omega_0 \, \Big(1 - (1/2)\, (v/c)^2 \Big)
%+ O \big( (v/c)^4 \big) \nonumber \\
%&\approx \big( \Delta m\, c^2 - m\, v^2 / 2 \big) / \hbar \\
%%\frac{m\, v^2}{2}   \\
%\mathbf{k}'_+ + \mathbf{k}'_- &= ...\, (\omega_0 / c) \, (\mathbf{v} / c) + O \big( (v/c)^2 \big) \nonumber \\
%&\approx ...\, \Delta m \, \mathbf{v} / \hbar
%\label{eq:two_photon12} .
%\end{align}
%\end{subequations}
\begin{align}
\omega_+ + \omega_- &= \omega_0 \, \Big(1 + (1/2)\, (v/c)^2 \Big)
+ O \big( (v/c)^4 \big) \nonumber \\
&\approx \big( \Delta m\, c^2 + \Delta m\, v^2 / 2 \big) / \hbar
\label{eq:two_photon12}\, .
\end{align}
Furthermore, the spatially dependent factors from the two beams no longer cancel out and one has an additional phase factor $e^{i \bar{\mathbf{k}}' \cdot \mathbf{x}'}$
%(and its complex conjugate)
with $\bar{\mathbf{k}}' = \mathbf{k}'_+ + \mathbf{k}'_-$ and
\begin{align}
\mathbf{k}'_+ + \mathbf{k}'_- &= - (\omega_0 / c) \, (\mathbf{v} / c) + O \big( (v/c)^3 \big) \nonumber \\
&\approx - \Delta m \, \mathbf{v} / \hbar
\label{eq:two_photon13}\, .
\end{align}
In fact, the total phase factor can be equivalently obtained by transforming to the comoving frame the phase factor in the laboratory frame:
%$e^{-i \omega_0 (t - t_\text{i})} = e^{-i \bar{\omega}' (t' - t'_\text{i})} \,
%e^{i \bar{\mathbf{k}}' \cdot (\mathbf{x}' - \mathbf{x}'_\text{i})}$.
$\exp \big( \! -i\, \omega_0 (t - t_\text{i})\, \big) = \exp \big( \! -i\, \bar{\omega}_\text{c} (\tau_\text{c} - \tau_\text{c}^\text{(i)}) + i\, \bar{\mathbf{k}}' \cdot (\mathbf{x}' - \mathbf{x}'_\text{i})\, \big)$.
In particular, the laser phase $\varphi$ is given by $\varphi = \omega_0\, t_\text{i} = \bar{\omega}_\text{c}\, \tau_\text{c}^\text{(i)} - \bar{\mathbf{k}}' \cdot \mathbf{x}'_\text{i}$ and the associated hypersurfaces of constant phase no longer correspond to simultaneity hypersurfaces in the comoving frame. Instead, a spatial separation $\Delta z$ along the beam direction implies a time difference
%$\Delta t_\text{i}' = \big( \bar{\mathbf{k}}' \cdot \Delta \mathbf{x}'_\text{i} \big)\, /\, \bar{\omega}' = - (v/c^2)\, \Delta z'$.
\begin{equation}
\Delta \tau_\text{c}^\text{(i)} = \big( \bar{\mathbf{k}}' \cdot \Delta \mathbf{x}'_\text{i} \big)\, /\, \bar{\omega}_\text{c} \approx - (v/c^2)\, \Delta z\label{eq:two_photon14} .
\end{equation}
%, which coincides with ...
As pointed out above, what determines the phases accumulated by the two clock states are the hypersurfaces of constant phase for the initialization pulse rather than the exact timing of its envelope.
For two wave packets with a separation $\Delta z$ of their central positions in a freely falling frame where they are both at rest this corresponds to the time difference given by Eq.~\eqref{eq:two_photon14} which is considered in Sec.~\ref{sec:ff_frame}.

As seen from Eq.~\eqref{eq:two_photon12}, we have $\bar{\omega}_\text{c} = \omega_0 - \delta$ with a detuning $\delta = - (\omega_0 / 2)\, (v/c)^2$, %$\delta = \Delta m\, v^2 / 2 \hbar$,
so that the two-photon process is no longer exactly on resonance when $v \neq 0$. One can still proceed similarly to the previous subsection and change to a \emph{corotating} internal frame through the following time-dependent unitary transformation:
\begin{equation}
\begin{aligned}
g(\tau_\text{c}) &\,\to\, \tilde{g} (\tau_\text{c}) = g(\tau_\text{c}) \, , \\
e(\tau_\text{c}) &\,\to\, \tilde{e} (\tau_\text{c}) = e^{i\, \bar{\omega}_\text{c}\, ( \tau_\text{c} - \tau_\text{c}^{(0)} ) }\, e(\tau_\text{c}) \, ,
\end{aligned}
\label{eq:two_photon15}
\end{equation}
which leads to an equation analogous to Eq.~\eqref{eq:two_photon3} but with the first diagonal matrix element replaced by $\delta$. For a square pulse this equation can be solved exactly and one finds a result analogous to Eq.~\eqref{eq:two_photon4} but with the modified Rabi frequency
\begin{equation}
\Omega_\text{eff} = \sqrt{\Omega^2 + \delta^2}
\label{eq:two_photon16}\, ,
\end{equation}
and Rabi oscillations with amplitude $(\Omega / \Omega_\text{eff}) < 1$. The oscillations are therefore largely suppressed far off resonance (for $\delta^2 \gg \Omega^2$), whereas nearly full-amplitude is recovered sufficiently close to resonance (for $\delta^2 \ll \Omega^2$).
As a quantitative example, for $\omega_0 = 2\pi \times \comment{400\, \text{THz}}$ and $v = 10\, \text{m/s}$ one has $\delta \approx 2\pi \times 0.2\, \text{Hz}$, which is much smaller than $\Omega = 2\pi \times \comment{25\, \text{Hz}}$, corresponding to a $\pi/2$ pulse of $10\, \text{ms}$ duration.

After inverting the unitary transformation in Eq.~\eqref{eq:two_photon15} and returning to the original Schr\"odinger picture, one is left with a state analogous to that in Eq.~\eqref{eq:two_photon5} but with $\delta$-dependent phase factors multiplying the ground and excited states.
%One the one hand, 
Firstly, a common factor $e^{- i  \hspace{0.2ex} \delta \, \Delta \tau_\text{c} / 2}$, where $\Delta \tau_\text{c} \approx \Delta t$ denotes the pulse duration in the comoving frame, multiplies both states.
%%$e^{- i (\delta/2) \Delta t}$
This will not affect differential phase-shift measurements of the two internal states.
%\comment{(In fact, it will not affect the separate phase shift for each internal state either because the factor is the same for both interferometer arms except for small differences in the pulse duration due to the finite speed of light.)}
%%Due to the finite speed of light, there are actually small differences in the pulse \comment{((timing / duration))} for the two branches of order $\Delta z / c$, but they give rise to an extra phase shift of order
%%$\Delta m\, v^2 (\Delta z / c) / \hbar \sim (v/c)\, \Delta m\, g\, \Delta z\, T / \hbar$ (assuming $v \sim g\, T$) which is largely suppressed by a factor $(v/c)$ compared to the signal of interest.
%One the other hand,
\comment{Secondly, a factor $e^{i  \, \delta \, ( \tau_\text{c}^\text{(f)} - \tau_\text{c}^\text{(i)} )}$, where $\tau_\text{c}^\text{(f)}$ denotes the time when the pulse ends, multiplies the excited state. Nevertheless, since $\delta \ll \omega_0$, the phase contribution from $e^{- i  \, \delta \, \tau_\text{c}^\text{(i)}}$ will be much smaller than from the factor $e^{i \, \omega_0 \, \tau_\text{c}^\text{(i)}}$, which decisively contributes to the signal of interest.
On the other hand, $e^{i  \hspace{0.2ex} \delta \, \tau_\text{c}^\text{(f)}}$ does not significantly affect the phase shift for the excited state either because the factor is the same for both interferometer arms except for small differences in the pulse duration %timing
of order $\Delta z / c$ due to the finite speed of light. Indeed, these give rise to an extra phase shift of order
$\Delta m\, v^2 (\Delta z / c) / \hbar \sim (v/c)\, \Delta m\, g\, \Delta z\, T / \hbar$ (assuming $v \sim g\, T$), which is largely suppressed by a factor $(v/c)$ compared to the signal of interest.}
Finally, there is an additional factor multiplying the \comment{excited} state that for $\delta \ll \Omega$ reduces to \comment{$e^{i \hspace{0.2ex} \delta / \Omega_\text{eff}}$}. In principle this does not affect the interferometer phase shift for the  \comment{excited} state provided that $\Omega_\text{eff}$ is the same for both branches, which requires that the laser intensity should be the same to a sufficiently high degree for a spatial separation $\Delta z$.
\comment{A quantitative estimate on how close $\Omega_\text{eff}$ for both branches should actually be is provided in the next paragraph as well as a simple method for relaxing this requirement.}

Similar conclusions will apply to other sources of detuning provided that they are sufficiently small. In particular, if we assume that the %(angular)
emission frequency of  the laser beam is $\omega_0 / 2$ in the laboratory frame, one will need to take into account the \emph{gravitational redshift} of the laser frequency at the position of the wave packet. For a weak and approximately uniform gravitational field, this implies a detuning %((contribution))
$\delta \approx \omega_0\, g\, L_z / c^2$, where $L_z$ is the height difference between the atomic wave packets and the laser emission (the effect of the small height difference between the two branches is addressed in the next paragraph), and it is of the same order as the detuning associated with the non-vanishing velocity (assuming $L_z \sim g\, T^2$ and $v \sim g\, T$ above). This will give rise to a phase-shift contribution $(\delta / \Omega_\text{eff})\, (\Delta \Omega_\text{eff} / \Omega_\text{eff}) \sim (\Delta m\, g\, L_z\, \Delta t / \hbar)\, (\Delta \Omega_\text{eff} / \Omega_\text{eff})$ for a change of the Rabi frequency $\Delta \Omega_\text{eff}$ due to differences \comment{between the laser %intensity
intensities for the two interferometer branches.}
The contribution will be much smaller than the right-hand side of Eq.~\eqref{eq:doubly_diff_ps} as long as $(\Delta \Omega_\text{eff} / \Omega_\text{eff}) \ll (\Delta z / L_z)\, (T / \Delta t)$. If we take $\Delta z = 1\, \text{cm}$, $L_z = 10\, \text{m}$, $T = 1\, \text{s}$ and $\Delta t = 10\, \text{ms}$, the condition becomes $(\Delta \Omega_\text{eff} / \Omega_\text{eff}) \ll 1/10$ and requires differences in the laser intensity well below the ten percent level.
Nevertheless, this requirement can be relaxed considerably by exploiting another obvious source of detuning: a change of the emission frequency by $\Delta \omega$
%$\Delta \omega / 2\pi$
leads to $\delta = - \Delta \omega$.
Indeed, by selecting $\Delta \omega = (\omega_0 / 2)\, \big( \! -v^2 / 2c^2 + g\, L_z / c^2 \big)$ one can cancel out the total detuning contribution due to non-vanishing velocity and gravitational redshift discussed above%
\footnote{\comment{This would also involve chirping the frequency shift $\Delta \omega$ to account for the time dependence of $v$ and $L_z$ during the pulse.}}.
%\footnote{There would not necessarily be a complete cancelation due to the finite duration $\Delta t$ of the pulse and small changes of velocity and height during the pulse, but any remaining contribution would be further suppressed by $(\Delta t / T)$.}.

%Note that
For the effects considered so far the detuning was the same for both interferometer branches. In contrast, the different gravitational redshift of the laser frequency for the two branches due to their vertical separation $\Delta z$ leads to detunings that differ by $\Delta \delta = \omega_0\, g\, \Delta z / c^2$. This gives rise to a phase shift of order $\Delta m\, g\, \Delta z\, \Delta t / \hbar$, which is suppressed by a factor $(\Delta t / T)$ compared to the signal of interest.
\comment{Moreover, it will cancel out in the doubly differential measurement when comparing the differential phase shifts for two different initialization times because it contributes in the same way to both of them.}
%\highlight{different ac Stark shift for the two branches due to intensity differences} %star

Finally, it should be noted that
the spatially dependent phase factor associated with the non-vanishing wave vector $\bar{\mathbf{k}}'$, given by Eq.~\eqref{eq:two_photon13}, imparts a small momentum kick $\hbar\, \bar{\mathbf{k}}'$ to the central trajectory of the wave packet of the excited state which corresponds to a recoil velocity $\Delta \mathbf{v} = - (\Delta m / m)\, \mathbf{v}$.
%Although the ((corresponding / resulting))
Although this residual recoil velocity is very small, it can give rise to a non-negligible phase-shift contribution. Nevertheless, it does not hinder the measurement scheme proposed in Sec.~\ref{sec:LPAI_redshift} because it affects both interferometer branches in the same way, as explained in Sec.~\ref{sec:residual_recoil}.
Furthermore, its contribution to the detuning $\delta$, which is quadratic in $\Delta m / m$, is negligible.

\appsection{Diffraction of atoms in internal-state superpositions}
%\appsection{Diffraction of internal superposition states}
\label{sec:diffraction_superpos}

\subsection{Magic-wavelength Bragg diffraction}

A natural way of diffracting atoms in a superposition of different internal states is by employing pulses of counter-propagating laser beams at the magic wavelength so that the resulting optical potential is the same for both internal states.
%This implies, in turn,
This implies that the Rabi frequency for the two-photon transition corresponding to Bragg diffraction of the atomic wave packets, which is proportional to the amplitude of the optical potential, is the same in both cases. Therefore, for a suitable duration and laser intensity such pulses can simultaneously act as $\pi/2$ pulses for the two internal states or any quantum superposition thereof.
On the other hand, since both internal states are diffracted by the same Bragg pulses, with a given $\mathbf{k}_\text{eff}$, the differential recoil discussed in Sec.~\ref{sec:diff_recoil}  due to the mass difference $\Delta m$ gives rise to a slightly open interferometer for at least one of the two internal states. Nevertheless, any phase-shift contribution associated with this cancels out in the doubly differential measurement of Sec.~\ref{sec:light-pulse_interf}
\comment{Moreover, the relative displacement between the interfering wave packets is too small to generate any significant loss of contrast.}

A disadvantage of this diffraction mechanism is that the two-photon Rabi frequency is rather small unless high laser intensities are employed. \comment{This is because the magic wavelength is typically largely detuned from any resonant transition and the contribution of any state (as a virtual state) to the Rabi frequency, which is given by $\Omega_1 \Omega_2 / 2 \Delta$ in terms of the corresponding single-photon Rabi frequencies $\Omega_1$ and $\Omega_2$, %$\Omega_\text{sp}$,
is suppressed by the frequency detuning $\Delta$ of the single-photon transitions.}
Although one could employ longer pulse durations, this is limited due to the velocity selectivity of the pulse because the maximum momentum width of wave packets that can be efficiently diffracted is inversely proportional to the pulse duration.

As an example, for $^{87}\text{Sr}$ a two-photon Rabi frequency of \highlight{$1\,\text{kHz}$ would require an intensity of about %$80\,\text{W/cm}^2$
$8\times10^1\,\text{W/cm}^2$ \cite{ye08}, so that more than $30\,\text{W}$ of laser power would be needed for a beam waist of $5\,\text{mm}$.} %size
%highlight{$120\,\text{kHz}$ would require an intensity of $10\,\text{kW/cm}^2$ \cite{ye08}, so that $XX\,\text{W}$ of laser power are needed for a beam diameter of $1\,\text{mm}$.} 
%highlight{$1\,\text{kHz}$ would require an intensity of $83\,\text{kW/cm}^2$ \cite{ye08}, so that $5\,\text{W}$ of laser power are needed for a beam waist of $2\,\text{mm}$.}
%As an example, for $^{87}\text{Sr}$ a two-photon Rabi frequency of \highlight{$1\,\text{kHz}$ would require an intensity of about $80\,\text{W/cm}^2$ \cite{ye08}, so that more than $5\,\text{W}$ of laser power are needed for a beam waist of $2\,\text{mm}$.} %size
Since such requirements on laser power are rather demanding, in the next subsection we will consider an alternative diffraction mechanism \highlight{with lower requirements but applicable only to fermionic isotopes.}

\subsection{Simultaneous single-photon clock transitions}

Single-photon transitions between the two clock states for group-II atoms such as Sr or Yb  are in principle forbidden by selection rules. However, for fermionic isotopes such as $^{87}\text{Sr}$ or $^{171}\text{Yb}$ they are actually weakly allowed due to hyperfine mixing of the excited clock state \cite{poli14,ludlow15}, %\cite{poli14,ludlow15,...}, %star
but still with a narrow linewidth of about \highlight{$1\,\text{mHz}$.}
%corresponding to a lifetime of \highlight{$1000\,\text{s}$. (??)}
Atom interferometers based on such transitions exhibit a number of interesting properties, including insensitivity to laser-phase noise in long-baseline gravitational antennas~\cite{graham13}. They have already been demonstrated experimentally for Sr atoms in a horizontal optical guide \cite{akatsuka17} and in free space \cite{hu17} (although in the latter case bosonic isotopes were employed and the transition was weakly allowed by applying an external magnetic field).

\begin{figure}[h]
\begin{center}
\includegraphics[width=8.0cm]{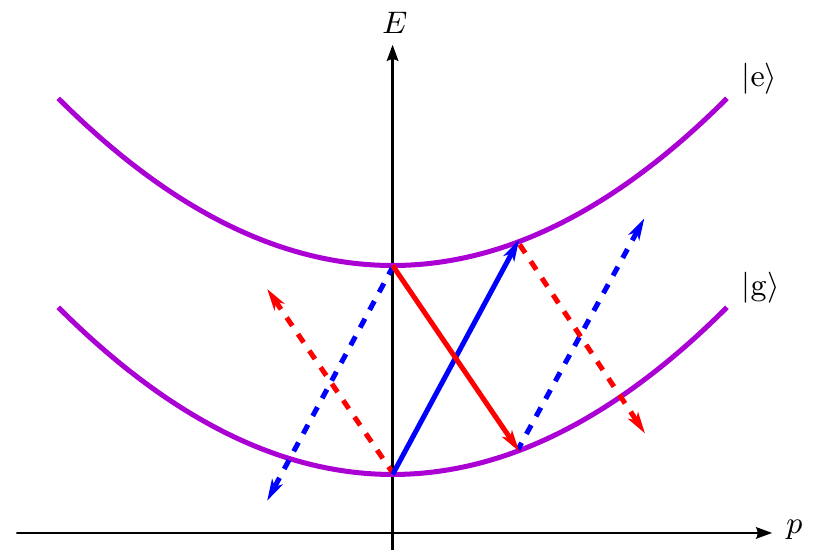}
%\vspace{-1.0ex}
\end{center}
\caption{Single-photon transitions between the clock states shown in an energy-momentum diagram that takes into account the internal energy and the COM motion (as described in the freely falling frame where the atoms are initially at rest). The frequencies of the two counter-propagating beams are shifted by $\omega_\text{rec}$ and $- \omega_\text{rec}$ from the clock frequency $\omega_0$ so that they drive resonant transitions of atoms initially at rest through absorption (blue arrow) and stimulated emission (red arrow) respectively.
%% !!!!!
\mbox{By applying} simultaneous $\pi /2$ pulses, one can generate an equal-amplitude superposition %of
involving an undiffracted wave packet plus a diffracted one with swapped internal states and a single-photon momentum transfer.
%In addition,
Relevant off-resonant transitions are additionally indicated with dashed lines.}
\label{fig:single-photon1}
\end{figure}

\begin{figure}[h]
\begin{center}
\includegraphics[width=8.0cm]{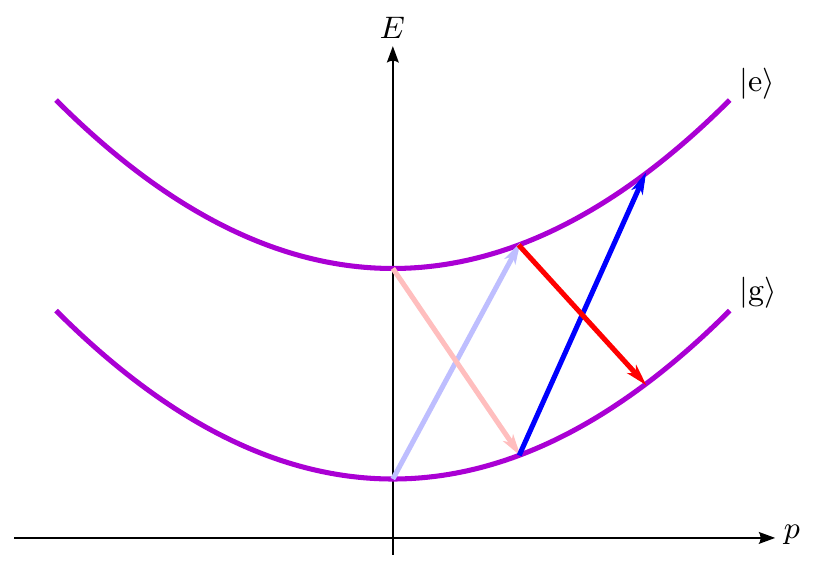}
%\vspace{-1.0ex}
\end{center}
\caption{If one shifts the frequencies of the two counter-propagating beams by $3\, \omega_\text{rec}$ and $-3\, \omega_\text{rec}$, one can resonantly address the diffracted atoms in Fig.~\ref{fig:single-photon1} with an additional single-photon momentum transfer in the same direction while swapping again the internal states.
When simultaneous $\pi$ pulses with these frequencies are applied right after the simultaneous $\pi / 2$ pulses of Fig.~\ref{fig:single-photon1}, the net result is to leave the internal states unchanged but generate an equal-amplitude superposition, as far as the COM motion is concerned, of an undiffracted wave packet and a diffracted one with a total momentum transfer $\hbar\, \mathbf{k}_\text{eff}$ corresponding to twice the single-photon momentum.}
%\comment{... comoving frame ...}
\label{fig:single-photon2}
\end{figure}

Interestingly, these single-photon transitions can also be exploited for quantum-clock interferometry as we explain next. %Indeed,
The proposed diffraction mechanism is based on a pair of counter-propagating laser beams that drive single-photon transitions between the two clock states with the momentum transfer in the same direction, as indicated by the red and blue arrows in Fig.~\ref{fig:single-photon1}. By sending simultaneous pulses for %each beam
the two beams and choosing the intensity and duration so that they both correspond to $\pi/2$ pulses, one can create an equal-amplitude superposition of an undiffracted atomic wave packet and a diffracted one with swapped internal states. After repeating the process with a subsequent pair of simultaneous $\pi$ pulses as shown in Fig.~\ref{fig:single-photon2}, the diffracted wave packet gets an additional momentum kick and its internal states are swapped again. The net effect is therefore a diffracted wave packet with the same internal state as the undiffracted one but a total momentum transfer $\hbar\, \mathbf{k}_\text{eff}$ corresponding to twice the single-photon momentum. 
This momentum transfer can be increased by applying a sequence of %further
additional pairs of $\pi$ pulses.

Besides driving a resonant transition, there are also off-resonant transitions associated with each pulse. The relevant ones for the first pair of pulses are indicated with dashed lines in Fig.~\ref{fig:single-photon1}. They are \highlight{off-resonant %detuned
by $2\, \omega_\text{rec}\,$,} where $\omega_\text{rec} = \hbar \mathbf{k}^2 / 2 m \approx (\omega_0 / 2)\, (\Delta m / m)$ is the recoil frequency in this case and $\hbar \mathbf{k}$ is the single-photon momentum. Hence, for a Rabi frequency of the resonant processes  sufficiently small compared to $\omega_\text{rec}$ and a Gaussian pulse envelope, these spurious transitions can be exponentially suppressed \cite{mueller08b,giese13}. This, in turn, requires a momentum width of the atomic wave packet which is narrow enough compared with $\hbar k$ to guarantee a high diffraction efficiency for all its Fourier components.

For the second pair of pulses their frequencies need to be \highlight{additionally shifted by $- 2\, \omega_\text{rec}$ and $2\, \omega_\text{rec}$} respectively to account for the non-vanishing momenta of the wave packets diffracted by the first pair. (Alternatively, this can be understood as the Doppler shift that arises when transforming to the reference frame where these wave packets are at rest and where the situation, including the effects of the off-resonant transitions, becomes identical to that discussed for the first pulse.)
Because of these frequency shifts the net momentum transfer experienced by the two internal states after the two pairs of pulses is actually slightly different with a relative difference of order $\Delta m / m$, but any phase-shift contribution caused by that will cancel out in the doubly differential measurement.
The recoil velocities, which are additionally affected by the differential recoil due to the mass difference $\Delta m$, will also exhibit a relative difference of the same order and give rise to slightly open interferometers, \comment{but the resulting relative displacement between the interfering wave packets is too small to generate any significant loss of contrast.}

Since the experimental set-up will typically involve a retro-reflection mirror, for each laser pulse there will be an additional one with the same frequency in the laboratory frame but propagating in the opposite direction. Nevertheless, if the pulses are applied when the atoms are far from the apex of the atomic fountain and hence moving with a velocity of several m/s, only one of the two pulses will be resonant in each case. That is because when transforming to the rest frame of the atoms, one of the two equal-frequency pulses
%propagating in opposite directions
is redshifted while the other is blueshifted.
It should also be noted that common laser-phase noise will affect the two resonant transitions in Fig.~\ref{fig:single-photon1} with an opposite sign and phase noise due to mirror vibrations will affect only one of the two. Fortunately, the subsequent pair of pulses in Fig.~\ref{fig:single-photon2}, which acts on swapped internal states, will have the reversed effect (except for high-frequency laser-phase and vibration noise with frequencies comparable to the inverse of the pulse duration, which are easier to mitigate), so that these phase contributions cancel out in the differential phase-shift measurements of the two internal states.

The ac Stark shift associated with off-resonant transitions such as those shown with dashed lines in Fig.~\ref{fig:single-photon1} can lead to different phase contributions for the the two interferometer branches. Interestingly, however, the main shifts experienced by the two clock states coincide and their contributions \comment{cancel out} in the differential phase-shift measurement. (This is also true for the shifts induced by the additional beams that are present in a retro-reflection set-up.) Furthermore, any phase-shift contributions from light shifts cancel out in the doubly differential measurement provided that the laser intensities are stable from shot to shot. A more detailed investigation of these questions will be presented elsewhere.

%%%%%
%
%The ac Stark shift associated with off-resonant transitions such as those shown with dashed lines in Fig.~\ref{fig:single-photon1} can lead to different phase contributions for the two internal states and the two interferometer branches that yield a non-vanishing net contribution to the differential measurement. They are % , however,
%suppressed by a factor of order $(\Omega / \omega_\text{rec})$ whereas those caused by the retro-reflection scheme are further suppressed because they involve Doppler shifts for velocities of m/s rather than cm/s.
%More importantly, such contributions do cancel out in the doubly differential measurements provided that the laser intensities are stable from shot to shot. A detailed quantitative investigation of these effects and the corresponding stability requirements for the laser intensities will be presented elsewhere.
%
%%%%%

%off-resonant processes / transitions ... \highlight{((spatially separate spurious paths not contributing)) ...}

\appsection{Gravity gradients and proper-time difference}
\label{sec:gravity_gradient}

\begin{figure}[h]
\begin{center}
\includegraphics[width=8.0cm]{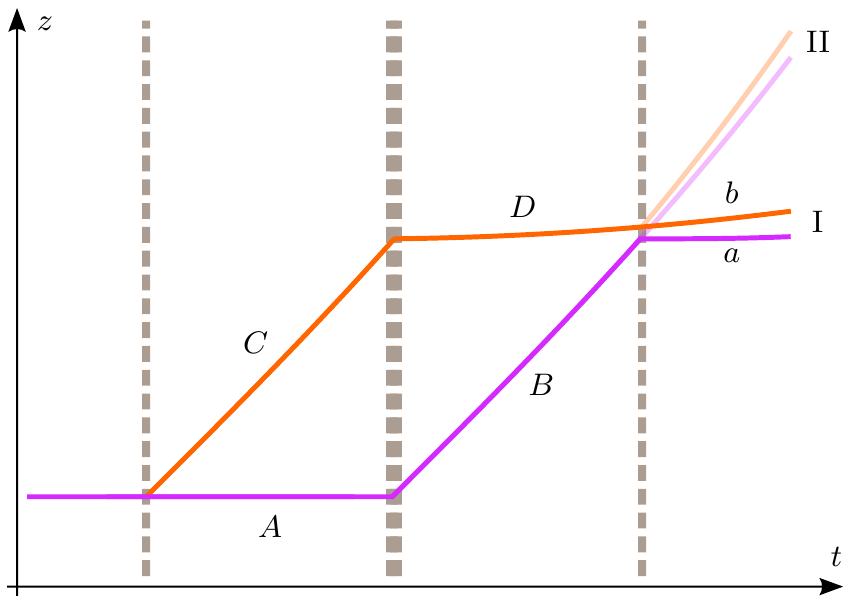}
%\vspace{-1.0ex}
\end{center}
\caption{Central trajectories for a Mach-Zehnder interferometer in the freely falling frame where the atomic wave packet is initially at rest at $z = 0$. They are no longer straight lines (except for segment $A$) due to the tidal forces associated with the gravity gradient. From the figure it is clear that the propagation phases for segments $B$ and $C$ are the same and so are the spatially dependent parts of the laser phases for branches $a$ and $b$.}
\label{fig:gravity_gradient}
\end{figure}

In this Appendix we analyze the effect of gravity gradients on the proper-time difference in a Mach-Zehnder interferometer. For that purpose it is particularly convenient to consider the freely falling frame where the atomic wave packet is initially at rest and where the main results can be understood in simple terms \cite{roura17b}. The central trajectories in this frame are shown in Fig.~\ref{fig:gravity_gradient}. %((depicted / displayed))

Before starting the analysis in the freely falling frame, it is important to understand the effect that the change of %reference
frame has on the laser phases. Given a laser phase
%$e^{i \varphi + i \mathbf{k}_\text{eff} \cdot \mathbf{x}}$
$\exp \hspace{-0.2ex} \big( i\, \varepsilon_j\, \varphi_j + i\, \varepsilon_j\, \mathbf{k}_\text{eff} \cdot \mathbf{X} (t_j) \big)$
for the $j$-th pulse in the laboratory frame and a non-relativistic frame transformation%
\footnote{\comment{The relativistic corrections would give rise to extra terms suppressed by powers of $1/c$.}}
characterized by $t \to t' = t$ and $\mathbf{x} \to \mathbf{x}' = \mathbf{x} - \mathbf{x}_0 (t)$, the laser phase takes an analogous form in terms of the new coordinates, namely
%$e^{i \varphi' + i \mathbf{k}_\text{eff} \cdot \mathbf{x}'}$
$\exp \hspace{-0.2ex} \big( i\, \varepsilon_j\, \varphi'_j + i\, \varepsilon_j\, \mathbf{k}_\text{eff} \cdot \mathbf{X}' (t_j) \big)$ with 
\begin{equation}
\varphi'_j = \varphi_j + \mathbf{k}_\text{eff} \cdot \mathbf{x}_0 (t_j)
\label{eq:gg1} .
\end{equation}
Thus, when working in the freely falling frame mentioned above, the information on the initial position and velocity of the atomic wave packet with respect to the laboratory frame is entirely contained in the phases $\varphi'_j$ associated with the various laser pulses.
For small time-independent gravity gradients the trajectory $\mathbf{x}_0 (t)$, which corresponds to the central trajectory that the atomic wave packet would follow in the laboratory frame in absence of kicks from the laser pulses, is well approximated by
%$\mathbf{x}_0 (t) \approx \left( \mathbf{x}_0 (0) + \mathbf{v}_0 (0)\, t \right) + \mathbf{g}\, t^2 / 2
%+ ( \Gamma \, t^2 ) \left( \mathbf{x}_0 (0) + \mathbf{v}_0 (0)\, t \right)$
\begin{align}
\mathbf{x}_0 (t) \, \approx \,\, & \big( \mathbf{x}_0 (0) + \mathbf{v}_0 (0)\, t \big) + \frac{1}{2} \mathbf{g}\, t^2
\nonumber \\
&+ \frac{1}{2} \big( \Gamma \, t^2 \big) \left( \mathbf{x}_0 (0)
+ \frac{1}{3} \mathbf{v}_0 (0)\, t + \frac{1}{12} \mathbf{g}\, t^2 \right)
\label{eq:gg2} ,
\end{align}
up to terms of higher order in $\big( \Gamma \, t^2 \big)$.
%where terms of higher order in $\big( \Gamma \, t^2 \big)$ have been neglected.
The contribution of the spatially independent part of the laser phases to the phase shift of the Mach-Zehnder interferometer is then given by 
\begin{align}
\delta \varphi' & = \varphi' (2T) - 2\, \varphi' (T) + \varphi' (0) \nonumber \\
& = \delta \varphi + \mathbf{k}_\text{eff} \cdot \mathbf{g}\, T^2 \nonumber \\
& \quad + \mathbf{k}_\text{eff}^\text{T}\, \big( \Gamma \, T^2 \big) \Big( \mathbf{x}_0 (0) + \mathbf{v}_0 (0)\, T + \frac{7}{12} \mathbf{g}\, T^2 \Big)
\label{eq:gg3} ,
\end{align}
where $T$ is the time between pulses.

%%\textcircled{\small 1}
%{\raisebox{.1pt}{\textcircled{\raisebox{-.4pt} {\small 1}}}}
%{\raisebox{.1pt}{\textcircled{\raisebox{-.4pt} {\small 2}}}}
%{\raisebox{.1pt}{\textcircled{\raisebox{-.4pt} {\small 3}}}}
%{\raisebox{.1pt}{\textcircled{\raisebox{-.4pt} {\small 4}}}}

Having discussed the transformation of the laser phases, let us now focus on the phase shift between the interfering wave packets at the first exit port~(I). 
In the freely falling frame, the contributions of the spatially dependent laser phases $\varepsilon_j\, \mathbf{k}_\text{eff} \cdot \mathbf{X}' (t_j)$ to the phase shift cancel out because $\mathbf{X}_b' (0) = \mathbf{X}_a' (T) = 0$ and $\mathbf{X}_b' (T) = \mathbf{X}_a' (2T)$ as can be easily seen in Fig.~\ref{fig:gravity_gradient}.
Therefore, besides $\delta \varphi'$ the phase shift is essentially due to the proper-time difference between the central trajectories of the two arms ($a$ and $b$) and the corresponding difference of the propagation phases.
This difference can, in fact, be easily determined to leading order in $\Gamma$. First of all, one notes that the segments $B$ and $C$ belonging to the two branches are equivalent and lead to identical propagation phases whose contributions to the phase shift cancel out. Hence, besides $\delta \varphi'$ the phase shift is entirely given by the difference of the propagation phases for segments $A$  and $D$,
which can be evaluated using Eq.~\eqref{eq:action2} with $U(\mathbf{X}') = - (m / 2)\, \mathbf{X}^{\prime \, \text{T}} \Gamma \, \mathbf{X}'$. Moreover, the contributions of the kinetic and potential terms clearly vanish for segment $A$, for which $\mathbf{X}'(t) = 0$. We are therefore just left with the contribution from segment $D$. To leading order in $\Gamma$ only the potential term evaluated along the unperturbed trajectory $\mathbf{X}'(t) = \mathbf{v}_\text{rec} \, T$ contributes and the total phase shift is \comment{given by}
\begin{equation}
\delta \phi \,=\, \delta \varphi' + \frac{\hbar}{2\, m}\,
\mathbf{k}_\text{eff}^\text{T}\, \Gamma \, \mathbf{k}_\text{eff} \, T^3
+ O \Big( \big( \Gamma \, T^2 \big)^2 \Big)
\label{eq:gg4} .
\end{equation}
Furthermore, the main conclusions (no contribution from the spatially dependent laser phases and non-trivial contribution only from segment $D$) hold to all orders in $\Gamma$.
In fact, they hold even in a fully relativistic treatment as long as the gravity-gradient tensor $\Gamma$ is time independent.

It is important to note that gravity gradients lead to a relative displacement $\delta \mathbf{X}' \approx \mathbf{v}_\text{rec}\, T\, \big( \Gamma\, T^2 \big)$ between the interfering wave packets at each exit port of a standard Mach-Zehnder interferometer%
\footnote{\comment{One can achieve a vanishing relative displacement, $\delta \mathbf{X}' = \mathbf{0}$, through the technique proposed in Ref.~\cite{roura17a} or by slightly modifying the time between the second and third pulse by a suitable amount $\delta T$, but this would also eliminate the contribution of interest in Eq.~\eqref{eq:gg4}.}}
as illustrated in Fig.~\ref{fig:gravity_gradient}. This implies an %extra
additional phase-shift contribution from the separation phase discussed in Appendix~\ref{sec:separation_phase}. However, in the freely falling frame considered here this contribution vanishes to leading order in $\Gamma$ for the first exit port because $\bar{\mathbf{P}}' = ( \mathbf{P}'_a + \mathbf{P}'_b ) / 2 \approx m \mathbf{v}_\text{rec} \big( \Gamma\, T^2 \big)$ plus higher-order corrections.
There will be, on the other hand, a non-vanishing contribution to second order in
%$\Gamma$
$\big( \Gamma\, T^2 \big)$
and higher which can still be interpreted in terms of the proper-time difference between the two branches as well as the relativity of simultaneity for different inertial frames, as explained in Appendix~\ref{sec:separation_phase}.

So far we have focused on the first exit port. Nevertheless, analogous results can be obtained for the second exit port (II) by considering an alternative freely falling frame where the atomic wave packet on branch $b$ is at rest after the first beam-splitter pulse.

We close this Appendix with some brief remarks on how the phase-shift in Eq.~\eqref{eq:gg4}, associated with the different tidal forces acting on the two interferometer arms, was actually measured in the experiments reported in Ref.~\cite{asenbaum17}. What was directly measured was the differential phase shift of two simultaneous interferometers at different heights and interrogated by common laser beams in a gradiometer set-up. In this way the laser phase noise and the effect of the vibrations of the retro-reflection mirror canceled out in the differential measurement.
However, the signal of interest, namely the second term on the right-hand side of Eq.~\eqref{eq:gg4}, would also cancel out. Therefore, for half of the measurements a stack of lead bricks was placed near one of the two interferometers to modify the gravity gradient experienced by the atoms in this interferometer. After subtracting the differential phase shifts for measurements with and without lead bricks, one is left with a contribution analogous to that in Eq.~\eqref{eq:gg4} but with $\Gamma$ replaced by the difference $\Delta \Gamma$ of the gravity gradients in both cases. %$\Delta \Gamma$.
While additional contributions analogous to the third term on the right-hand side of Eq.~\eqref{eq:gg3} but with $\Delta \Gamma$ instead of $\Gamma$ remain too, these can be subtracted out by modeling the initial position and velocity for that interferometer as well as the change of $\Gamma$ caused by the lead bricks. Furthermore, independently of that and due to its quadratic (rather than linear) dependence on $\mathbf{k}_\text{eff}$, the contribution of interest
%, corresponding to the second term on the right-hand side of Eq.~\eqref{eq:gg4},
in Eq.~\eqref{eq:gg4}
can be clearly identified by performing a series of experiments where $\mathbf{k}_\text{eff}$ is changed while leaving the remaining parameters unchanged.

\appsection{From QFT in curved spacetime to single-particle quantum mechanics}
%\appsection{From quantum field theory in curved space to single-particle quantum mechanics}
\label{sec:QFTtoQM}

%Our
The description of the external dynamics of the two-level atom
employed in the rest of the paper
can be regarded as single-particle relativistic quantum mechanics in curved spacetime and it can be derived, under appropriate conditions, from QFT in curved spacetime. In this appendix we briefly outline the key aspects of such a derivation
%that such a derivation would entail
and the conditions that need to be fulfilled.

QFT in curved spacetime is intrinsically a many-body theory with non-trivial particle-creation effects, even for free fields, and the absence of a preferred vacuum (and the associated notion of particles as quantum excitations thereof) for generic spacetimes. A single-particle description is therefore only possible in a suitable regime where non-trivial second-quantization effects are not important. Indeed, provided that the %\comment{reduced}
Compton wavelength \comment{$\lambda_m = h / m c$} %\comment{$\lambdabar_m = \hbar / m c$} 
associated with the rest mass of the atom is much smaller than the curvature radius $\ell$, one can consider \emph{adiabatic vacua} where vacuum ambiguities and particle creation are exponentially suppressed \cite{birrell94}. Furthermore, since similar effects arise for accelerated observers even in flat space \cite{unruh76,birrell94}, an additional condition is required when considering accelerated central trajectories: the so-called Unruh temperature $T_\text{U}$, which is proportional to the acceleration, should be much smaller than the rest mass, i.e.\ $k_\text{B }T_\text{U} \ll m c^2$. This can be equivalently rewritten as the following restriction on the acceleration: \comment{$a\, \lambda_m / c^2 \ll 1$.} %\comment{$a\, \lambdabar_m / c^2 \ll 1$.}

The two internal states can be represented by two different fields whose masses differ by $\Delta m = \Delta E / c^2$, and the electromagnetic coupling driving transitions between them
%both states
can be modeled by a nonlinear interaction term involving the electromagnetic field and the fields of the two internal states.
This leads to an additional requirement on the curvature radius $\ell$, which should be much larger than the photon wavelength $\lambda_\text{ph}$ corresponding to the transition between the two internal states. The requirement is actually stronger than for the Compton wavelength because $\Delta m \ll m$, and guarantees that an adiabatic vacuum can be defined for the electromagnetic mode. It also guarantees that any spontaneous excitation (or decay) induced by the time dependence of the effective coupling in the curved background spacetime is negligible.
Similarly, demanding the absence of such transitions due to the Unruh effect imposes the stronger restriction \comment{$a\, \lambda_\text{ph} / c^2 \ll 1$} on the acceleration of the central trajectory, so that $k_\text{B }T_\text{U} \ll \Delta E$.

If we focus for simplicity on spin-zero particles (such as $^{88}\text{Sr}$ atoms), scalar fields can be employed for the second-quantization description of the atoms. These fields satisfy the Klein-Gordon equation,
\highlight{which may include an external potential $V_\text{ext}(x^\mu)$ directly related to the potential $V(x^\mu)$ considered in Sec.~\ref{sec:propagation_forces} and Appendix~\ref{sec:external_forces}
in order
to account for external forces.}
Among the solutions of the Klein-Gordon equation one can consider the subspace %subset
generated by the positive frequency modes, identified up to exponentially suppressed ambiguities as long as the conditions mentioned above are fulfilled. Given then any initial wave packet within this subspace with size $\Delta x \ll \ell$ and non-relativistic momentum width,
%corresponding to a %non-relativistic
%velocity spread $\Delta v \ll c$, 
one can show that the Klein-Gordon equation governing its evolution reduces in the corresponding Fermi-Walker frame to the Schr\"odinger equation for the centered wave packet derived in Appendix~\ref{sec:wp_propagation}.

For non-vanishing spin one can proceed analogously by considering higher-spin fields and their corresponding equations of motion (Dirac, Proca %\highlight{, Rarita-Schwinger, and
\highlight{or Bargmann-Wigner in general}). This will generically lead to spin-curvature coupling terms similar to those appearing in the Mathisson-Papapetrou equation, which describes the motion of a spinning test particle in general relativity and leads to deviations from geodesic motion \cite{mathisson37,papapetrou51,dixon74,leclerc05}.
Nevertheless, for spins which are a small multiple of $\hbar$ the contribution of such terms is highly suppressed, typically by a factor \comment{$(\lambdabar_m / \Delta x) \lesssim 10^{-12}$}, compared to the tidal forces associated with the usual gravity-gradient term in the equation of motion.

When considering the interaction with the electromagnetic field, the details of the internal dynamics can be encoded in a form factor which involves the transition-matrix element between the two internal states of the  electric-dipole operator and which couples to the electric field. This holds for E1 transitions, but it can be generalized to magnetic-dipole transitions as well as higher electric and magnetic multipoles.
Note that the calculation in curved spacetime of the internal energy eigenstates, their energies and the relevant transition-matrix elements gives rise to small corrections proportional to the Riemann tensor of order $(a_\text{B} / \ell)^2$, where $a_\text{B}$ is the Bohr radius \cite{parker80}.
\highlight{Moreover, vacuum fluctuations will lead to a modified Lamb-shift including also small corrections proportional to the Riemann tensor
%(and its derivatives) divided by the corresponding powers of the electron mass
divided by the square of the electron mass, %$m_\text{e}$
which are of order  $(\lambda_e / \ell)^2$.}

\comment{Finally, one should also keep in mind that there is no unique relativistic definition of the COM for a composite system \cite{moeller49,dixon70}. Several proposals exist which feature each desirable properties, but they are mutually incompatible and in some cases frame dependent. Nevertheless, the corresponding worldlines for the different proposals are confined within a \emph{worldtube} with a radius comparable to the Compton wavelength of the composite particle.}

\vspace{1.5ex}

%The next line should be commented out for the titles to appear.
\bibliographystyle{apsrev4-1}
\bibliography{literature2}

%\newpage
%\mbox{}

\end{document}